\documentstyle[12pt,epsf,epsfig,amsmath]{article}
\newfont{\thiplo}{msbm10 scaled\magstep 2}
\newfont{\gothic}{eufb10 scaled\magstep 2}
\newfont{\unc}{eurb10} 
\newskip\humongous \humongous=0pt plus 1000pt minus 1000pt
\def\caja{\mathsurround=0pt}\def\eqalign#1{\,\vcenter{\openup1\jot \caja
        \ialign{\strut \hfil$\displaystyle{##}$&$ 
        \displaystyle{{}##}$\hfil\crcr#1\crcr}}\,}
\newif\ifdtup

\def\eqright #1\cr{\noalign{\hfill$\displaystyle{{}#1}$}}
\def\eqleft #1\cr{\noalign{\noindent$\displaystyle{{}#1}$\hfill}}

\def\oldreffmt#1{\rlap{[#1]} \hbox to 2\parindent{}}

\def\figfmt#1{\rlap{Figure {#1}} \hbox to 1in{}}

\def\sectioneq{\def\theequation{\thesection.\arabic{equation}}{\let
\holdsection=\section\def\section{\setcounter{equation}{0}\holdsection}}}%

\newcounter{holdequation}

\def\auto{\eqno(\refstepcounter{equation}\theequation)}
\def\begineq #1\endeq{$$ \refstepcounter{equation}\eqalign{#1}\eqno
	(\theequation) $$}
\def\contlimit{\,{\hbox{$\longrightarrow$}\kern-1.8em\lower1ex
\hbox{${\scriptstyle (a\rightarrow0)}$}}\,}
\def\centeron#1#2{{\setbox0=\hbox{#1}\setbox1=\hbox{#2}\ifdim
\wd1>\wd0\kern.5\wd1\kern-.5\wd0\fi
\copy0\kern-.5\wd0\kern-.5\wd1\copy1\ifdim\wd0>\wd1
\kern.5\wd0\kern-.5\wd1\fi}}
\def\centerover#1#2{\centeron{#1}{\setbox0=\hbox{#1}\setbox
1=\hbox{#2}\raise\ht0\hbox{\raise\dp1\hbox{\copy1}}}}
\def\centerunder#1#2{\centeron{#1}{\setbox0=\hbox{#1}\setbox
1=\hbox{#2}\lower\dp0\hbox{\lower\ht1\hbox{\copy1}}}}
\def\lsim{\;\centeron{\raise.35ex\hbox{$<$}}{\lower.65ex\hbox
{$\sim$}}\;}
\def\gsim{\;\centeron{\raise.35ex\hbox{$>$}}{\lower.65ex\hbox
{$\sim$}}\;}

\def\super#1{\ifmmode \hbox{\textsuper{#1}}\else\textsuper{#1}\fi}
\def\textsuper#1{\newcount\holdspacefactor\holdspacefactor=\spacefactor
$^{#1}$\spacefactor=\holdspacefactor}

\def\getcite#1,{\advance\citenumber by1
\def\getcitearg{#1}\def\lastarg{@}
\ifnum\citenumber=1
\ref{#1}\let\next=\getcite\else\ifx\getcitearg\lastarg\let\next=\relax
\else ,\ref{#1}\let\next=\getcite\fi\fi\next}

\def\pom{{\rm P\kern -0.53em\llap I\,}}
\def\spom{{\rm P\kern -0.36em\llap \small I\,}}
\def\sspom{{\rm P\kern -0.33em\llap \footnotesize I\,}}

\relax
\def\contlimit{\,{\hbox{$\longrightarrow$}\kern-1.8em\lower1ex
\hbox{${\scriptstyle (a\rightarrow0)}$}}\,}
\def\upon #1/#2 {{\textstyle{#1\over #2}}}
\relax
\renewcommand{\thefootnote}{\fnsymbol{footnote}} 

\def\mainhead#1{\setcounter{equation}{0}\addtocounter{section}{1}
  \vbox{\begin{center}\large\bf #1\end{center}}\nobreak\par}
\sectioneq
\def\subhead#1{\bigskip\vbox{\noindent\bf #1}\nobreak\par}

\def\til#1{\centeron{\hbox{$#1$}}{\lower 2ex\hbox{$\char'176$}}}
\def\tild#1{\centeron{\hbox{$\,#1$}}{\lower 2.5ex\hbox{$\char'176$}}}
\def\sumtil{\centeron{\hbox{$\displaystyle\sum$}}{\lower
-1.5ex\hbox{$\widetilde{\phantom{xx}}$}}}

\begin{document} 

\begin{titlepage} 

\begin{center} 
  
{\large\bf New Physics at the Tevatron and the LHC May Relate to 
Dark Matter Visible in UHE Cosmic Rays.} 

\medskip

Alan. R. White\footnote{arw@hep.anl.gov }
 
\vskip 0.6cm 
\centerline{Argonne National Laboratory}
\centerline{9700 South Cass, Il 60439, USA.}
\vspace{0.5cm}

\end{center}
\begin{abstract} 
A two flavor color sextet quark sector added to QCD yields the {\it uniquely} unitary Critical Pomeron at high energy while also producing electroweak symmetry breaking. In this paper it is argued that a number of experimental phenomena in Cosmic Ray and hadron collider physics can be interpreted as evidence for the sextet sector, as follows. 
\begin{enumerate} \openup-1\jot{\it
\item{The majority of UHE cosmic rays are
Dark Matter sextet neutrons that are strongly interacting at high energy but appear as WIMPs at low energy.} 
\item{The cosmic ray spectrum knee reflects both an incoming
sextet neutron threshold and a proton production threshold for sextet
states.} 
\item{The enhancement of high multiplicities and small $ p_{\perp}$
at the LHC is related to a sextet generated triple pomeron coupling.}
\item{Tevatron and LHC events with a $Z$ pair and a high multiplicity of small  $p_{\perp}$ particles are
associated with sextet electroweak symmetry breaking.}
\item{Top quark production is via the $\eta_6$ sextet quark pseudoscalar
resonance - interference with the background will produce an asymmetry.}
\item{Longitudinal $Z$ pairs, produced as sextet pions, provide a high mass excess cross-section that includes the $\eta_6$ at the $t\bar{t}$ threshold mass.}
\item{Enhanced $W$ pair production would produce an excess in the $W +$dijet cross-section. }}
\end{enumerate}
Combining the sextet sector and the electroweak interaction without short-distance anomalies requires QUD - a {\it unique} underlying weak coupling massless SU(5) gauge theory. Remarkably, it appears that the Standard Model might be reproduced
(in a radical conceptual change) by the QUD bound-state S-Matrix. Infra-red divergent gauge bosons coupled to massless fermion anomalies produce a ``wee parton vacuum'' that confines the
elementary fermions.  
All S-Matrix particles have dynamically generated masses, with sextet baryons as the only new particles - beyond the Standard Model. Anomaly color factors produce large sextet amplitudes and there is no Higgs boson.
\end{abstract}

\renewcommand{\thefootnote}{\arabic{footnote}} \end{titlepage}

\mainhead{1. Introduction}

There are a significant number of results from Cosmic Ray experiments that suggest the existence of new strong interaction physics at very high energy. 
In this paper\cite{arwdm}, I will propose an explanation for these results via a two flavor color sextet quark sector of QCD and will argue that 
elements of the same physics have also been seen at the Tevatron and at the LHC. The sextet sector not only complements (very naturally)
the familiar triplet sector but also dramatically changes the dynamics of QCD in a manner that already has substantial experimental support. 

The explanations I give, for a variety of disparate phenomena, imply that physics ``Beyond the Standard Model'' is very different to what has been generally anticipated, both theoretically and experimentally. Cosmic rays provide insight into a very high energy, low luminosity, strong interaction kinematic regime that is widely assumed to be orthogonal to the high luminosity, rare new physics, focus of the LHC. However, if the physics to be discovered is as I describe, a significantly different LHC program may ultimately be required. There is no new short-distance physics at the electroweak scale, contrary to the focus of all planning. Instead, there is a new QCD strong interaction that negates the fundamental assumption of a clear separation between ``interesting new large $p_{\perp}$ physics'' and ``uninteresting small $p_{\perp}$ physics'' that has been the basic justification for high luminosity LHC running.

Theoretically, the most important new element is the necessary introduction of
QUD\footnote{Quantum Uno/Unification/Unique/Unitary/Underlying 
Dynamics.} - a unique\cite{kw,arw10} underlying weak coupling massless SU(5) gauge theory. Amazingly, this theory lies in a ``conformal window'' but has a bound-state S-Matrix that might provide an aesthetically and philosophically appealing, if paradigm changing, origin for the Standard Model. A remarkably economic dynamical unification of the strong and electroweak interactions results 
from a combination of gauge boson high-energy infra-red divergences and massless fermion anomalies. In QCD there is confinement and chiral symmetry breaking, that coexists with both a Regge pole pomeron and the parton model (in contrast to conventional QCD), as well as large cross-sections for the high mass sextet sector. In addition, a very different understanding of the third quark generation connects top quark production to a pseudoscalar sextet quark resonance, the $\eta_6$, that should also appear in the Z pair cross-section.

Experimentally, given the recent negative search results that are now almost complete, the most important feature is that there is no ``Higgs boson'' to be discovered\footnote{As it relates to the $W^{\pm}$ and $Z^0$, the $\eta_6$ is
the ``sextet Higgs''. It is, perhaps, already discovered via top quark production - with an asymmetry due to pseudoscalar interference with the background~!} at the LHC. Instead, the reggeon diagram construction of the high-energy S-Matrix (outlined in \cite{arw10}) implies that all the particles are bound-states with dynamically generated masses\footnote{A physically well-defined bound-state mass is in strong
contrast with 
the elusive concept of a lagrangian parameter mass - that the Higgs has been asked to provide. QUD has no lagrangian mass parameters. The dynamical mass spectrum arises from the removal of initial reggeon mass and $p_{\perp}$ cut-off regulators via a symmetry restoration that involves the Critical Pomeron.}
(including neutrinos).
The spectrum has the Standard Model form, with the only 
new states being the sextet quark baryons that play a vital role in my explanation of cosmic ray phenomena. In particular, sextet neutrons (that I call {\bf neusons}) provide a very special form of dark matter. I will argue that neusons are strongly interacting as very high energy cosmic rays but behave as standard WIMPs at low energy.

Electroweak symmetry breaking is a QCD effect of the sextet 
sector and once high-enough energies are obtained (hopefully, the eventual LHC energy will be sufficient) this new QCD interaction will, pre-eminently, require exploration at all momentum scales. For the present, within the current
LHC operation mode, the ``new physics'' evidence to be searched for is, mostly, hard to isolate connections between strong interaction physics and electroweak vector boson physics. As I will discuss, the inclusive Z pair
cross-section is the most direct place to look. This cross-section will provide strong supporting evidence for the role of the $\eta_6$ if, as it 
appears\cite{CDF4l7,CDFEPS} may be the case, it does indeed contain a resonance at the mass threshold for top quark/antiquark 
production. 

In general, unfortunately, the search for present evidence is seriously hindered by both experimental and theoretical factors that are consequences of the narrowness of the existing expectations and preparations for a discovery. 
On the experimental side, the major focus of the LHC 
detectors on the kinematic regions where short-distance Higgs-like physics has been expected to appear, certainly makes it difficult to see much of the new physics.
Much worse is the current high luminosity pile-up of interactions that is a consequence of the intense pursuit of ``ultra-small cross-section'' physics. 
While the immediate reward may be the, obviously deeply significant and crucial (particularly from our perspective), demonstration that the Higgs boson 
does not exist, in the longer term the pile-up makes it close to impossible to study new physics that mixes electroweak-scale
high $p_{\perp}$ physics with small $p_{\perp}$ hadronic physics and that may have
larger cross-sections. For example, I expect the associated production of high multiplicity soft hadron states accompanying multiple electroweak vector bosons to be a major phenomenon of this kind that is spread across a large part of the rapidity axis. However, the detectors will only be able to see a fraction of the electroweak
boson states and the pile up will make it extremely difficult to determine any properties of the soft hadronic part. 

I will suggest that, in fact, the Tevatron has actually seen a glimpse of this physics and that it  
might have seen much more if, for at least part of the program, the experiments had covered a wider rapidity range with a limited  luminosity.
Disappointingly, the rapidity coverage of the LHC experiments is, currently, even smaller
relative to the full available rapidity and the luminosity pile-up is, of course, much worse.

It should be mentioned that special low luminosity runs are scheduled
for TOTEM and it is just possible that
a definitive discovery could be made if there could be a collaboration with
CMS (that continues to be anticipated for the future) to look for the double pomeron production of vector boson pairs. Most likely, however, higher energy will   
be required to see this process.

On the theoretical side, a potential problem is the development\cite{dmg} of ``QCD phenomenology'' aimed at extending both the precision and the domain of perturbation theory, by the inclusion of ``non-perturbative'' ingredients. While this phenomenology has the intention of increasing the potential for the discovery of new short distance physics, it has the effect of making ``perturbative QCD'' fits to short-distance cross-sections look much more comprehensive than is really the case. Unfortunately, 
the success of this phenomenology could well obscure important QCD elements of the new sextet quark physics that are appearing.

\subhead{1.1 The Sextet Sector, the Critical Pomeron, and the QUD S-Matrix}

While my understanding of the role of QUD is relatively recent, I have long argued\cite{arw82,arw04} (as I will discuss in more detail later) that the sextet sector 
should appear at high energy in
order to produce the  Critical Pomeron\cite{cri}. 
The discovery of the Critical Pomeron can be viewed\cite{arw00} as the summit of abstract S-Matrix Theory.
It is formulated as a renormalization group fixed-point solution of pomeron Reggeon Field Theory  that, uniquely, satifies 
multiparticle t-channel unitarity and produces rising total cross-sections. It has also been shown to satisfy all
multiparticle unitarity constraints in s-channel
scattering regions. It was formulated without reference to any underlying theory
and, as far as is known, provides a uniquely unitary
possibility for the high-energy behavior of an S-Matrix.

The existence of the sextet sector has also long been 
advocated\cite{wm} as  
providing a particularly attractive solution to the problem of electroweak symmetry breaking, with the sextet chiral scale identified as the electroweak scale.
As I will return to shortly, it is the
consistent combination of the sextet quark sector with the electroweak interaction that requires the embedding in QUD. In addition to
impacting the new QCD physics introduced by the sextet sector (and also modifying
other areas of QCD), the underlying massless theory ensures that Critical Pomeron asymptotic scaling is present in conjunction (non-trivially) with massive particle states. 

By constructing (``infinite momentum'') multi-regge amplitudes, I have argued\cite{arw10} that QUD has a massive bound-state S-Matrix. 
The essence of the dynamics is that infra-red divergent 
anomalous color parity wee gluons, that couple via massless fermion chirality transition anomalies, 
provide a ``wee parton vacuum'' that selects and strongly enhances the physical high-energy amplitudes. The resulting amplitudes are directly determined by the charge conjugation and parity properties of the gauge field couplings to the fermion representation and it is truly remarkable, given the uniqueness of the field theory, that
only the interactions and states required by the Standard Model seem to appear, 
apart from the new sextet states and their amplitudes. If, in fact, QUD does produce both the Critical Pomeron and the Standard Model S-Matrix, it will surely revive the ``ancient'' idea of a unique\footnote{That the existence of 
a non-perturbative S-Matrix in any four-dimensional field theory 
\newline $~~~~~~$ remains unproven is often regarded as just another  
``inconvenient truth''.} unitary particle S-Matrix.

Regrettably, my construction of the QUD multi-regge 
S-Matrix is still at a very formative stage. The highly selective anomaly dynamics is novel and many details are either not yet given or remain to be better understood. 
Consequently, while it is possible to deduce general properties, the development of a complete calculational framework is a long way away - even though the reggeon diagram formalism that I use is basically perturbative. This is why, in most of this paper, the arguments I give will be descriptive only.
Although this makes the content seem even more speculative than it necessarily is, I hope it is understandable on a qualitative level. 

The underlying massless theory does not appear directly in most of my discussion, even though it's existence is crucially responsible for the anomaly dominance of couplings that produces extra large sextet cross-sections. In principle at least, an effective theory closer to the Standard Model might be obtained from QUD by integrating out the ``elementary leptons''
contained in the theory 
- leaving massive bound-state physical leptons (including neutrinos), massive 
triplet quarks (via self-energy interactions involving the elementary leptons), and massless sextet quarks. However, I have not studied this possibility in any detail.

I will give a brief outline of the formal multi-regge 
construction of S-Matrix amplitudes, that I plan to give more details of in the near future, in the later Sections of the paper. 
Since QUD has no explicit parameters and also has no possibility
for extension or further embedding, it is very fortunate that solutions to the well-known ``Beyond the Standard Model'' problems of dark matter and neutrino masses also appear to be provided. 
Indeed, it has to be emphasized that if any established element of the Standard Model is not reproduced by the QUD S-Matrix, the uniqueness of the theory implies that it must be wrong. Also, since the S-Matrix necessarily has no off-shell amplitudes, if it is to have any hope of gaining general acceptance as a replacement for the Standard Model, significant changes in the currently accepted theory paradigm will be essential. 

\subhead{1.2 Cosmic Ray Results and Models}
 
The most recent Cosmic Ray results suggesting new physics are from the Auger collaboration. Their results\cite{AUGER} on the penetration and fluctuation of air showers imply,  as shown in Fig.~1, that the very highest energy 
cosmic rays do not look like protons. 
\begin{center}
\epsfxsize=2.6in \epsffile{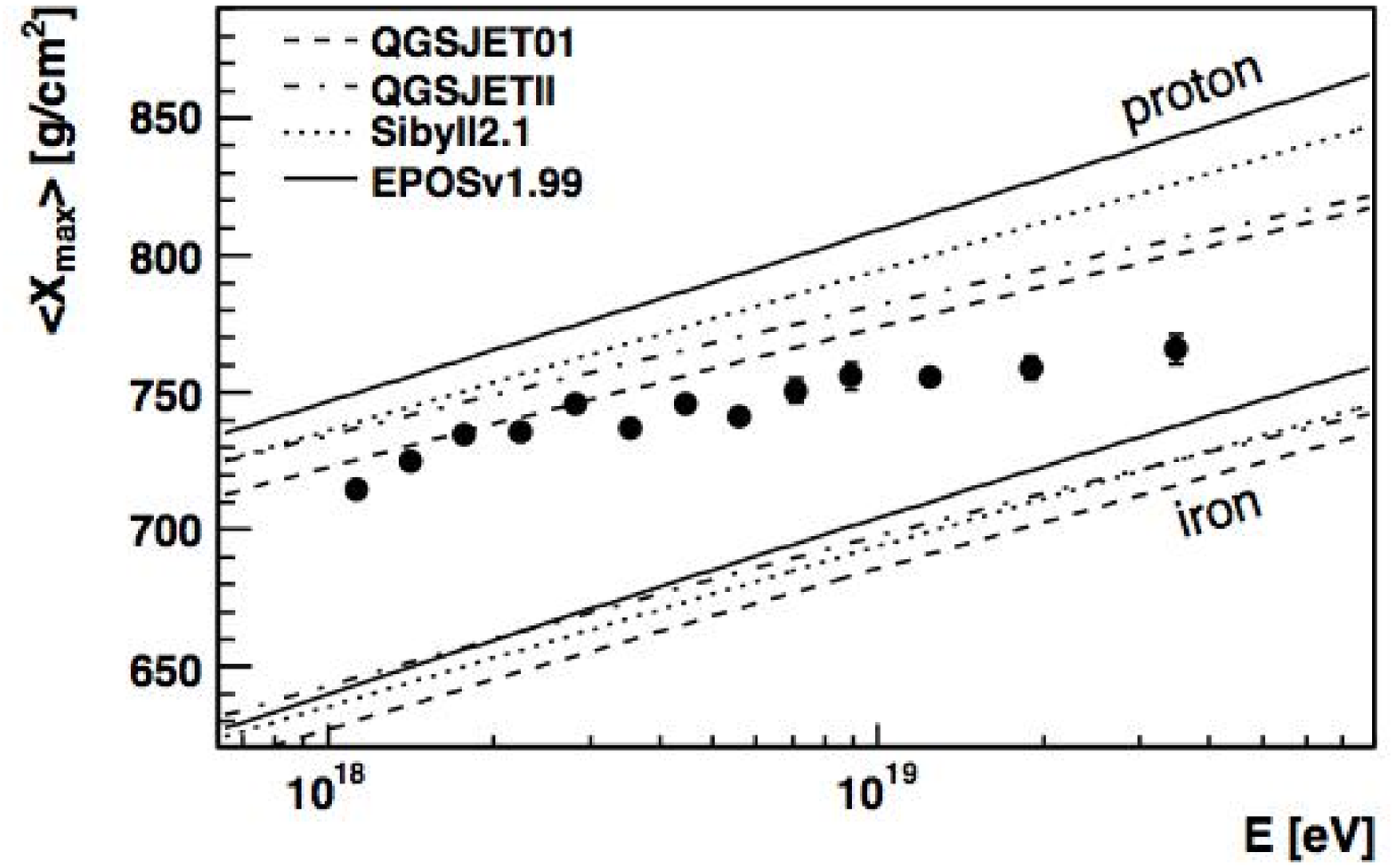}
\hspace{0.2in}
\epsfxsize=2.6in \epsffile{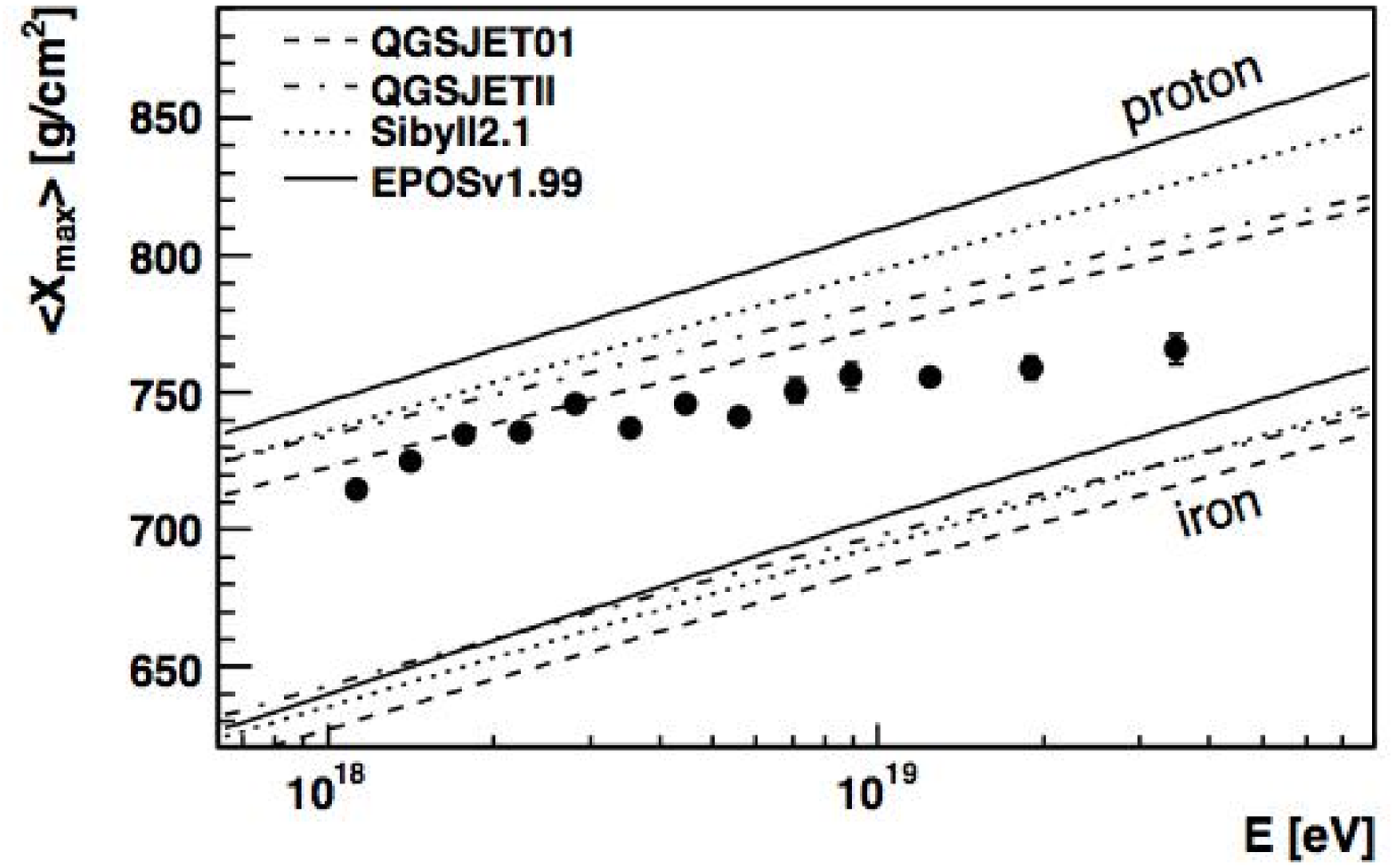}

Figure 1. Auger High-Energy Results for the Average Depth of the Shower Maximum, together with the Corresponding Fluctuations 
\end{center}
Within established strong interaction phenomenology, they can only be heavy nuclei - primarily iron nuclei. This is astonishing; 
not only is there no known mechanism for producing such high energy nuclei, it also defies all intuition that they would be produced, and survive, in preference to protons. 

The Auger experiment is in a league of it's own with repect to high statistics data, detector size, and level of analysis. A comparison\cite{AUGER}  of the accuracy of the Auger results with that of the highest energy HiRes results is shown in Fig.~2(a). 
\begin{center}
\parbox{2.8in}{\vspace{0.05in}
\epsfxsize=2.6in\epsffile{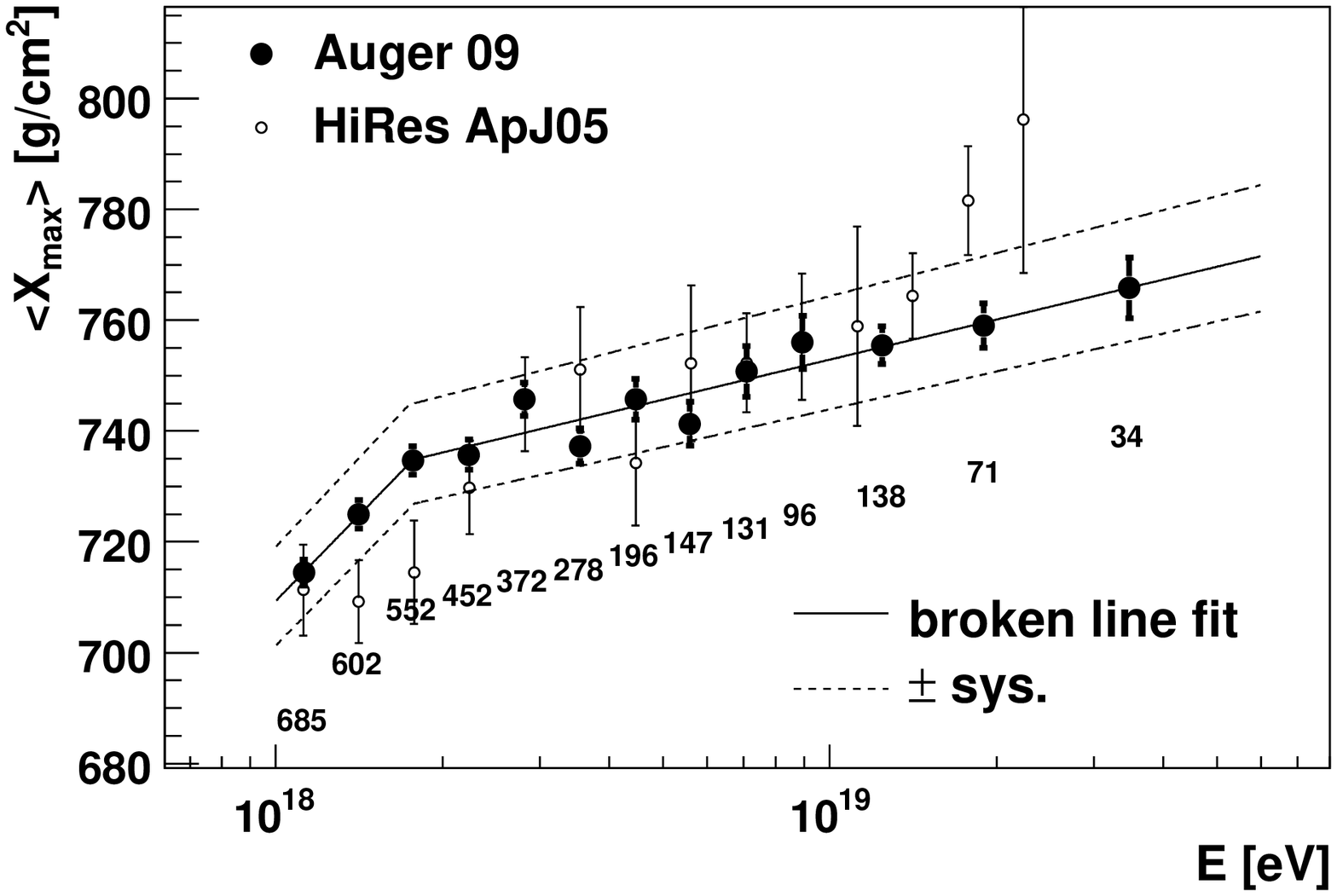}
\newline $~$
\centerline{(a)}}
$~$
\parbox{3in}{\epsfxsize=2.8in\epsffile{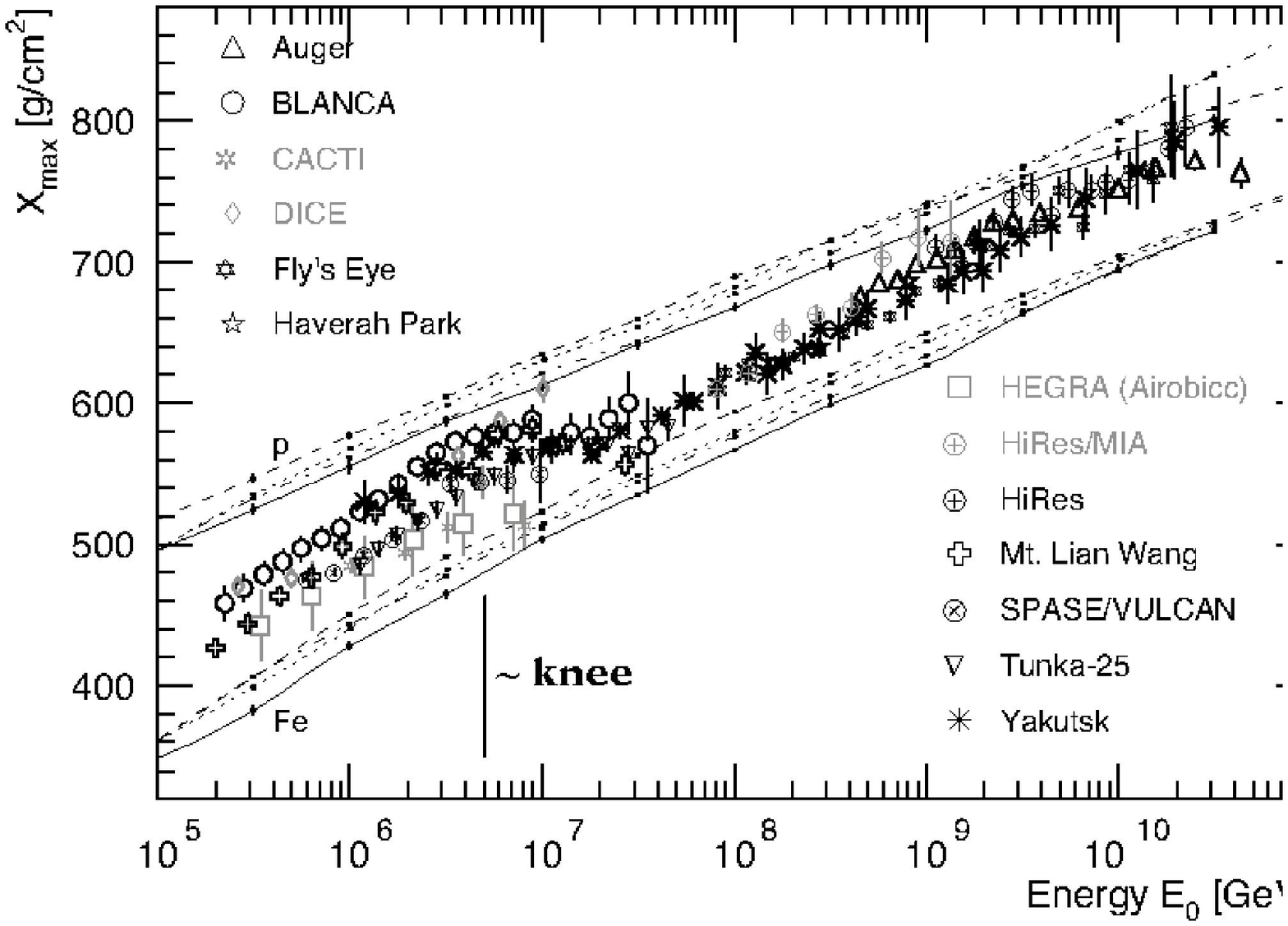}
\centerline{(b)}}

Figure 2. (a) Comparison of Auger and HiRes Results (b) 
Results for the Average Depth of Shower Maximum, From All Experiments, Over a Wide Energy Range
\end{center}
I will comment later on the change of slope (in the region of the spectrum ``ankle'') that the accuracy of the data exposes. 
Also shown, in 
Fig.~2(b), is the wide variation of results from other 
experiments\cite{beh} over a large energy range that includes the interval in which the spectrum knee occurs. 
Although there is much disagreement about absolute values, the different experiments all show a movement towards the iron composition line that begins before the knee and continues for a substantial energy interval above it. This will be important for our discussion. Significantly, the knee occurs between the energy of the Tevatron and the current LHC energy.
  
In spite of the improved accuracy of the Auger results,
to explicitly relate the experimental shower measurements to the nature of the incoming cosmic ray it is still necessary to use a strong interaction model that, in principle, should be based directly on QCD. In practise, QCD is not well understood in either the appropriate kinematic regime or 
at the needed level of complexity and so ``QCD motivated'' phenomenological models have to be used. It is well known that the models are not able, in general,
to simultaneously reproduce well all the key properties of the showers, 
with the muon content being particularly elusive. The models also give inconsistent answers when different experimental methods are compared.
This has led to a variety of confusing and contradictory results
on the elemental composition, from energies below the knee
up to the highest energies - with the Auger results, perhaps, capping the confusion.
Although the sole source of the problems could be that the models are failing to reproduce QCD, as is the general consensus, it is also plausible that something beyond established QCD physics is being seen. 

I will argue that much of the present confusion in
cosmic ray physics, including both the existence and properties of the knee and the apparent presence of very high energy heavy nuclei, is caused by the existence of the two flavor color sextet quark sector of QCD. 

\subhead{1.3  QCD Monte Carlos, Small ${\bf p_{\perp}}$, and the Critical Pomeron.}

As I will enlarge upon in Section 4, the relation between QCD Monte Carlo models and the new LHC minimum bias data on multiplicity and transverse momentum distributions
is somewhat similar to the cosmic ray situation. It is a very fortunate outcome of the initially reduced
luminosity that the LHC detectors have been able to take data that is much more accurate and records the small transverse momentum region far more extensively than has been done previously. As a result, serious deficiencies of the commonly utilised ``QCD based'' Monte Carlo models have been exposed. Most importantly, it is clear that central plateau multiplicities are severely
underestimated, particularly in small transverse momentum events. Again, it could be a simple failing of the models, as the experimenters conclude, or it could be that QCD does not conform to conventional expectations at high energy and small transverse momenta.

In the absence of any conflicting experimental data from the Fermilab Tevatron
or HERA, it has been widely assumed that a semi-hard dynamical version of QCD, built on BFKL dynamics, is realized\cite{klrw} in nature (at high energy).
The large transverse momentum of the interactions producing the BFKL pomeron 
greatly reduces the dynamical significance of the small transverse momentum region in a manner that could, at least partly, justify the parton model starting point
for the most commonly used Monte Carlo models. However, a dominant dynamical role for large transverse momenta is surely in conflict\cite{rmk} with the strong enhancement, 
increasing with energy, of the small transverse momentum (central plateau) region that is seen at the LHC. 

Contrastingly, the Critical Pomeron is a small transverse momentum phenomenon
in which central region cross-sections are dominated by high multiplicity small transverse momentum states associated with a strongly self-interacting
regge pole. The LHC minimum bias data show that such states are being produced with a significant cross-section that could be a signal of the new strong interaction scale. I will suggest that the increasing central region multiplicities are a consequence of a large triple pomeron coupling appearing at the electroweak scale that is due to the sextet sector. 

A priori, QCD does not produce a regge pole pomeron. 
However, in massless QCD the addition of two sextet flavors to six triplet flavors
(giving ``QCD$_S$'')
introduces an infra-red fixed-point that enhances massless quark anomaly interactions coupled to wee gluons, with the result that the BFKL dominance of large transverse momentum is overwhelmed. 
A regge pole pomeron appears as a reggeized gluon in a color compensating 
anomalous wee gluon condensate and the Critical Pomeron results from pomeron self-interactions. 
Fortunately, as I emphasize many times in this paper, 
the embedding of massless QCD$_S$ in QUD allows effective quark masses to be acquired in the formation of bound states, without disturbing the high-energy behavior of the massless theory. 

After discussing top quark physics in the last Section of the paper, I will elaborate on how the sextet quark sector could also affect 
the application of perturbative QCD calculations to jet cross-sections and other
short-distance hadronic physics.

\subhead{1.4 Electroweak Physics, the QUD S-Matrix, and the Theory Paradigm}

That Z pairs can be produced in association with a high multiplicity of small $p_{\perp}$  particles is suggested by a
spectacular four electron event that was seen at the Tevatron before the luminosity-induced pile-up, and (to a lesser extent)
the very first LHC Z pair event seen by CMS, that was also recorded before pile-up. Events of this kind would be expected
if the sextet sector is responsible for electroweak symmetry breaking. The high 
multiplicity states contribute to the  
electroweak scale pomeron, as discussed in the previous sub-section, and their strong coupling to electroweak states is a necessary reflection (as I will discuss further in Section 9) of the direct coupling of the pomeron to electroweak  states. Indeed, I will suggest that a large cross-section for multiple Z's and W's  produced {\it across a wide rapidity range, generally with a high multiplicity of small $p_{\perp}$ particles,} will be a major component of the new physics that is ultimately discovered at the LHC. 

If it should become clear that this physics is present, the inescapable conclusion would be that the electroweak scale is a new strong interaction scale and should be investigated as such. At the same time,
the resulting interaction change would be finally understood as
the origin of the cosmic ray spectrum knee. Unfortunately, the current detectors may not be able to see very much of the physics that is involved, which is surely not what they were
designed for. In addition, the enormous pile-up will make the recognition of those events that can be seen even more difficult.
Most likely, this
will ultimately lead to a radical redirection of 
both the luminosity program and the focus of the detector data taking. 
 
For the immediate future, all we can hope to see is the central region, large $p_{\perp}$, production 
of multiple vector bosons. The Z pair cross-section involving only charged leptons,
although relatively small, is very clean and will be the 
easiest to access. Very encouragingly, both the Tevatron and the LHC detectors may have now seen physics of the kind that I am proposing in this cross-section. 
If the $\eta_6$ resonance is present at the mass of the top/antitop threshold, as it appears might be the case, together with an additional excess cross-section that is (essentially) rapidity independent, this would provide very strong support for QUD. 
The strongest evidence for the presence of the $\eta_6$ has come, so far, from CDF. Since it is a pseudoscalar it is, presumably, possible that the $pp$ cross-section for the production of this resonance at the LHC is smaller than the $p\bar{p}$ cross-section seen at the Tevatron. This would be very unfortunate, given 
the current demise of the Tevatron. 
Nevertheless, as the realization that there is no Higgs' boson to be discovered becomes widespread and fully accepted, it is likely that the focus on the ZZ cross-section as the ``new physics window'' will become intense at the LHC.

The consistent combination of the sextet quark sector with the electroweak interaction (referred to earlier) involves requiring both asymptotic freedom and
the cancelation of all short-distance anomalies. These requirements  
lead\cite{kw} uniquely  
to QUD. I will briefly review QUD in the penultimate Section of the paper. The last Section discusses the link between top quark physics and the $\eta_6$, for which properties of QUD  are essential. That the QUD short-distance anomaly
cancelation involves both the sextet and triplet sectors is a very important element. 

As I will review, the QUD high-energy S-Matrix in which all the elementary ``quarks'' and ``leptons'' are confined is an outcome of the same regge region anomaly dynamics that gives the Critical Pomeron in QCD$_S$. 
The multi-regge construction of bound-states implies there will be generations of physical leptons and hadrons, exactly as in the Standard Model. Even though much remains to be done to obtain the specifics of the hadron and lepton spectra, the extraordinary possibility that the QUD S-Matrix could provide an underlying origin for the established elements of the Standard Model, while also solving the core ``Beyond the Standard Model'' problems, seems to be very real. 

Nevertheless, it has to be emphasized that not only is there an enormous amount of structural and algebraic detail that still has to be given, but also a dramatic proposal is involved which is surely a revision of the current theory paradigm. This is that a very weak coupling field theory can
produce a high-energy S-Matrix (without off-shell amplitudes), in which the interaction strengths are those of the Standard Model, via the infra-red divergent enhancement of anomalies resulting from massless fermion chirality transitions. 
Moreover, even though the physical states are all massive, the underlying massless theory is responsible for anomalies that dominate high-energy couplings and, in particular, enhance both small transverse momentum production cross-sections involving  sextet states and the sextet generated triple pomeron coupling.

\subhead{1.5  Anomaly Dynamics and QCD}

That I have started with the Critical Pomeron and arrived at a potentially unique underlying unification for the Standard Model, when over three decades of GUT research has failed to arrive at a unified theory, is due (I believe) in large part to a gradual realization of the crucial role that infra-red anomalies, due to 
underlying massless fermions, must play in 
high-energy amplitudes. After a very long search for the Supercritical Pomeron, I have become convinced that the anomaly interactions of (confined) massless fermions are essential
for a gauge theory to produce a solution of t-channel unitarity at high-energy.
Very unfortunately, the deep constraints imposed by t-channel unitarity are currently ignored by not only the model/theory building community but also by almost all theorists working on the high-energy behavior of QCD.

Although much remains to be understood about the anomaly dynamics, there are many desirable consequences for QCD at energies where the sextet sector is not directly evident. While the arguments
are only outlines, as I have so far presented them, an origin for both confinement and the parton model
is provided and their co-existence is clear. The essential regge pole nature of the pomeron, which is well-established experimently, is also a direct outcome. In addition, the anomaly dynamics produces only chiral symmetry breaking quark states and so glueballs, which have not been seen experimentally, are absent. Together with the top quark and jet physics discussed in the last Section of this paper, these phenomena are part of the significant existing experimental evidence referred to at the outset.

Most important for the present discussion are, of course, the consequences of the anomaly dynamics for the Cosmic Ray and LHC physics that I will focus on in the following.
Although I have long believed that the color sextet sector of QCD should be evident in high-energy cosmic rays, and although I realized\cite{arw94} very early on that the spectrum knee must in some way be 
the threshold for the new physics, it is only in the last few years that I have understood just how the anomaly dynamics of QCD is introduced by the sextet sector and what the most important experimental consequences should be. 
In later Sections, I will give a brief description of how anomaly chirality transitions determine bound-states and scattering amplitudes, first in massless QCD$_S$ and then in the underlying SU(5) theory. In the bulk of the paper, however, I will give only brief justification for phenomena that I will claim are an outcome of the anomaly dynamics. 

Two particularly significant phenomena should be emphasized at the outset. First, the S-Matrix interactions of the SU(5) theory are {\it only} those of the Standard Model. Second, because the bound-states necessarily contain Goldstone bosons associated with chiral symmetries, there are no mixed triplet/sextet states. As a consequence 
of these properties, triplet and sextet fermion numbers are separately conserved and there are sextet baryons that are stable, just as the proton and anti-proton are stable.

\subhead{1.6 Outline}

In the next Section I will discuss the occurence of sextet baryons in cosmic rays. In Section 3, I will briefly summarize the current state of (contradictory) experimental knowledge of the knee.
I will then describe the LHC minimum bias data in Section 4 and, in Section 5, make the connection between the 
the appearance of high multiplicities in the central plateau and a large triple pomeron coupling due to sextet quark anomalies.
In Section 6, I will describe why, and how, the knee is evidence for the large cross-section physics of the sextet sector. In Section 7, I will discuss a particularly significant high multiplicity $Z^0Z^0$ event that was seen at the Tevatron, as well as why other similar events may have been missed. In Section 8, I will argue that the first $Z^0Z^0$  event seen at the LHC and the Tevatron event, belong to the class of sextet sector generated events that should eventually
be a dominant component of new physics at the LHC. I will then discuss the striking properties of the $Z^0Z^0$ pair cross-section 
seen at the Tevatron that may also be emerging at the LHC, most importantly
the possible presence of the $\eta_6$ resonance. I will also discuss,
more generally, what might be seen at the LHC in the near and long-term future. In Section 9, I will give a brief theoretical review of massless QCD$_S$ and in Section 10 will give a similar review of QUD. Finally, in Section 11, I will argue that top quark production at the Tevatron is via the neutral pseudoscalar $\eta_6$ and that this naturally explains the observed asymmetry. I will also discuss why the top
mass scale should appear in QCD jet physics and how this is obscured by current
``QCD phenomenology''.

\mainhead{2. Dark Matter in Ultra High Energy Cosmic Rays}

The quantum numbers of the two flavor sextet sector are such that the physical states are analagous to those of the u/d triplet sector. The ``sextet pions'' become the longitudinal components of the electroweak vector bosons and, in doing so, solve the fundamental problem of electroweak symmetry breaking. 
In addition, the ``sextet baryons'' may solve an
equally fundamental problem by providing a very particular form of ``Dark Matter''. 

\subhead{2.1 Neusons and Prosons}

The core element of my explanation of the Auger results will be that the majority of the highest energy cosmic rays are actually {\bf dark matter} sextet baryons, i.e. they are
stable massive ``sextet neutrons''  composed of three color sextet quarks. I will call these particles {\bf neusons} 
\footnote{Sextet baryon names are obtained via {\bf tr}~\{t\} $\equiv$ {\it triplet} $\rightarrow$ {\bf s}~\{s\} $\equiv$ {\it sextet}.}. They have crucial properties that could lead to their appearance as high-energy cosmic rays. 

There will also be ``sextet antineutrons'', i.e. {\bf antineusons},
that are similarly composed of three sextet antiquarks. 
Because of SU(3) triality, the 6$^*$/6 representation corresponds to the 3/3$^*$ 
representation when when we build SU(5) representations. 
Therefore, we identify ``sextet quarks'' as transforming under the 6$^*$
representation, so that they carry the same electroweak quantum numbers as triplet quarks. This becomes 
much more than a question of nomenclature if we identify neusons as the dark matter that is dominantly present in the universe and antineusons as having 
only the same significance as antiprotons. However, since there is no current
understanding of why protons, rather than antiprotons, make up ordinary matter, this identification could well be wrong. 

Because sextet quarks retain their zero current mass, it is the neuson and antineuson that are stable. The electromagnetic mass contribution is sufficient to make 
the charged ``sextet proton'' (that I will call a {\bf proson}), analagously composed of three color sextet quarks, heavier than the neuson to which it decays. This is in contrast to the u/d triplet quark sector, where the neutral baryon (the neutron) is heavier than the charged baryon (the proton) because of triplet current quark masses. (Note that massless sextet quarks are necessary for sextet pions to be massless and produce electroweak symmetry breaking.)
Very attractively, therefore, the existence of a large dark matter component in the universe is a very simple and natural complement to the predominantly charged, normal matter, triplet sector.

That very high energy cosmic ray particles must be accelerated by ``cosmic accelerators'' requires, according to current understanding, that they be charged, which (by definition!) dark matter particles are not. Indeed, the focus on the GZK cut-off has strongly cemented the conviction amongst many physicists that 
the highest energy cosmic rays must be charged particles. 
However, stable cosmic ray neusons can originate as the decay product of unstable, but relatively long-lived, charged
prosons that have been accelerated by conventional cosmic accelerators. 

\subhead{2.2 Sextet Sector Interactions}

The self-interactions of the sextet sector will be those of a stronger coupling, higher mass spectrum, version of the triplet sector. The pomeron provides the
high energy interaction within each sector and also between the two sectors. It contains dynamical gluon reggeons (one in the simplest approximation)
together with zero transverse momentum anomalous wee gluons that neutralize the 
exchanged color. As we noted earlier, the wee gluon anomaly color factors that are involved imply that the coupling of the pomeron to sextet states 
is much stronger than the coupling to triplet states. Equally important, is the
large sextet quark anomaly contribution to the triple pomeron coupling.

High multiplicity states with large rapidity and small transverse momenta that are the intermediate states of the exchanged pomeron will also have, as we already remarked  in the Introduction, a strong coupling to sextet states. This is because, as discussed further in Section 9, the amplitudes involved also contain anomaly coupled  
wee gluons. Consequently, the production of sextet states will dominate high energy QCD cross-sections.
In particular, at high enough energy, these cross-sections will dominate particle production in the rising, expanding, central plateau\footnote{Note that a rising {\it universal} central plateau is a defining property\cite{arw84} of the Critical Pomeron.} that is the main characteristic of the high-energy strong interaction. Indeed, as I have already suggested, and discuss further below, the sextet sector may be appearing at the LHC, in combination with a
large increase in the height of the plateau due to the 
production of small transverse momentum particles associated with
the large triple pomeron coupling.

The high-energy dominance of sextet states, potentially, provides an explanation for the dominance of dark matter production in the ultra-high-energy processes responsible for early universe formation.
It also implies that prosons will be preferentially produced
in the present universe, relative to protons, by the highest energy strong interaction processes. 
Hence, after cosmic acceleration, neusons produced by prosons (and antineusons produced by antiprosons) will be the dominant cosmic rays, at the highest energies.

There are no meson or nucleon exchanges that can provide a low-energy interaction between the sextet and triplet sector states. Consequently, this interaction occurs only {\it at high enough energy} for the intermediate states to appear that produce the gluon exchange pomeron in the total cross-section. Indeed,
the absence of a low-energy interaction between the two quark sectors  is an essential part of my explanation of cosmic ray phenomena. It is closely related to the, conventionally surprising but experimentally verified\cite{cm}, absence of glueballs in the hadronic spectrum. If glueballs existed, their exchange would provide a straightforward interaction between neusons and protons. 
Instead, as is generally believed to be an essential property of dark matter,  neusons will have no strong interaction with normal matter at low energy. A neuson will appear, therefore, as a conventional WIMP (weakly interacting massive particle) in most dark matter experimental searches.

The currently popular understanding of QCD dynamics predicts\cite{bpst} that glueballs should be a major component of the hadronic spectrum and that they should be associated with a multigluon, BFKL generated, semi-hard pomeron. Whereas, because fermion chirality transitions are an essential element in the formation of bound-states, the anomaly dynamics produces both a much more limited particle spectrum and a much simpler pomeron. All physical states are formed from anomaly pole chiral Goldstone bosons that necessarily contain quarks and so there are no glueballs. Moreover, the pomeron is a self-interacting regge pole, as t-channel unitarity requires,
involving dynamical gluon reggeons that are produced by small transverse momentum intermediate states and that are color neutralized by (anomalous) wee gluons.
The pomeron is not associated with any bound-states, glueballs or otherwise.

\subhead{2.3 Cosmic Ray Neusons}

As cosmic rays, neusons will have a high energy threshold for atmospheric interactions that produce strong interaction air showers. The most distinctive feature of the produced ``dark matter showers'' will be that the initiating 
neuson will disappear from within the shower. It will simply slow down until it no longer interacts with the atmosphere and will proceed (probably) to the center of the earth, where there must be, presumably, an accumulation of ``dark matter''.
The observed shower will appear, therefore, to be less penetrating than a proton shower, while still having a very high (although underestimated) energy, and so will look suggestively like a heavy nucleus shower.

A neuson interaction that initiates a shower 
will have the general form shown in Fig.~3, where the corresponding, shower producing, proton interaction is also shown.
The conservation of sextet quark quantum numbers determines that in the neuson ``fragmentation'' region there will be a sextet state with neuson quantum numbers. The remainder (the major part) of the produced state will, at the lowest energies, simply be a soft (low transverse momentum) state composed of triplet quark 
hadrons spread across a large rapidity interval in (essentially) a universal manner. The ``wee parton'' produced hadrons involved will be the same in proton initiated showers and will be the same as those that build up the pomeron in the hadron total cross-section. Communication between a sextet state and a large rapidity triplet state has to involve gluon(s) (in a background of divergent wee gluons) and a, perhaps minor, issue that we will return to is whether, in general, this should be reflected in a reduced soft particle rapidity density in the immediate (rapidity) neighborhood of the sextet state. 

The longitudinal components of the $W^{\pm}$ and $Z^0$ vector mesons are the ``pions'' of the sextet sector and their large cross-section production (first pairs and then higher multiplicities) will be the dominant feature as the energy increases.  In neuson scattering, this will take place first 
in the fragmentation region and then in an increasingly wide central region. 
In proton-proton 
scattering, sextet states will be produced with a large cross-section only within the (widening) central plateau or, equivalently, only when  there are large rapidity states on either side, as illustrated in Fig.~3. There will be no fragmentation region production. 
Ultimately, at the highest energies,
there will be many rapidity regions of multiple vector meson production
separated by, pomeron producing, large rapidity soft hadronic states.
\begin{center}
\epsfxsize=5.8in\epsffile{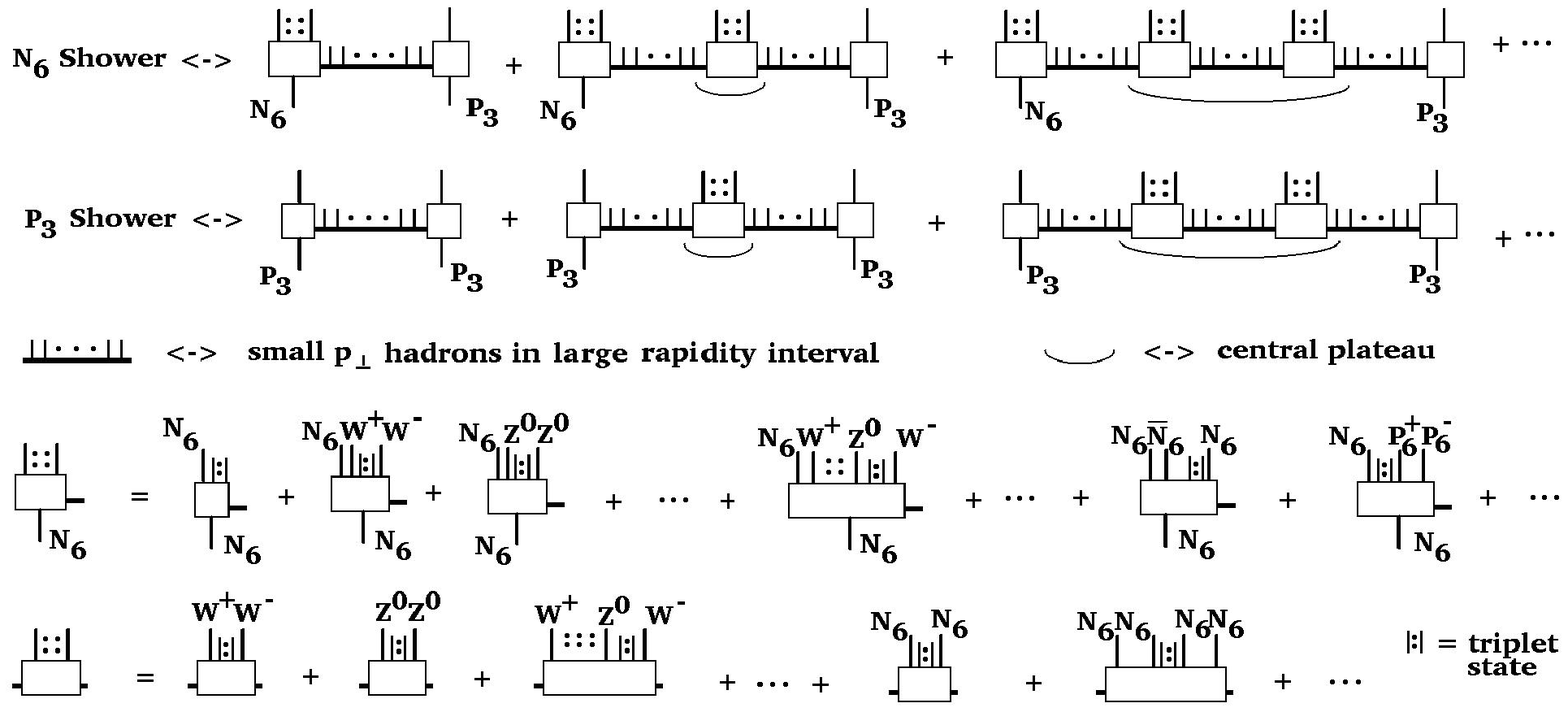}

Figure 3. The Production of Sextet States in Neuson ($N_6$) and Proton ($P_3$) showers via Large Rapidity States 
\end{center}

Dark matter neusons (as well as prosons) will be produced in both the neuson fragmentation region and the central plateau. However, since they can not be singly produced, and since a best guess for the neuson/proson mass is $\sim$ 500 Gev, I anticipate that this production will require more energy and, relative to multiple vector meson production, will increase only slowly with energy. At the highest energies, however, the neuson production will also be spread across almost all of the rapidity axis via the multiple pomeron forming interactions,
in both neuson and proton interactions, as illustrated in Fig.~3.

The decay of the vector bosons to wide-angle pairs of quarks and leptons will significantly spread the shower and so, relative to the energy, will reduce the depth and multiplicity. The shower will also spread, relative to the energy, 
if (as we discuss further later) 
the relative multiplicity density increases in rapidity regions close to the vector meson production.
Moreover, the production of high energy neutrinos will result in a substantial missing energy component in many showers. 
Within the conventional analysis of neuson showers, these properties will add to the absence of the initiating neuson in leading to the identification of the shower as originating from a heavy nucleus. 
Obviously, the underestimation of the energy will steadily increase
as the original neuson energy increases. This will be very important when we discuss the origin of the knee, and also discuss proton initiated showers. At the same time we will discuss the additional, also very important, underestimation of all shower energies resulting from the enhancement of central region small transverse momenta and high multiplicities, seen at the LHC, that we discuss in Section 4.
 
\mainhead{3. Properties of the Knee}

As shown in Fig.~4, the ``knee'' in the cosmic ray spectrum occurs between Tevatron and LHC energies. 
It is both well-known and remarkably well-established, yet it's origin is still not understood. 

\subhead{3.1 The Knee as a Dark Matter Threshold?}

Soon after it's discovery, it was suggested\cite{nik} that the knee could be the threshold for
a new interaction that produces 
neutral particles not observed in the ground level detectors.
This would produce an underestimation of the shower energy above the threshold and would lead to
a pile-up of events below the threshold energy which,
together with the depletion of the spectrum above the threshold, would be 
observed as the ``knee'' that appears to be clearly visible when the data
is presented as in Fig.~4(a). 
\begin{center}
\parbox{2.65in}{\epsfxsize=2.6in\epsfbox{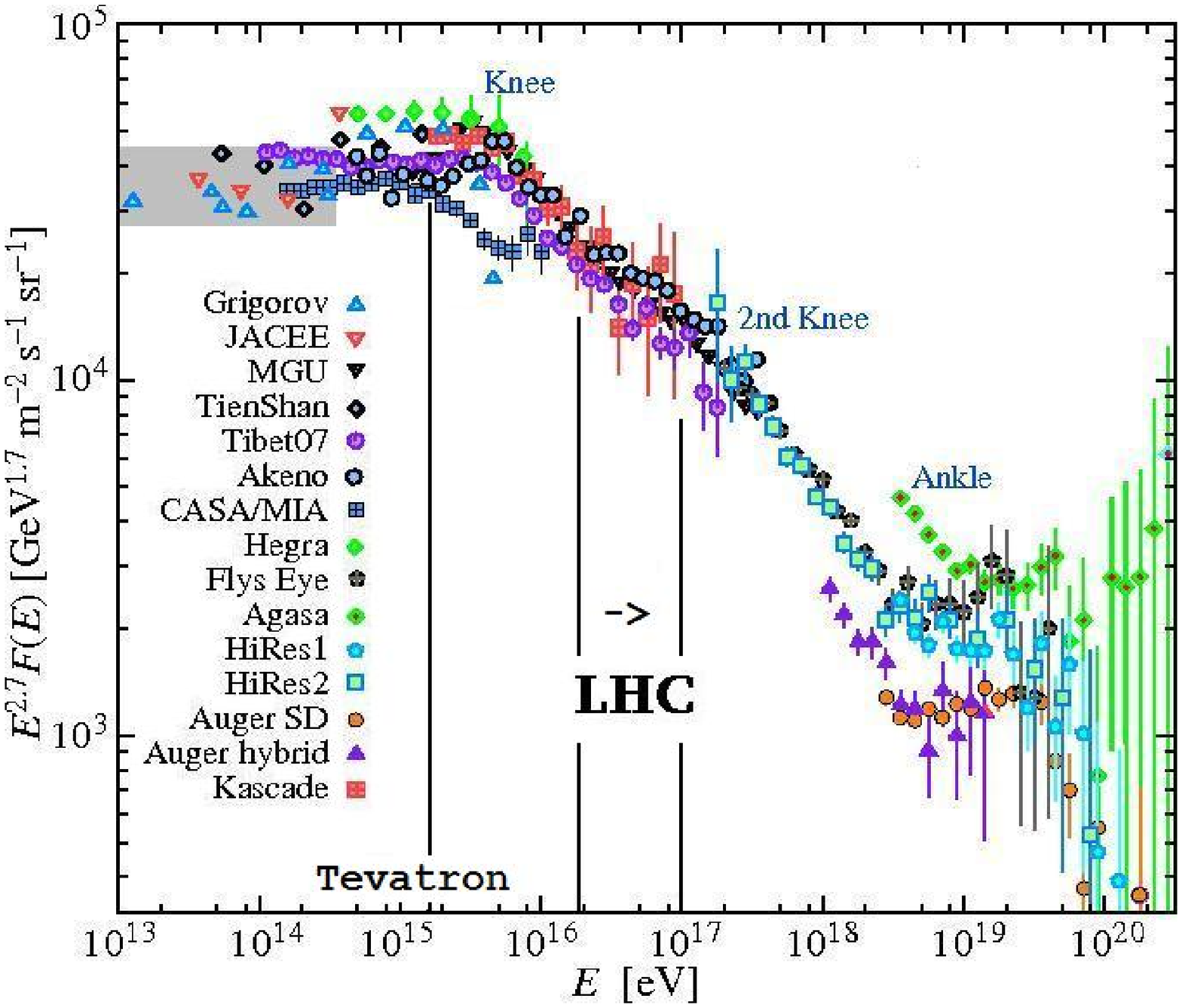}}
\parbox{3.15in}{\epsfxsize=3.1in\epsffile{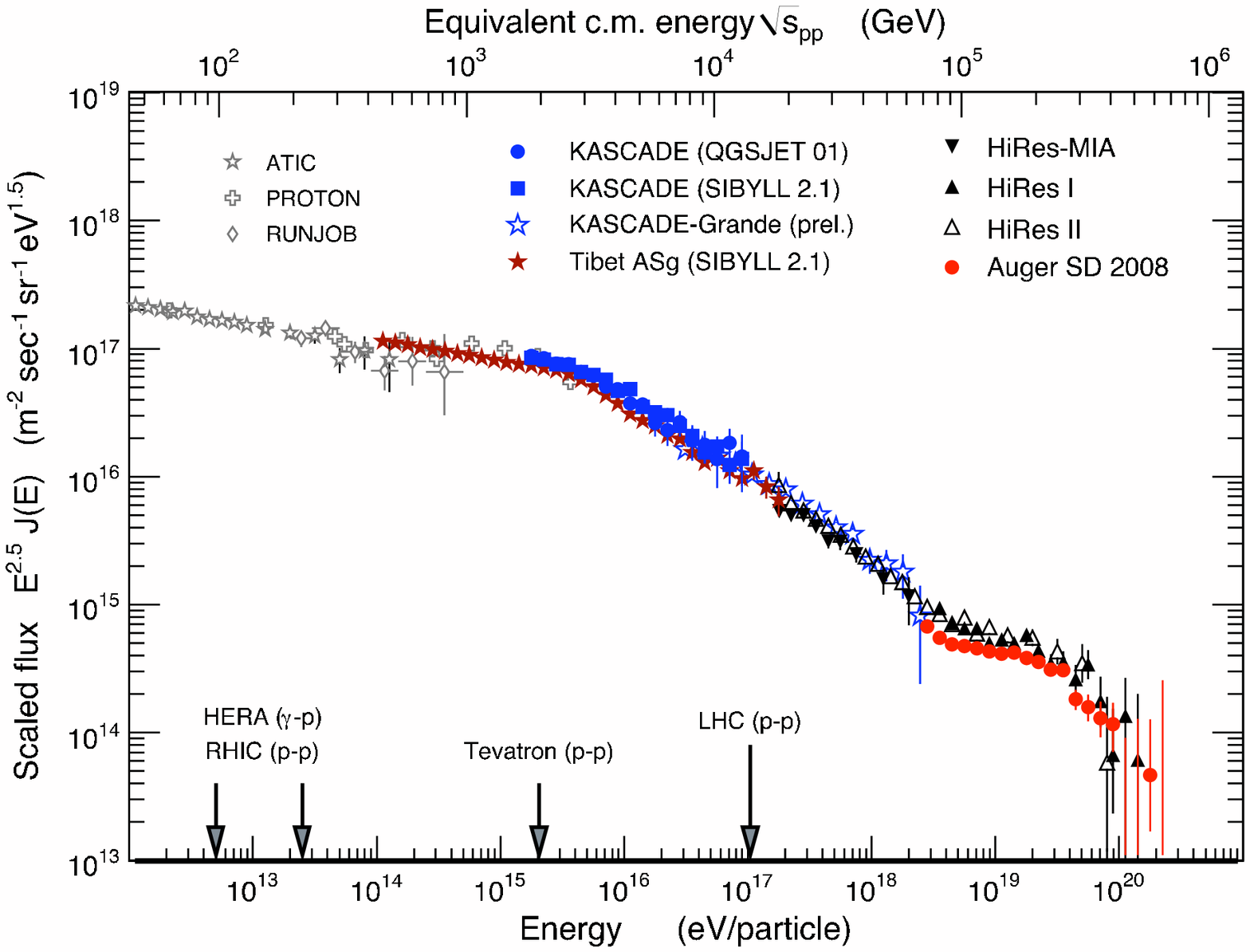}
\newline $~~$}
  
(a) \hspace{2.8in} (b)

Figure 4. The Cosmic Ray Knee (a) Data From All Experiments\cite{gs}
(b) A Selection of Data Based on Chosen Model and Composition Varying 
Analyses\cite{beh}.
\end{center}
However, Fig.~4(a) is an unselective presentation of all experiments. A 
presentation 
that exploits recent model and composition varying analyses, and is much more selective with respect to both the experiments and the models used, is shown in Fig.~4(b). It is now less obvious that there is a ``knee'', rather than a more straightforward change of slope. 
It is clear, from both Fig.~4(a) and Fig.~4(b), that if the knee were simply a neutral particle threshold in the hadronic interaction, with no related change in the 
incoming primary composition, then almost all
the hadronic cross-section would have to be affected by
the threshold at the highest energies. 
Also, although the different experiments contributing to Fig.~4(a) are
not very consistent, the threshold would, surely, have to be remarkably sharp.
Since there was no serious idea what the neutral particle(s) could
be (dark matter was essentially unknown at the time), and there was no reason to expect such a dramatic effect in the strong interaction,
particularly after the discovery that this interaction is described by QCD, there was no general acceptance of the threshold proposal.

Given the current knowledge of, not only the existence of dark matter, but also 
it's predominance in the universe, it might seem surprising that the ``dark matter threshold'' explanation of the knee has not been revisited. A major reason is that, within the 
currently popular proposals for dark matter, there is no possibility for the cross-section to be as large and as dominant as would be required. An even more important reason may be that the current consensus is (as we noted earlier) that dark matter does not strongly interact with normal hadronic matter. 

\subhead{3.2  The Confusion of Composition Analyses}

Even though the knee is, today, widely believed to be astrophysical in origin, the
only consensus is that there is no consensus on what the cause could be.
A priori, even if very special sources are involved, it seems almost inconceivable that a conspiracy of external phenomena could produce such a pronounced local effect in a cosmic ray spectrum that, 
naively at least, is arriving from all directions and all distances 
of the universe. From this point of view, it is far more plausible that the cause is in the 
atmospheric interaction. Indeed, a recent phenomenological analysis, 
has shown that the break in the Tibet III full spectrum is easily fitted\cite{djmm} by a missing energy threshold. On a more detailed level, it has also been argued that relative features of the muon and electron spectra in the neighborhood of the knee are such that it must be a consequence of an interaction change\cite{ys}.

It is remarkable that, while the number of experiments and papers
studying the knee have increased at an incredible rate in the more than fifty years since it's discovery (there are nearly 300 papers on Spires with ``knee'' explicitly in the title), there is more disagreement today, than ever, on it's nature. This is due in part to the current conflict between results obtained using different experimental methods, and/or using different hadronic models and, most importantly, the primary composition assumed. That the elemental composition is completely unknown, and so can be assumed to be energy dependent - almost without constraint, leads to dramatically different conclusions when each experiment fits the particular parameters it can measure with hadronic interaction models. 

Consider, first, the variation with energy of the average logarithmic mass of cosmic rays derived\cite{beh} by two different methods, as shown in Fig.~5. 
The general form
of the results displayed in Figs.~5(a) and 5(b) is so distinct that the only conclusion
that can be drawn is that the models used must surely be either 
misrepresenting, or missing, significant elements of the physics. 
\begin{center}
\epsfxsize=2.6in
\epsffile{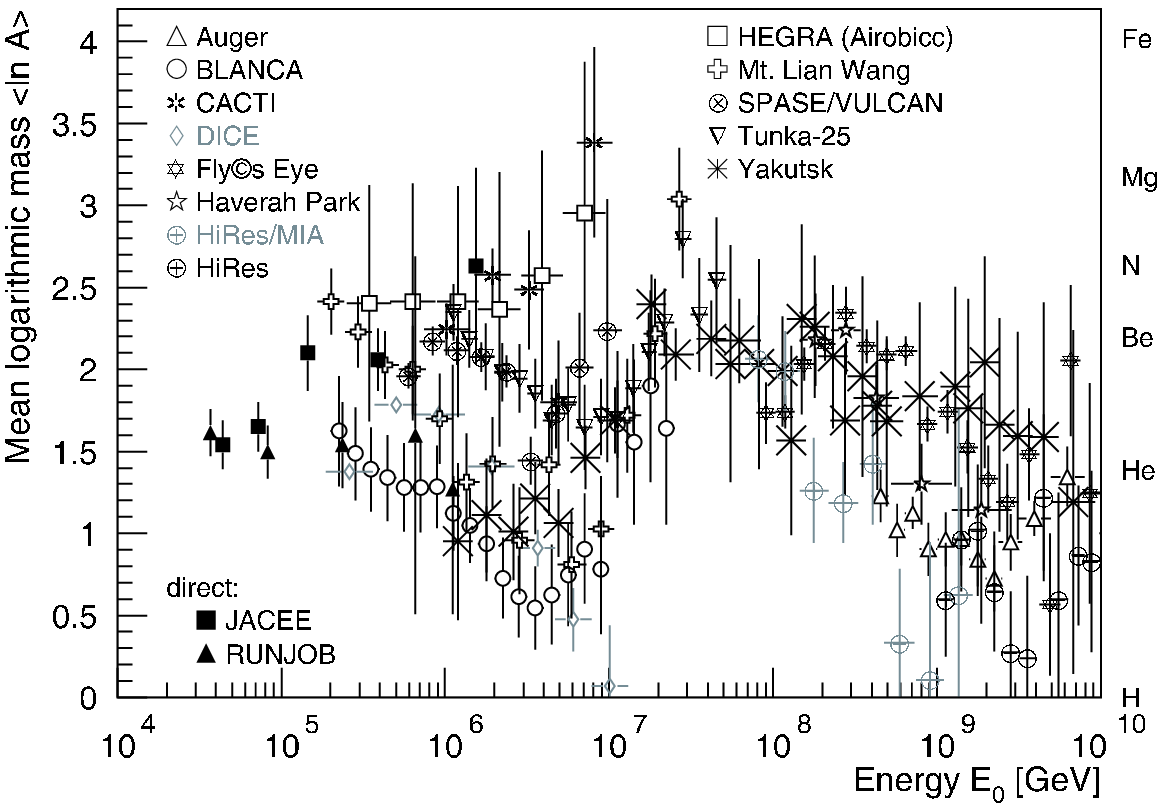}
$~$
\epsfxsize=2.6in
\epsffile{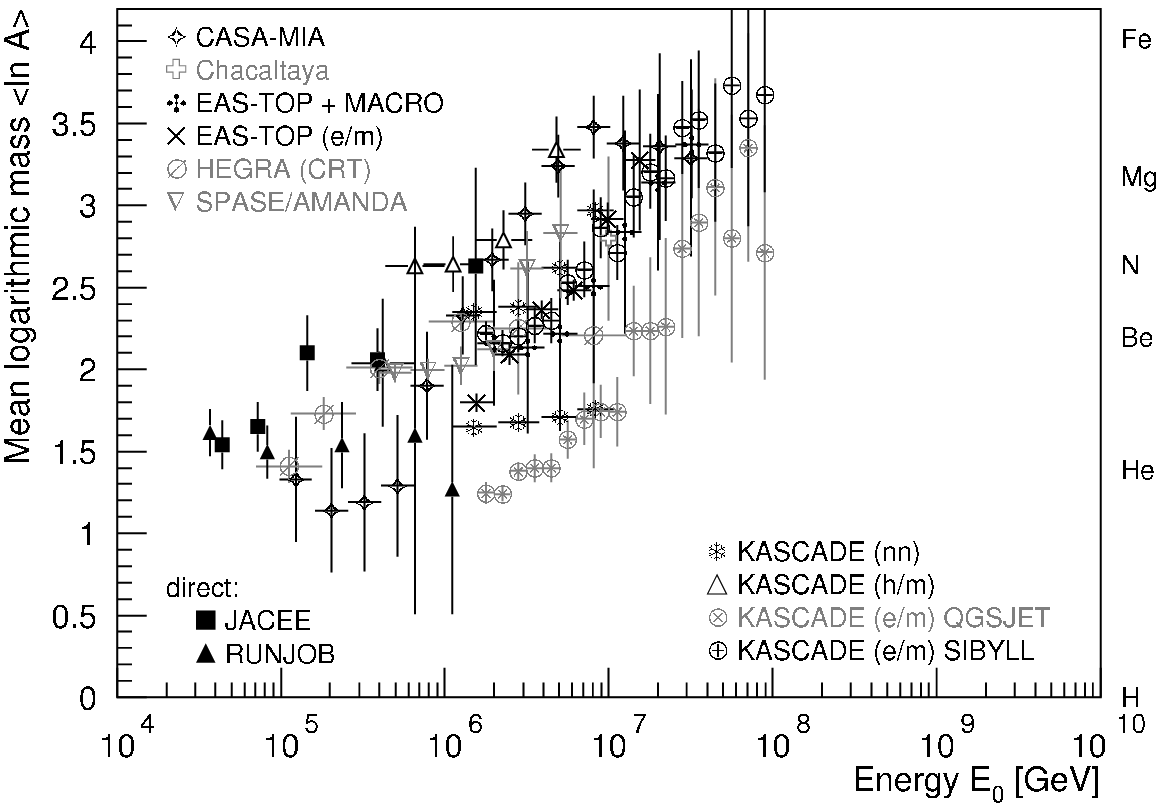}

(a) \hspace{2.7in} (b)

Figure 5. Mean Logarithmic Mass Derived Via (a) Depth of the Shower Maximum (b) Electrons, Muons and Hadrons at Ground Level
\end{center}

Moving on to the energy dependence of the elemental groups, we can compare the results obtained\cite{beh} by different experiments using a range of experimental methods and with a variety of hadronic models. The results for protons and for iron are shown in Fig.~6. 
All the experiments see the relative increase of the heavy 
component, as the knee is approached, referred 
to in the previous Section. 
However, the spread of the results is 
considerable and, as an outcome, the different experiments actually draw widely varying conclusions as to the main contribution to the knee.

Amongst the biggest experiments, 
KASCADE concludes that the knee is due to the extreme drop-off of the proton contribution that they see, with the heavy nuclei component not changing significantly. In contrast, as is apparent from 
Fig.~6, the Tibet AS experiment does not see the drop-off and, instead, claims that the heavy component must be responsible for the structure of the knee. Clearly, a straightforward variation of the composition is not universally seen and so, again, it seems likely 
that the physics involved is either not well represented or not well understood.

Results\cite{gr3} from GRAPES-3 are shown in isolation in 
Fig.~7. They lie between the extremes of the other two big experiments. 
There is a clear knee in the proton component, while the heavy component falls and then rises (relatively) as the knee is approached. Both features can be enhanced or weakened (together) by an appropriate model choice. The GRAPE-3 results fit most directly with our argument.
\begin{center}
\epsfxsize=3.8in\epsffile{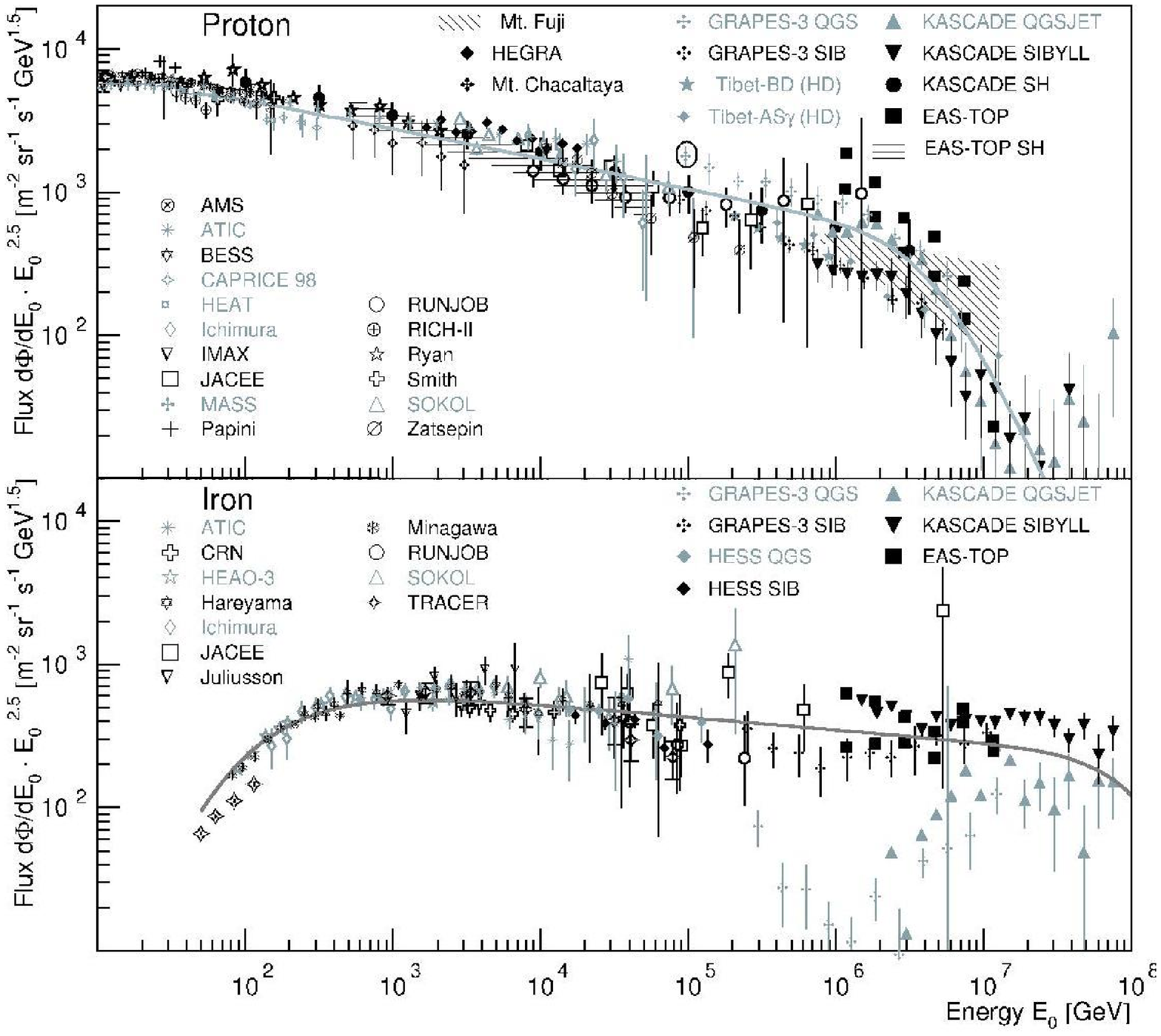}

Figure 6. The Energy Spectrum for Proton and Iron Showers 
\end{center}
\begin{center}
\epsfxsize=2.7in
\epsffile{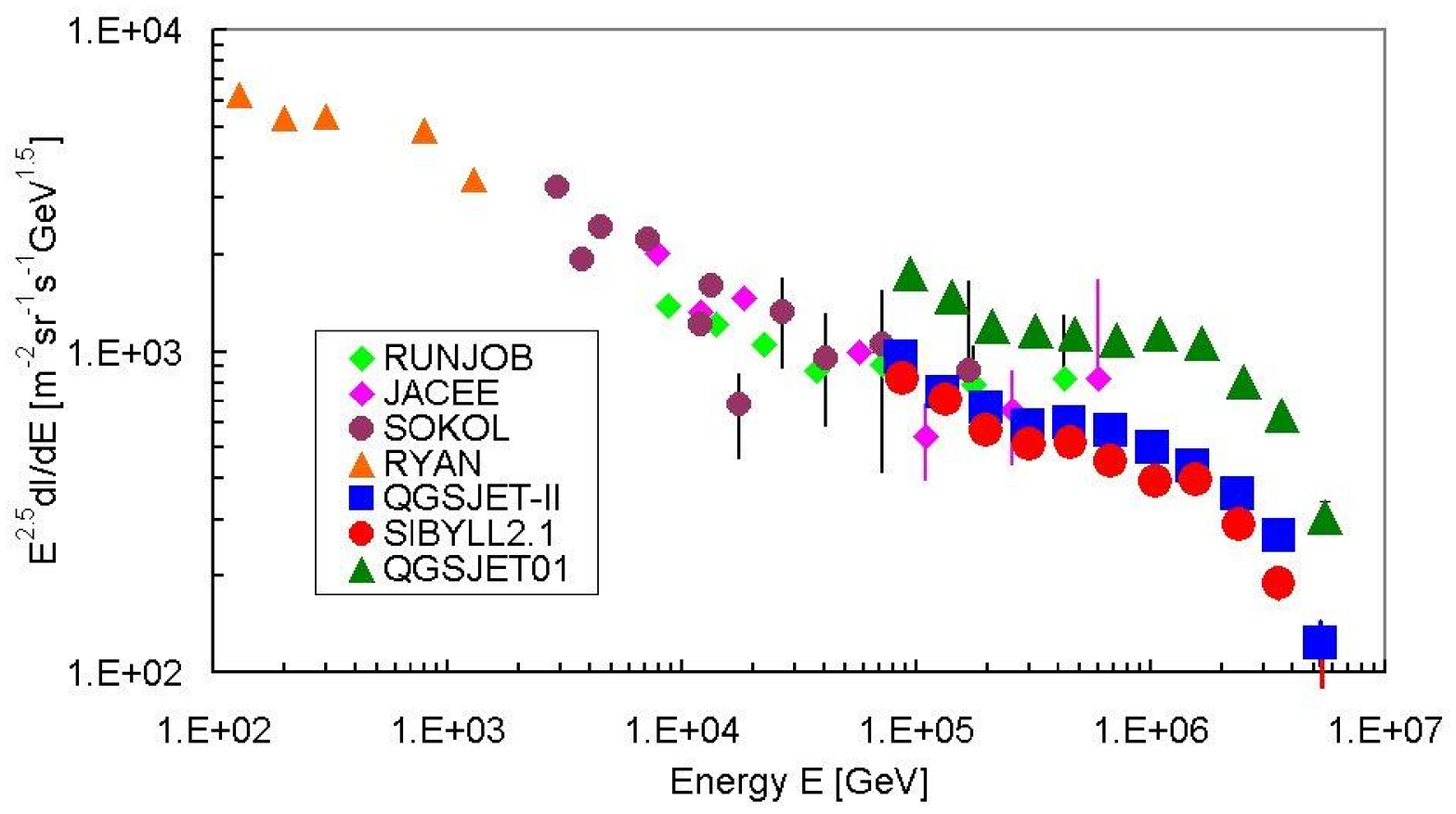}
$~$
\epsfxsize=2.7in
\epsffile{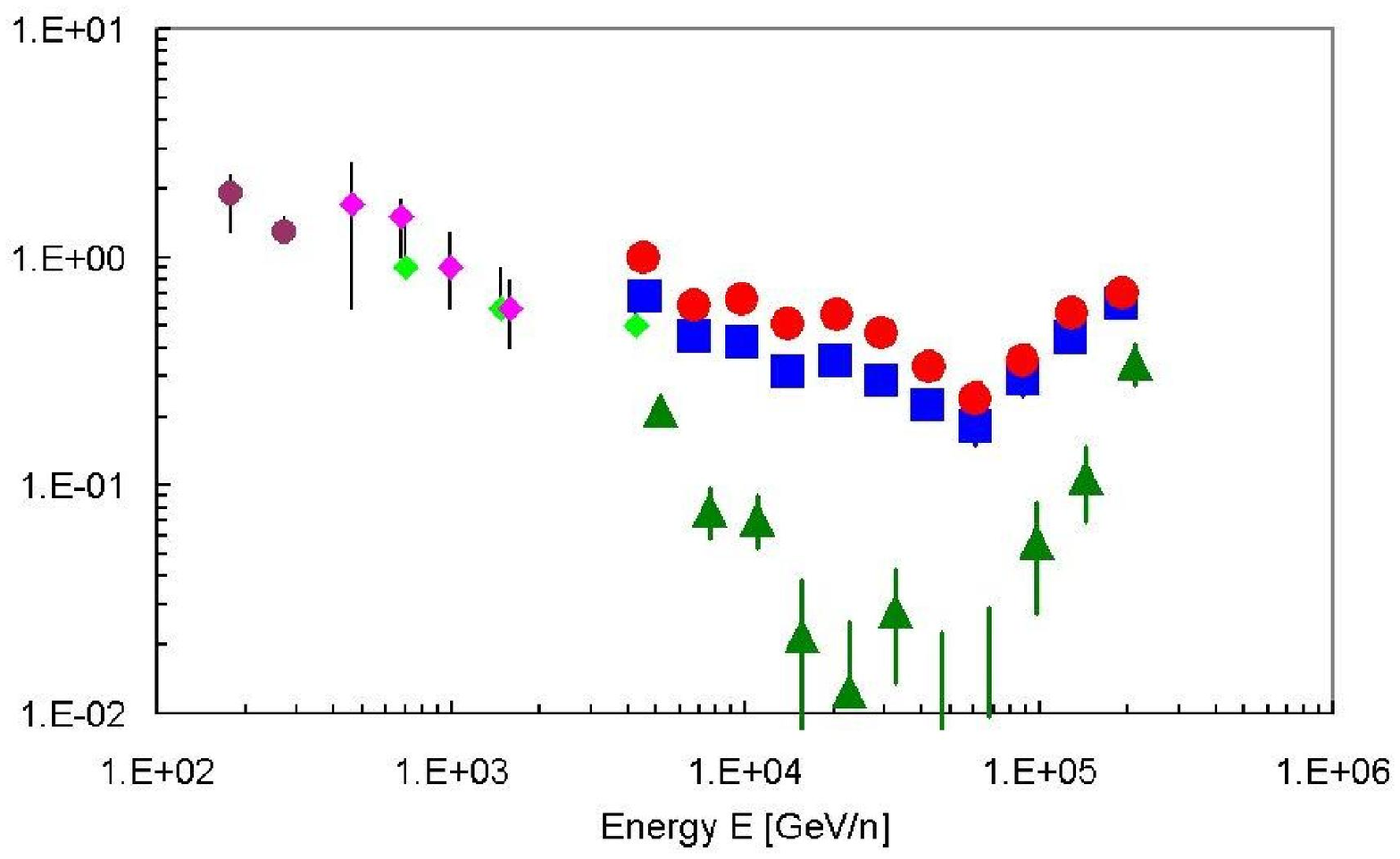}

(a) \hspace{2.7in} (b)

Figure 7. GRAPES-3 Results for (a) Protons (b) Iron
\end{center}

\mainhead{4. Minimum Bias Physics At the LHC}

As we noted in the Introduction, the recent LHC data imply that there are far more events with small $p_{\perp}$ and high central region mutiplicity than is predicted by the QCD Monte Carlo models that have been tuned to 
moderate and high $p_{\perp}$ events at Fermilab. This is seen by both ATLAS and CMS and, in principle, could also have been seen at the Tevatron if the experiments had looked for small $p_{\perp}$ particles. It is, however, a phenomenon that is increasing fast with energy, as is evident in the rapidly rising central plateau that is produced. 

\subhead{4.1 Small $p_{\perp}$ Multiplicity Distributions}

We show first, in Fig.~8(a), the ATLAS multiplicity distribution\cite{atmp} at the highest 
energy, as a function of $p_{\perp}$.  
The models shown are mostly based on Tevatron data 
with varying modifications for the LHC. In particular, the AMBT1 model has been tuned by ATLAS to their own low multiplicity (moderate $p_{\perp}$) data. As can be seen, at very small $p_{\perp}$ close to
50\% of the events are missed. Correspondingly, at large $p_{\perp}$ there is an over-estimation of the number of events that can be as high as 50\%. The highest energy ATLAS multiplicity distribution\cite{atmp} as a function of the number of charged particles
in the event is shown in Fig.~8(b). Now it is 
the higher multiplicities that are very badly 
reproduced by the models, from multiplicities of 80 upwards.
In both of the plots in Fig.~8, the $p_{\perp}$ cut-off is 100 MeV . ATLAS also has similar plots to those shown with the $p_{\perp}$ cut-off at 
500 MeV. The Monte Carlo models do much better, particularly at the lower 
energies where ATLAS also has\cite{atmp} a complete set of plots. This is hardly surprising, since this is where they were tuned. 
\begin{center}
\epsfxsize=2.2in\epsffile{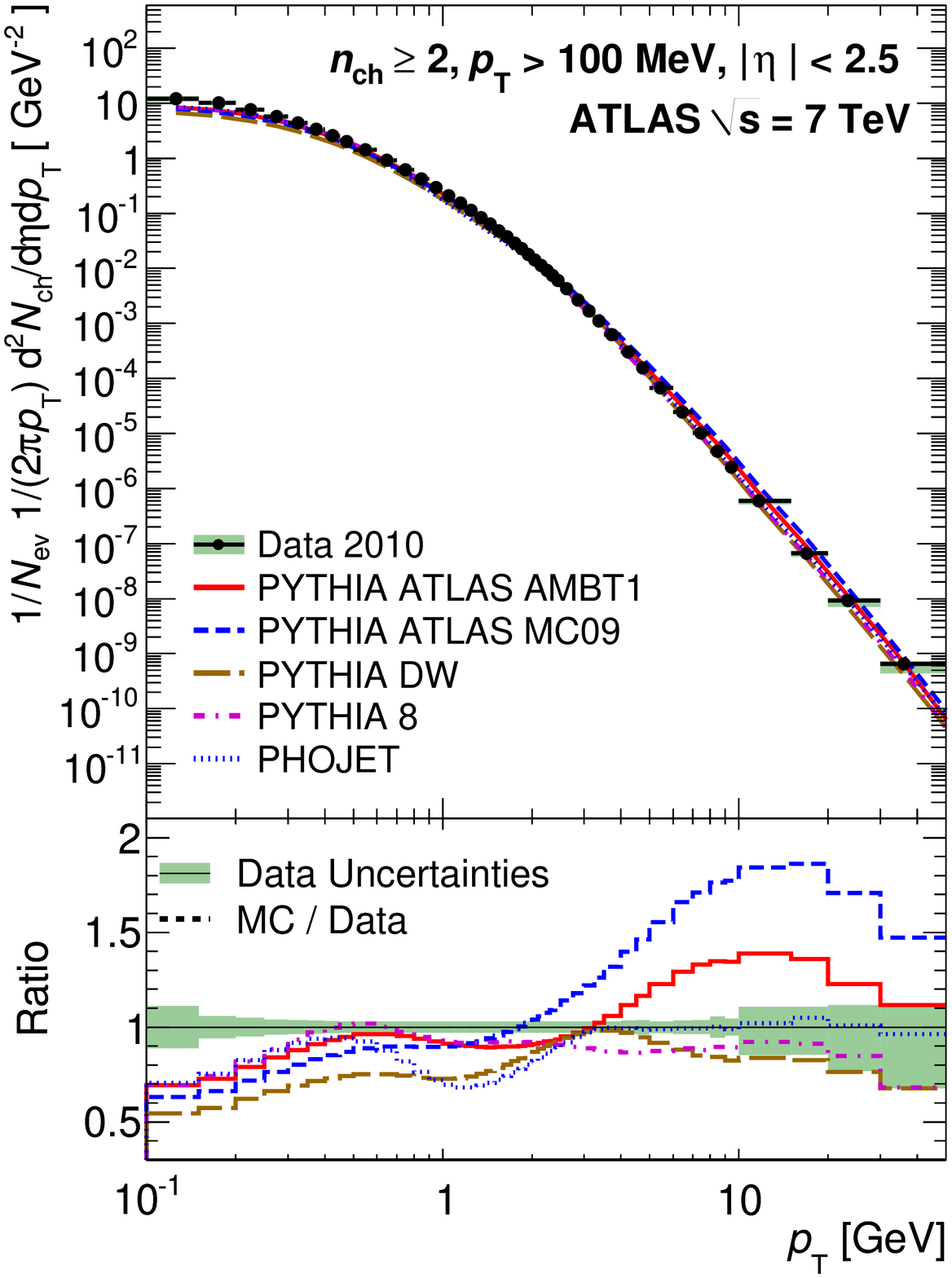}
$~~~~$
\epsfxsize=2.2in\epsffile{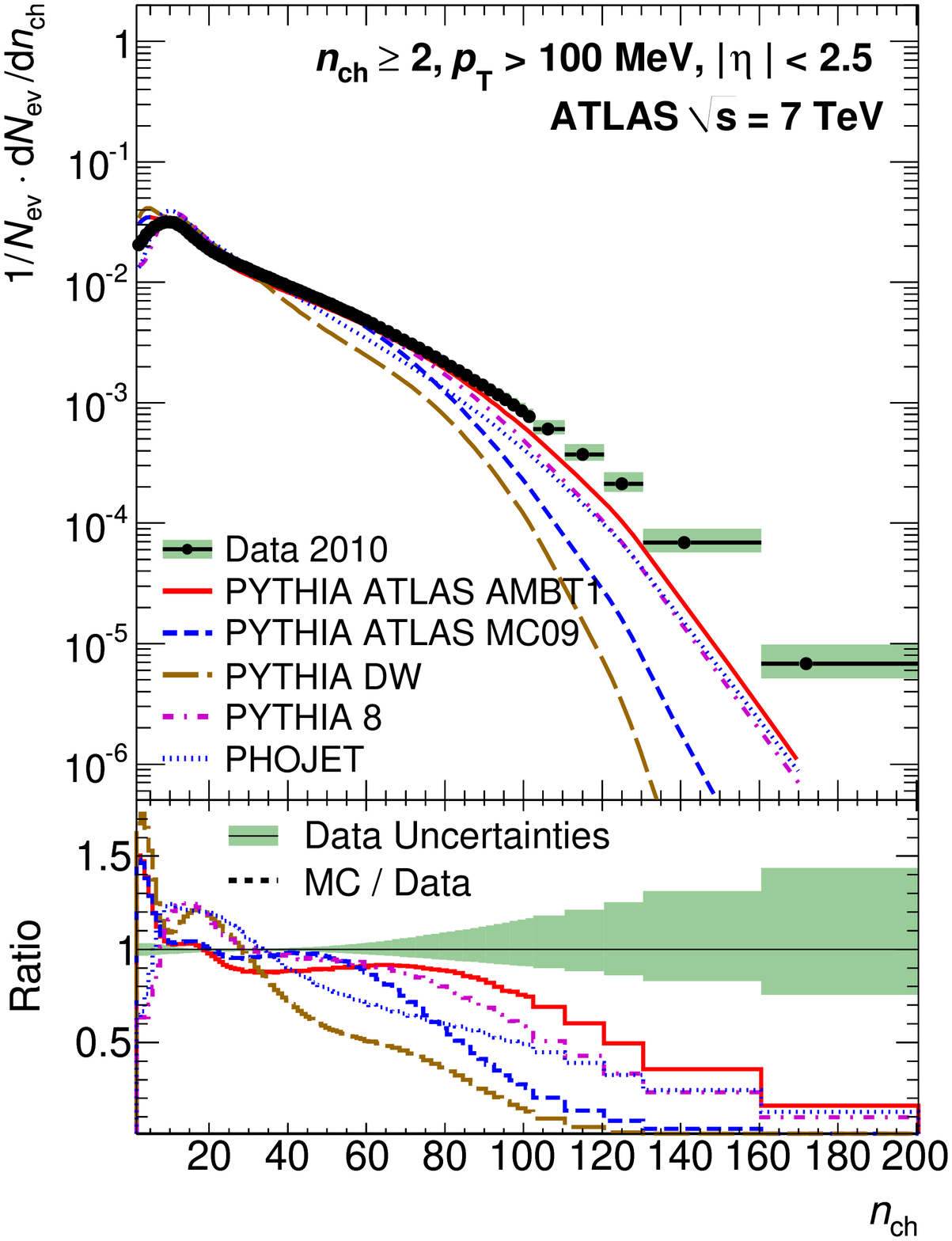}

(a) \hspace{2.5in} (b)

Figure 8. ATLAS Multiplicity Distributions at $\sqrt{s} = 7$ TeV 
\newline (a) as a Function of $p_{\perp}$ and (b) as a Function of $n_{ch}$.
\end{center}

The overall effect of the $p_{\perp}$ cut-off can be seen very easily in the CMS plots\cite{cmmp} shown in Fig.~9. 
Multiplicity distributions 
are shown at three energies, both with no $p_{\perp}$ cut-off and with a cut-off of 500 MeV. 
The high multiplicity failure of the models based on Fermilab data, at the highest energy and with no cut-off, is very evident. Again, the models 
do much better when the $p_{\perp}$ cut-off is imposed. As a further observation from Fig.~9, we note that the model which appears to best fit the no cut-off distributions, grossly overestimates the number of high multiplicity 
events when the cut-off is imposed. 
\begin{center}
\epsfxsize=2.2in\epsffile{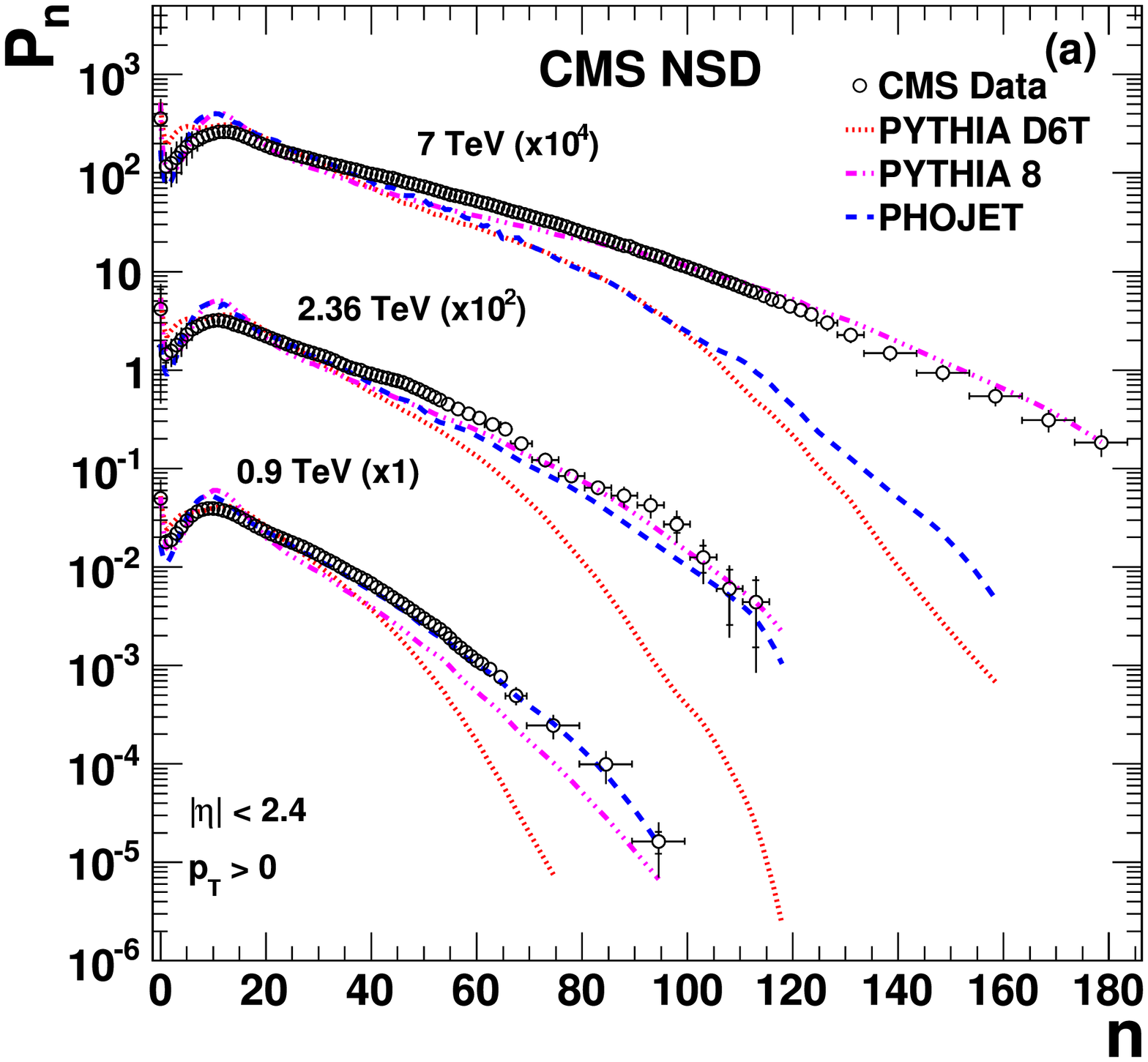}
$~~~~$
\epsfxsize=2.2in\epsffile{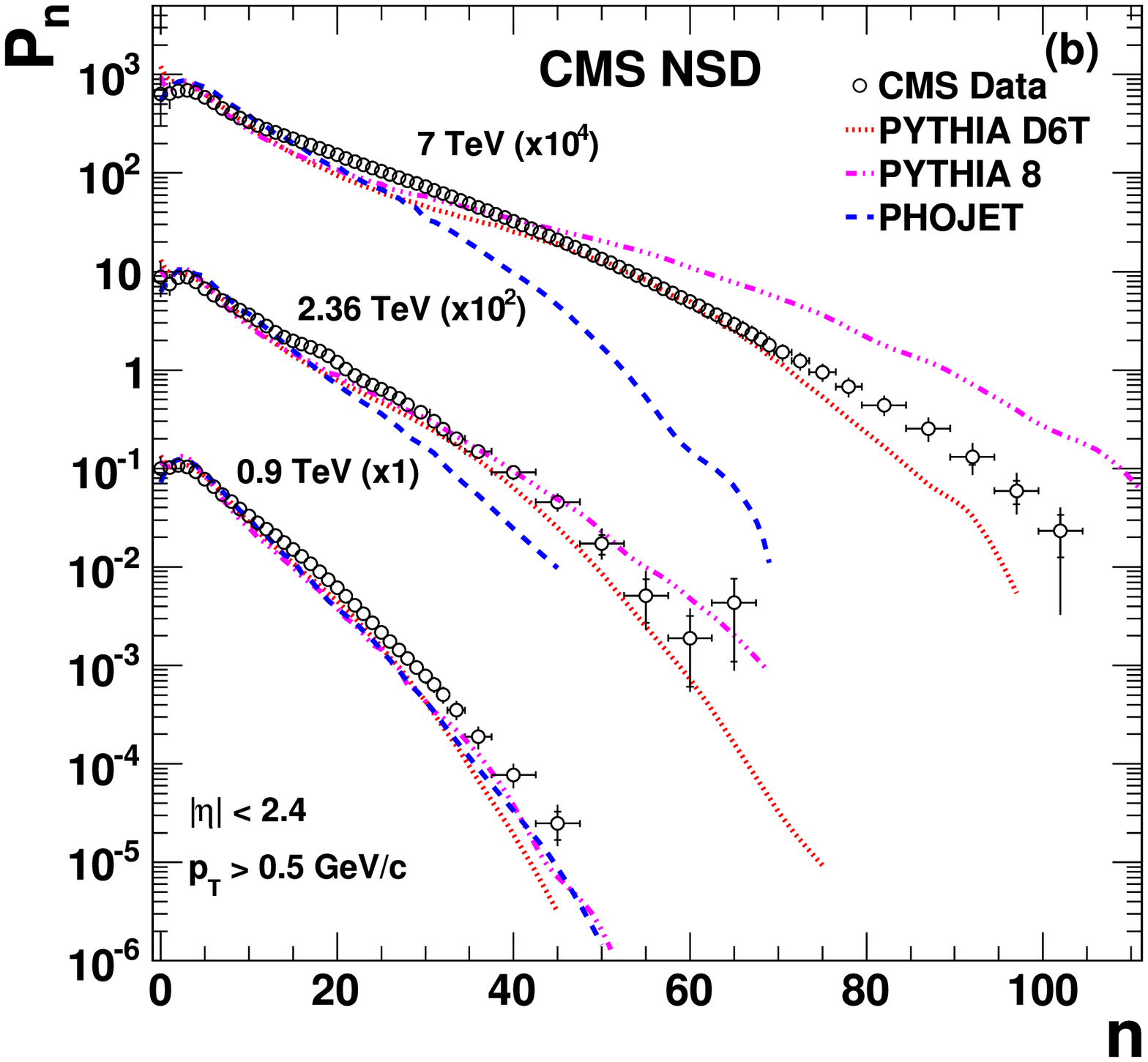}

(a) \hspace{2.5in} (b)

Figure 9. CMS Multiplicity Distributions as a Function of $n_{ch}$  
\newline (a) with no $p_{\perp}$ Cut-off and (b) with $p_{\perp} > $ 500 Mev
\end{center}

\subhead{4.2 The Central Plateau and Associated Transverse Particles}

Next, we show ATLAS results\cite{atmp} that illustrate how
the contribution of small $p_{\perp}$
large  multiplicities is dramatically evident in the rise of the central plateau. 
Results for $dN_{ch} /d\eta$, with and without small $p_{\perp}$ particles, are shown in Figs.~10 and 11. 
\begin{center}
\epsfxsize=1.7in\epsffile{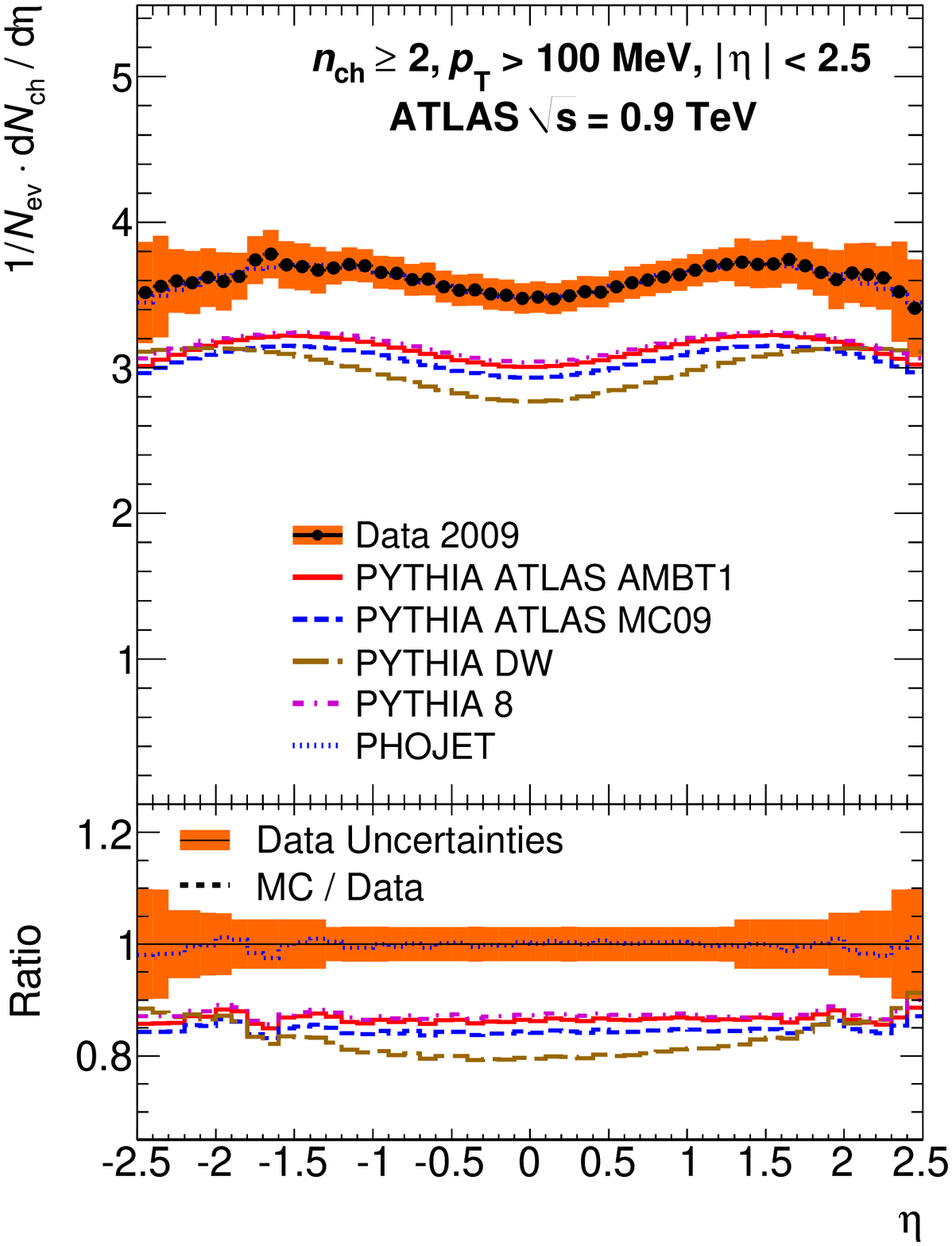}
$~$
\epsfxsize=1.7in\epsffile{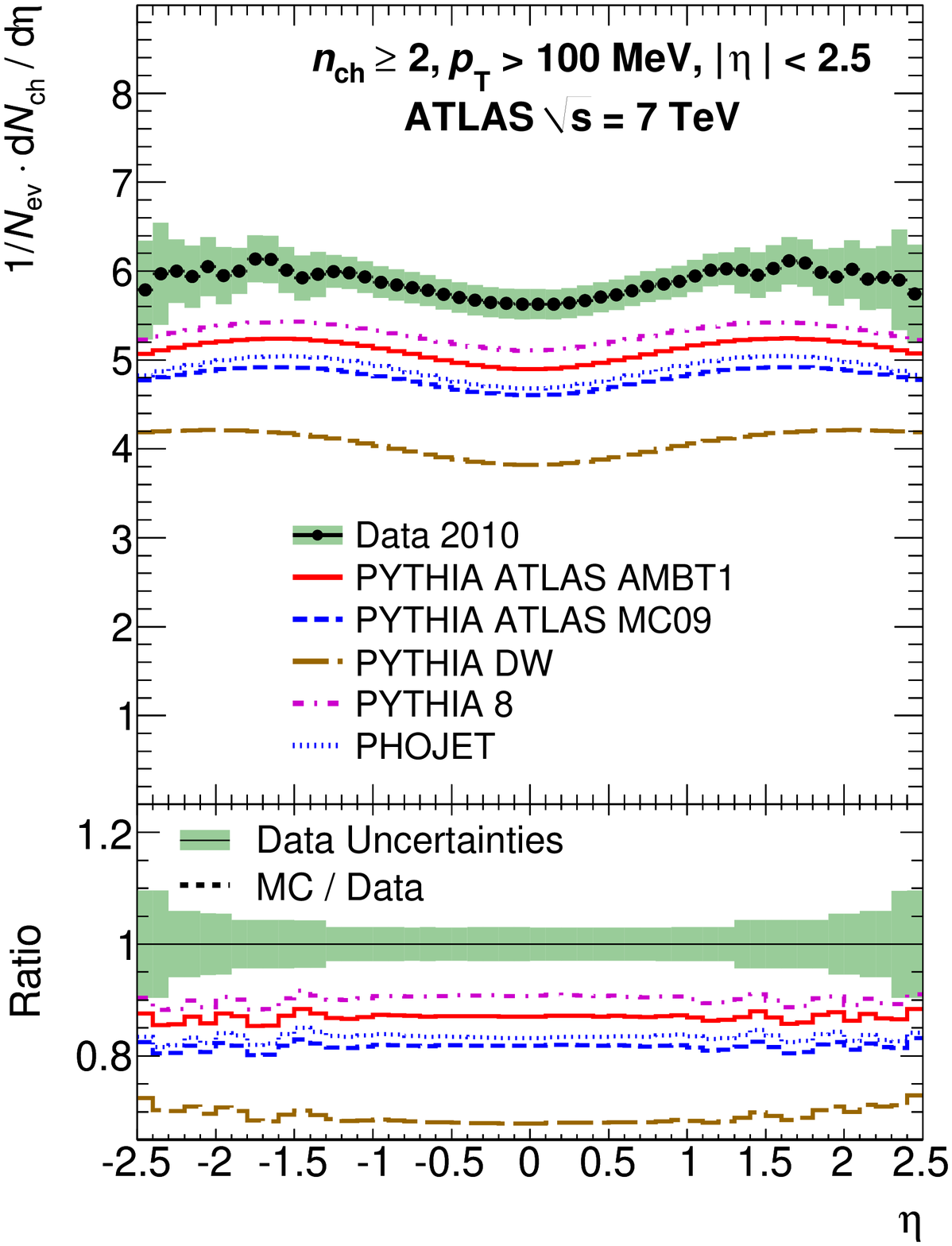}
$~$
\epsfxsize=1.7in\epsffile{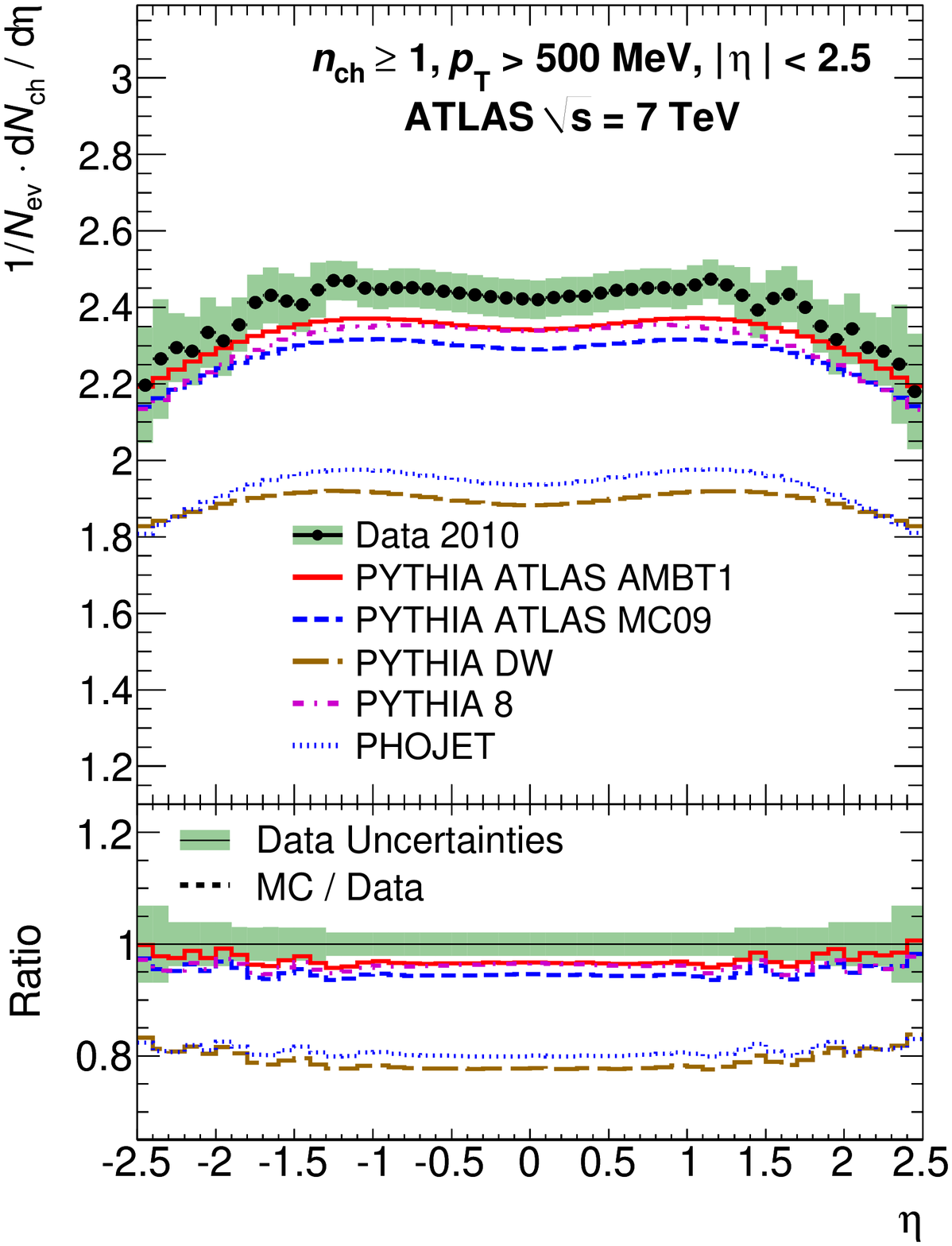}

$~~~~$ (a) \hspace{1.7in} (b) \hspace{1.8in} (c)

Figure 10. ATLAS measurements of $dN_{ch} /d\eta~~~$ 
(a) $p_{\perp} \geq$ 100 MeV, $\sqrt{s} = 0.9$ TeV 
\newline (b) $p_{\perp} \geq$ 100 MeV,  $\sqrt{s} = 7$ TeV (c) $p_{\perp} \geq$ 500 MeV, $\sqrt{s} = 7$ TeV
\end{center}
Not surprisingly, the models are again seen to 
seriously underestimate the small $p_{\perp}$ cross-sections and only fit well when a high $p_{\perp}$ cut-off is imposed, as illustrated in both Fig.~10(c) and Fig.~10(d). 
\begin{center}
\epsfxsize=2.8in\epsffile{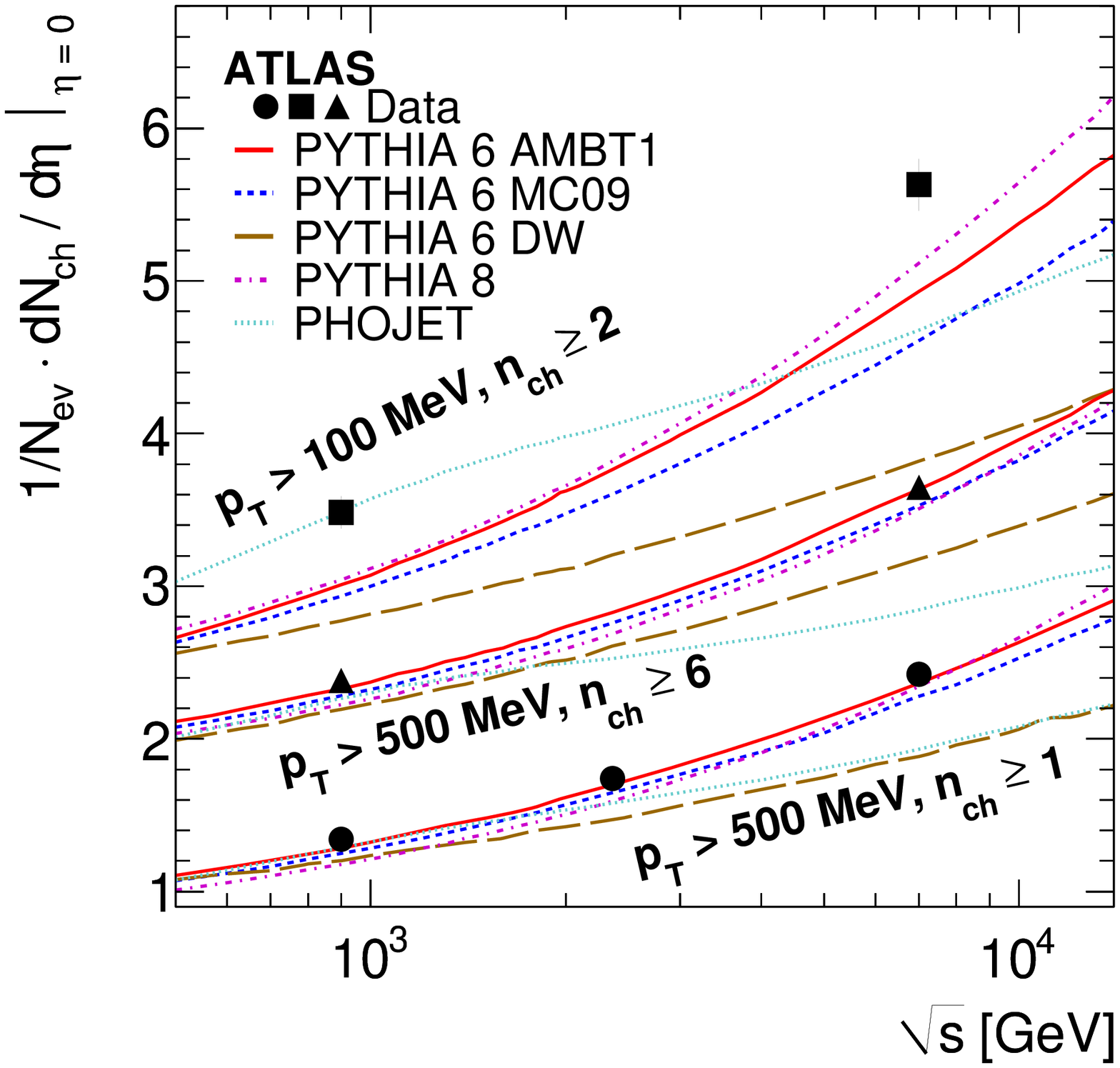}

Figure 11 The ATLAS  s-dependence of $dN_{ch} /d\eta$ at $\eta = 0$
\end{center}

Finally, we show results\cite{atul} that point directly towards the physics that we discuss in the next Sections.
In Fig.~12 we show $<d^2N_{ch} /d\eta d\phi>$, as a function of the $p_{\perp}$ of the leading particle, in the angular region transverse to the direction of this particle. 
\begin{center}
\epsfxsize=2.5in\epsffile{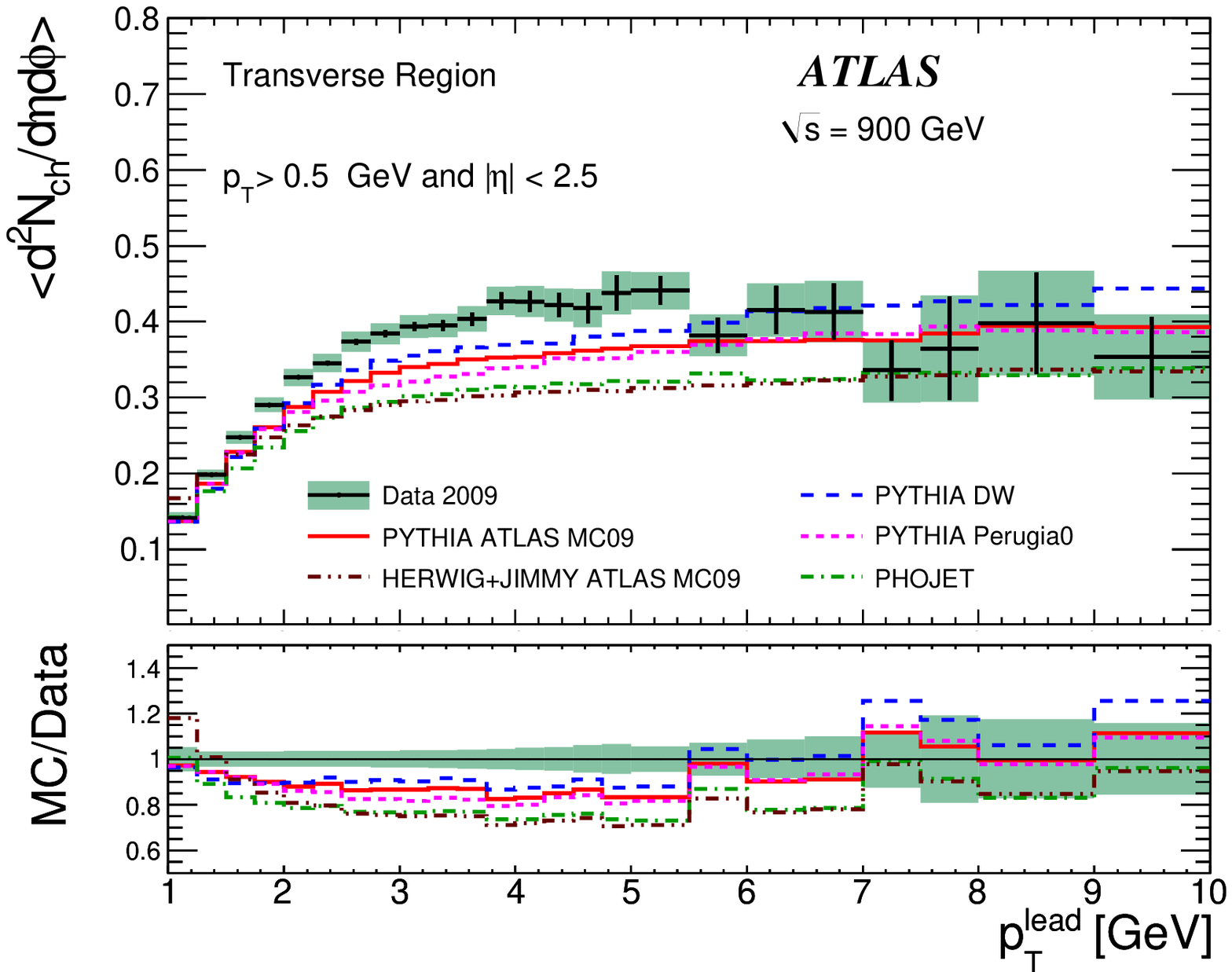}
$~~$ \epsfxsize=2.5in\epsffile{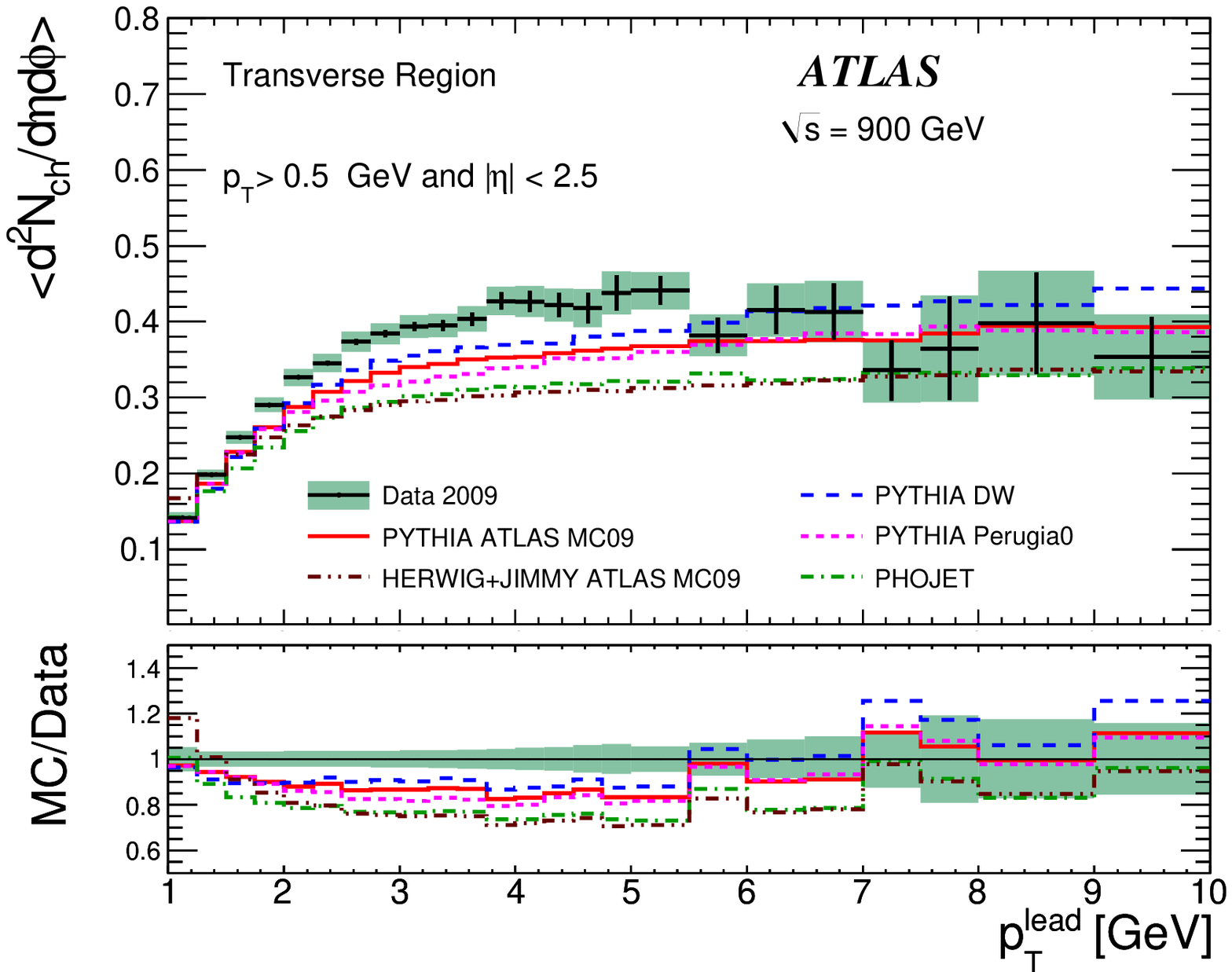}

(a) \hspace{2.8in} (b)

Figure 12 $<d^2N_{ch} /d\eta d\phi>$, as a function of the leading particle $p_{\perp}$, in the transverse region (a) $p_{\perp} \geq$ 100 MeV, $\sqrt{s} = 7$ TeV  (b) $p_{\perp} \geq$ 500 MeV, $\sqrt{s} = 900$ GeV
\end{center}
Not only is this density more than 50\% above 
the model predictions, it also shows no sign of decreasing as the leading particle
$p_{\perp}$ reaches values for electroweak boson production. Again, when the cut-off is raised to $p_{\perp} \geq $ 500 MeV and the energy is lowered, the data falls back on top of the Monte Carlo predictions.

ATLAS also has plots for other angular regions and 
at low and high energy which confirm that disparities with the models develop rapidly with energy
and are most striking in the transverse region and when small $p_{\perp}$ particles are included.  

\subhead{4.3  What is Wrong?}

At the highest energy in particular, it is plausible that there is a major physical phenomenon that is being missed entirely by the models.
All of the models have a QCD -based parton model starting-point in some way or other. As I have implied earlier, 
such models could be expected to do reasonably if the small $p_{\perp}$ physics 
involved is no more than a direct extension of semi-hard BFKL pomeron physics.
However, I have argued for many years that because the QCD
interactions that provide the BFKL framework drive the dominant physics towards large $p_{\perp}$, a solution 
of $t$-channel unitarity at small $p_{\perp}$ can not possibly be obtained this way.
Instead, a much different picture of small $p_{\perp}$ physics at high energy (that is unitary) is obtained in the very special version of QCD obtained by adding the sextet sector. Obviously, I would like
to associate, at least part of, the large multiplicity rising plateau with the pomeron intermediate states involved in the production of sextet states, as suggested in Section 2, and the sextet produced triple pomeron coupling, as discussed in the next Section. 

\mainhead{5. Multiplicity Fluctuations in Triple Pomeron RFT}

Reggeon Field Theory (RFT) formulated with a unit intercept regge pole pomeron and a triple pomeron coupling gives the Critical Pomeron directly.  The corresponding ``cut RFT'' describes the multiplicity
fluctuations in large rapidity hadron scattering that will dominate the asymptotic
Critical Pomeron. (Higher-order couplings, although present in general, are ``irrelevant'' for the critical behavior.)

\subhead{5.1 Multiplicities and Cut Pomerons}

Consider the contribution 
to the total cross-section of a large rapidity soft hadron state of the kind that appears in Fig.~3. As illustrated in Fig.~14, increased multiplicity densities, relative to a minimal average density, are represented by a corresponding increase in the number of exchanged cut pomerons.
The asymptotic Critical Pomeron is produced by summing, to all orders, diagrams of the form of those in Fig.~14. The result is a central region multiplicity density increasing with a power of the energy that is a critical exponent. Given that the theory is known to be critical, a large triple pomeron coupling implies that the contribution of low-order diagrams will give a reasonable approximation to the asymptotic behavior.  

The complete set of cut pomeron diagrams includes diagrams (that we have not shown) in which some pomerons are uncut and so describe rapidity gap interactions. Nevertheless, it is clear from the diagrams shown that a large triple pomeron coupling produced by sextet quark anomalies (as will be discussed in Section 9) will obviously result in increasingly many events with large central region multiplicities. The scale at which the asymptotic amplitudes appear is, at present, something that we can only discuss empirically. That it is related to the electroweak scale would
obviously explain the consequent appearance of large multiplicity small transverse 
momentum states at the LHC.
\begin{center}
\epsfxsize=5in\epsffile{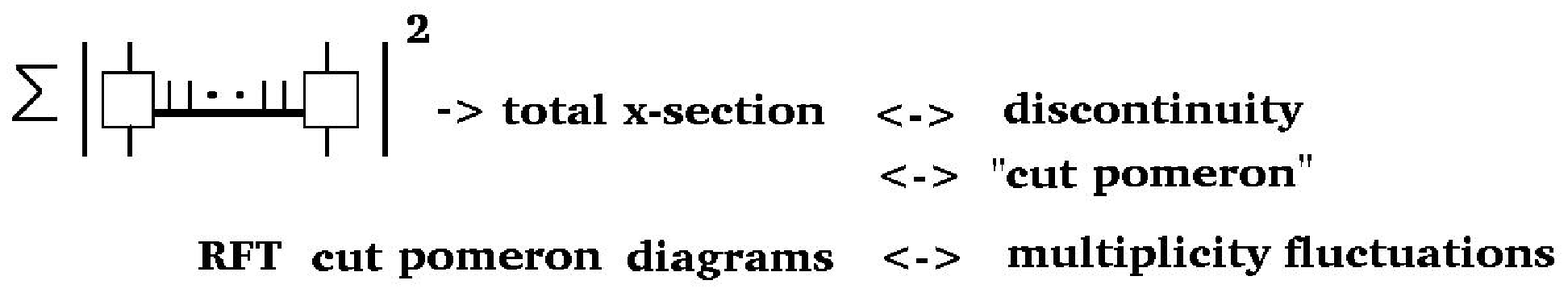}

\epsfxsize=5.5in\epsffile{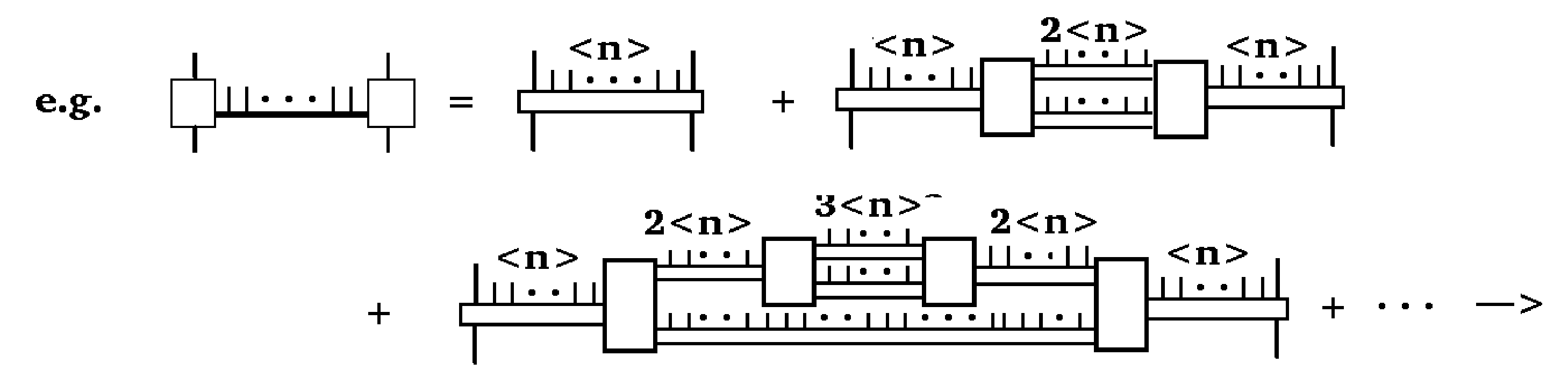}

\epsfxsize=5.5in\epsffile{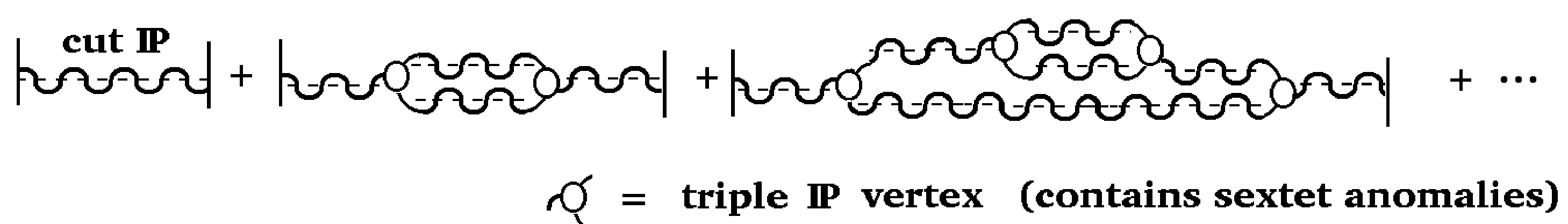}

Figure 14. Multiplicities Associated With Cut Triple Pomeron Diagrams
\end{center}

\subhead{5.2 Cut Pomerons and Central Plateau Sextet States}

The most important consequence of the activation of the 
sextet sector at the LHC should be the central region production of sextet states, as illustrated in Fig.~3. The plateau production of sextet states
and the corresponding description by cut pomeron diagrams,
as illustrated in Fig.~14, will be identical in both proton and neuson interactions. It will be predominantly multiple electroweak
bosons, with the number increasing rapidly with energy. 
The relative multiplicity density should be large in rapidity regions on either side of the vector meson production as a consequence of the strong coupling of soft hadron states to the sextet states. As the energy increases,
multiplicity fluctuations within the soft hadronic state will also become increasingly significant, producing an even more dramatic rise in the central region multiplicity density. 

As I discuss further in Sections 7 and 8, the detection of multiple vector meson 
states produced across a wide part of the rapidity axis, in association with high multiplicity soft hadronic states, will be 
a major challenge for the LHC experiments. This is because of the limited rapidity coverage of the detectors and the consequent difficulty of identifying all the leptons involved.
\begin{center}
\epsfxsize=5.5in\epsffile{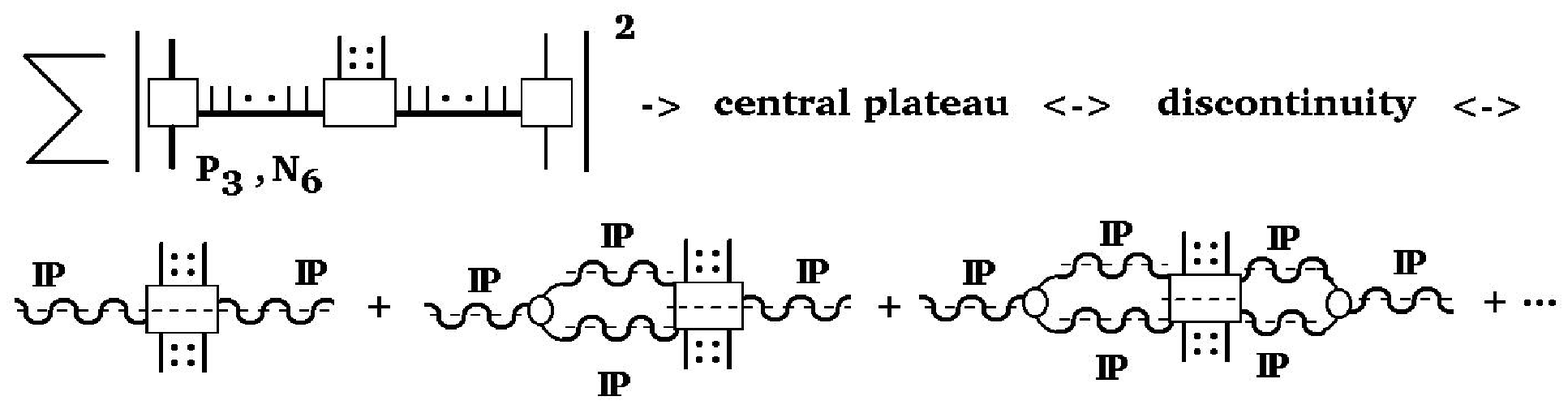}

Figure 14. Plateau Sextet States Produced in Neuson and Proton Interactions 
\end{center}

\mainhead{6. The Origin of the Knee}

Obviously, to be responsible for the knee, the sextet sector has to produce a major change in the strong interaction. If the cross-sections are as large as the anomaly dynamics arguments of later Sections will suggest, the change with energy of the neuson and proton interactions, could be exactly what is needed.

The strong increase with energy of small transverse momenta central region multiplicities that we have attributed (indirectly) to the sextet sector, clearly
was unknown before the new LHC data. As a consequence it has not been included 
in the models used for shower analysis. The resulting spread of the shower will be a
major factor in underestimating the energy via the depth of shower maximum.
Most likely, however, the central region production of sextet states will be equally significant in spreading the shower. 

\subhead{6.1 Thresholds and the Spread of Showers}

As discussed for neusons in Section 2, 
the effect of vector boson decays to wide-angle pairs of quarks and leptons will be to cause a spread and reduce the  
depth and multiplicity dramatically for both proton and neuson showers , surely causing the energy to be substantially underestimated. In addition,
the production of high energy neutrinos will result in showers with a large missing energy component. Also, as the energy increases,
multiplicity fluctuations within the soft hadronic state, towards the central region, will become increasingly
significant (as we discussed in the last section). In general,
there will obviously be a very significant underestimation of the highest shower energies.  

As described in Section 2, it is because the production of showers by neusons requires a pomeron-related large rapidity state that there is an effective threshold. Taking into account the underestimation of neuson shower energies, it would be consistent with pomeron phenomenology and the anticipated large neuson mass if this threshold is in the energy range (not far below the knee) where, as I noted in the Introduction and discussed more in Section 3, the experiments have all seen an apparent increase with energy of the heavy nuclei component relative to the proton component. Although less so than the Auger result, this relative increase also seems counter-intuitive. If it is largely a consequence of the interaction threshold for dark matter neuson showers, this would be consistent with such showers being in the majority at the highest energies seen by Auger. Indeed, the more recently apparent ``ankle'', that is present in the very high energy end of the full cosmic ray spectrum shown in Fig.~3, could be directly associated with the increasing dominance of neuson showers 
in the incoming spectrum that I interpret the Auger data as implying. 
As presented in Fig.~2(a), the Auger data show a clear 
discontinuity in the energy interval around the ankle.

Because they can not be singly produced, and because of their anticipated large mass, the production of dark matter neusons (as well as prosons) will require more energy and, relative to multiple vector meson production, will increase only slowly with energy, just as in neuson showers. At the highest
energies, the vector meson and neuson production will spread across the rapidity axis, via multiple pomeron forming interactions, in an identical manner in both proton and neuson showers, as described in Section 2 and illustrated in Fig.~3. 

If the atmospheric interaction threshold for arriving dark matter neusons is at a lower energy than the knee, then
they will make a major contribution in the knee region. This  threshold  should be lower than that required for sextet interactions
in proton collisions because only one large rapidity interaction is needed rather than two. While the much larger neuson mass will raise this threshold, the underestimation of the neuson shower energy will, effectively, lower it.

\subhead{6.2 Schematic Spectra}

A schematic illustration of how the two kinds of showers 
might contribute to the full spectrum in the neighborhood of the knee, which takes into account all the properties that we have discussed, is shown in Fig.~16. 
I have taken the
contribution of neuson and proton showers to be comparable directly above the knee, since this is (very crudely) what is suggested by Fig.~2(b). I have already suggested that, at much higher energies, the Auger data
(and perhaps the presence of the ankle) could be due to the increasing presence of neusons in the incoming spectrum.

If the sextet sector is indeed responsible for the phenomena I have described, then a major revision of high-energy cosmic ray phenomenology is called for. This is, however, a challenge far more daunting than providing a consistent triplet sector phenomenology, the difficulties of which have already been apparent in the previous Sections. It should be emphasized, therefore, that there is no calculational, or even phenomenological, framework involved in producing the spectra shown in Fig.~15. I have found, however, that consistency with all of the above discussion does not allow as much freedom as might be thought. 
\begin{center}
\epsfxsize=4.5in\epsffile{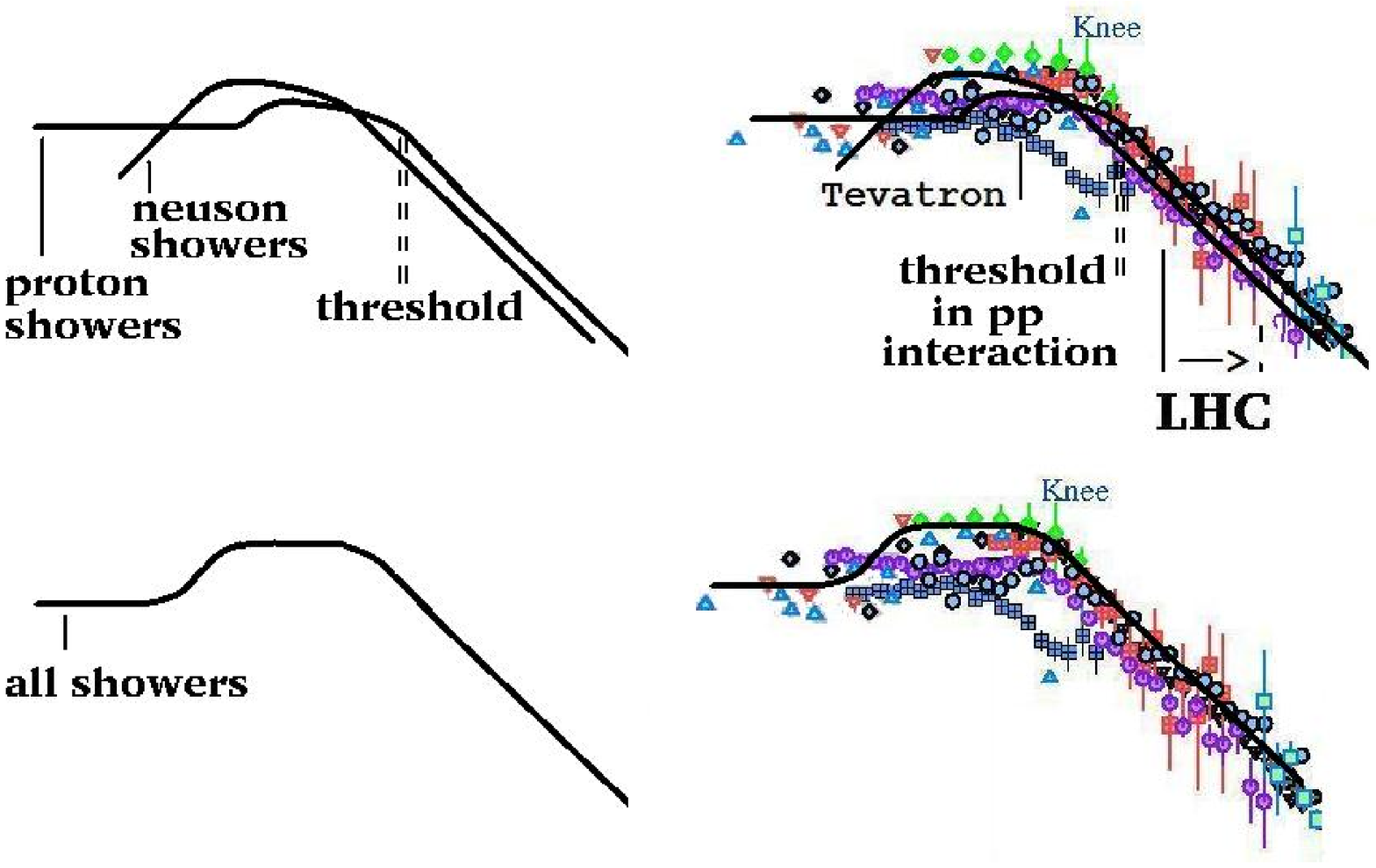}

Figure 15. Spectra Potentially Produced by Neuson and Proton Showers
\end{center}

The central issue is clearly, how big must the proton cross-sections for sextet state production be at the LHC? The implication of Fig.~15 is that the threshold for the appearance of the sextet sector in the proton interaction is just below
the current LHC energy. It is obviously hard to postpone it much further,
although it could easily be lower, implying that the phenomena involved
should have begun to appear at the Tevatron, as I will suggest in the next Section. In either case, there is no escaping the conclusion that the phenomena involved must become increasingly apparent as the LHC energy is increased. The central question is, how much of the physics can be seen by the detectors, as currently configured, and with the accelerator running at such a high luminosity? I will discuss this further, after I have discussed what has 
been seen, as well as what might have been missed, at the Tevatron. 

\mainhead{7. At the Tevatron - Seen (and Perhaps) Unseen}

A significant number of Z pair events have been recorded at the Tevatron.
Initially, the cross-section appeared to be consistent with the Standard Model. 
However, this consistency has been slowly eroded via a number of 
analyses\cite{CDF4l1,CDF4l2,CDF4l7} performed using a gradually increasing data sample. Currently, the cross-section appears to 
be\cite{CDF4l7,CDFEPS} about twice the Standard Model value, 
largely due\cite{CDF4l7} to the presence of a dramatic high mass cluster of events.
Unfortunately, the analyses can not be easily compared because, as the event selection and tracking procedures 
have improved, the kinematic details of some events have changed and other events have moved in to or out of the final sample chosen. The CDF analysis\cite{CDF4l7} that produces the high mass events is the most recent and so, I assume, uses maximally improved procedures.

In general, the selected ZZ events
satisfy very strict requirements, as is illustrated by 
a four electron event that was discarded until the most recent analysis but is of particular importance for our discussion. Very fortunately, this event was
recorded sufficiently long ago that there is no pile-up (multiple vertex) problem in the event display - from which we learn a lot. 

\subhead{7.1 The CDF $Z^0Z^0$ Event}

The event\cite{hkg} ({\small R/E 147806/1167222}) of interest is, in fact, a high multiplicity event of the kind anticipated in Section 5 and illustrated in Fig.~3. It was recorded in 2004 and is shown in Fig.~16. 

There were two central positrons and two plug electrons which, when combined as indicated ($e^+_{a,b}$ with $e^-_{a,b}$), give unambiguous $Z^0$ masses that are given in \cite{hkg} as 91 and 92 GeV, with the pair mass, given as 194 GeV, being just above threshold. 
The event was initially counted in the CDF $Z^0$-pair sample but it was subsequently discarded because one of the electrons is insufficiently isolated. It was not included in cross-section estimates until the analysis of \cite{CDF4l7}. (It is the first event in the table of Fig.~19 - with the kinematic details updated in only
a minor way.) Given the accumulated luminosity at the time, a four electron event would be a very rare event within the Standard Model, suggesting that it might well be part of a cross-section that is being missed altogether.

From our viewpoint, this event has several important properties. Firstly, when the 
$E_T$ cut-off is set at 500 MeV, as in Fig.~16(c), only a small number of accompanying particles is seen. When this cut-off is lowered to 100 MeV, as in Fig.~16(d), 
a very large number of particles ($>$ 70) appears, covering a large part of the rapidity axis away from the produced electrons. Clearly, a high multiplicity, small 
$p_{\perp}$, event of the kind discussed in previous Sections
is accompanying the vector boson production. As we have emphasized, a large cross-section for events of this kind, is just what is predicted
when the sextet quark sector is responsible for electroweak symmetry breaking. So could such a cross-section have been missed at the Tevatron,
in part because of the general lack of interest in the small $p_{\perp}$ component of events?
\begin{center}
\parbox{2.25in}{
\begin{center}
\epsfxsize=1.9in\epsffile{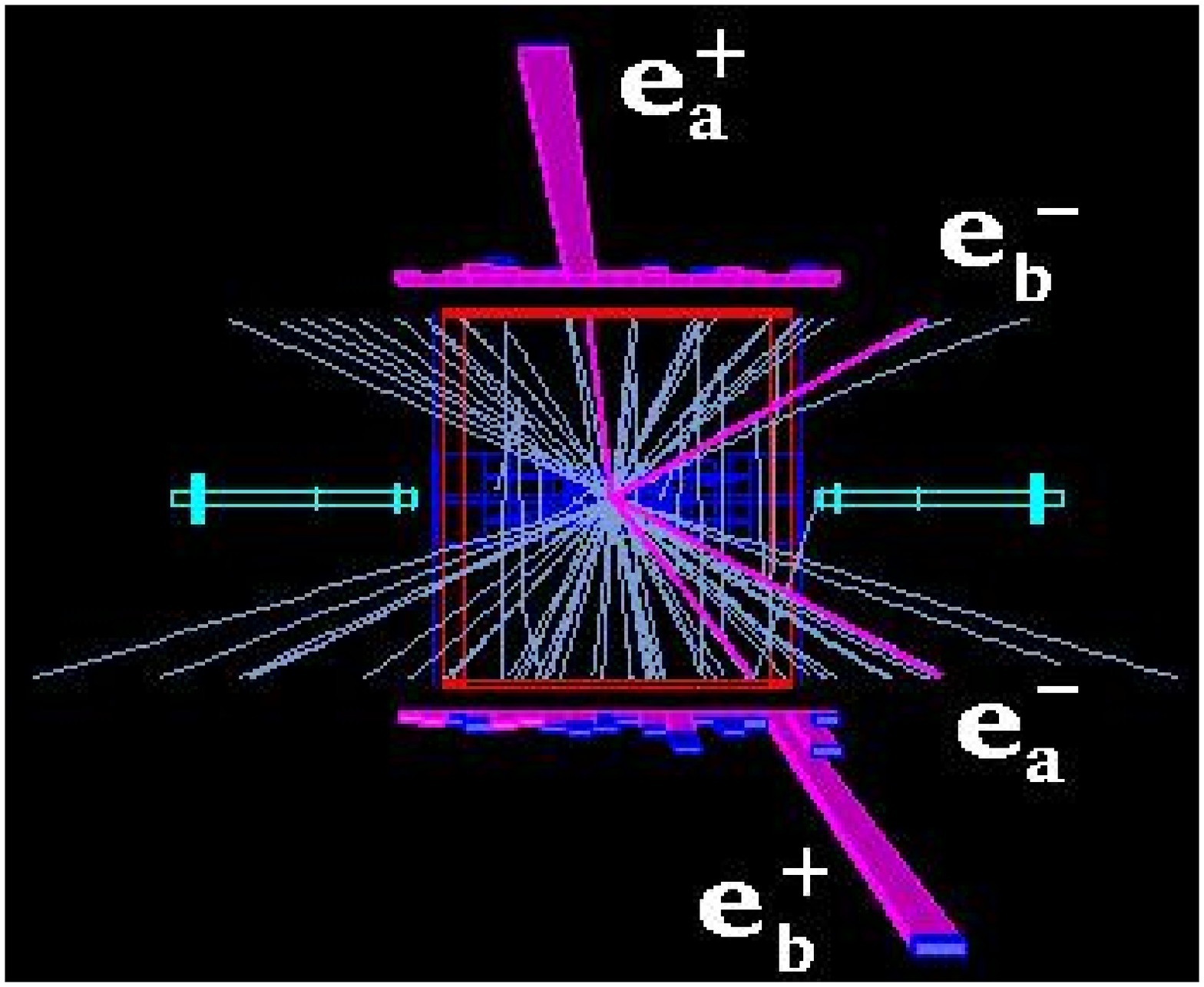}

\vspace{0.1in}(a)

\vspace{0.1in}
\epsfxsize=2in \epsffile{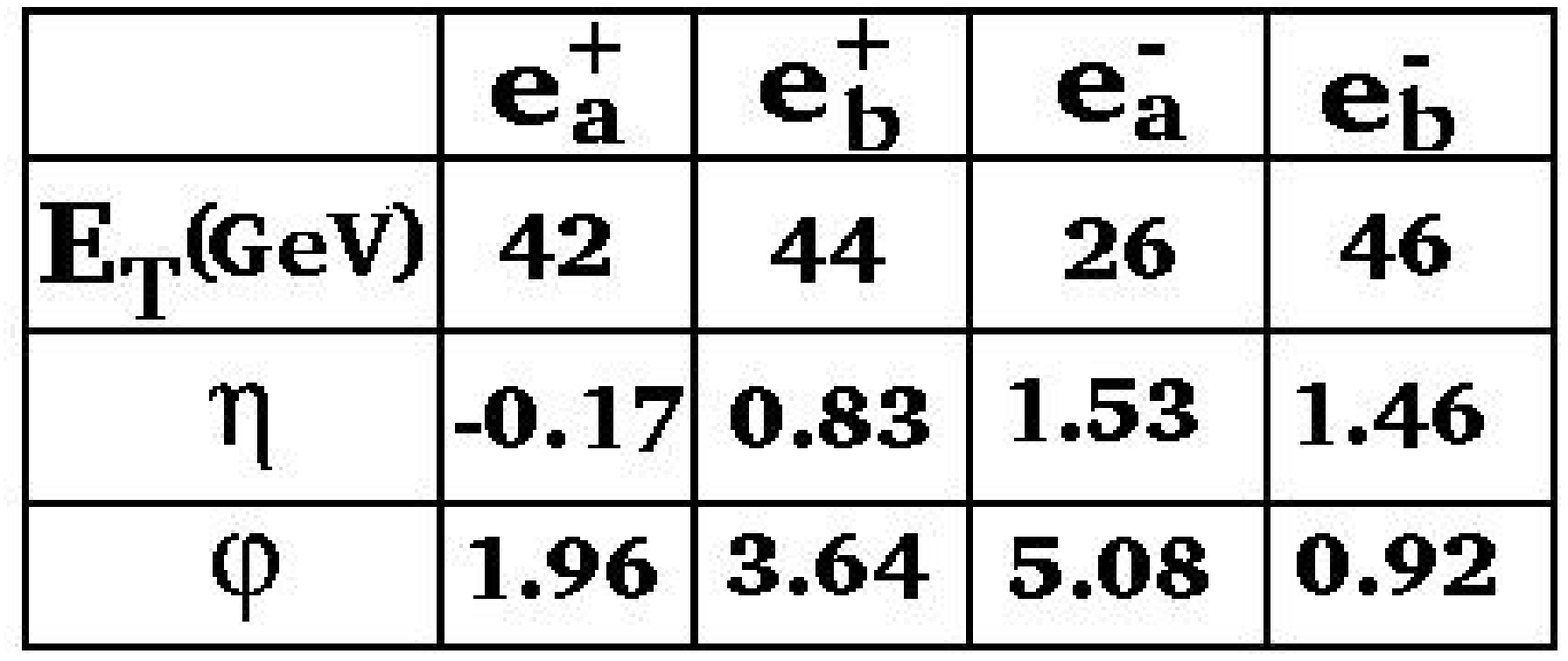}

(b)
\end{center}}
$~~~~$ \parbox{2.7in}{
\epsfxsize=2.6in \epsfysize=1in \epsffile{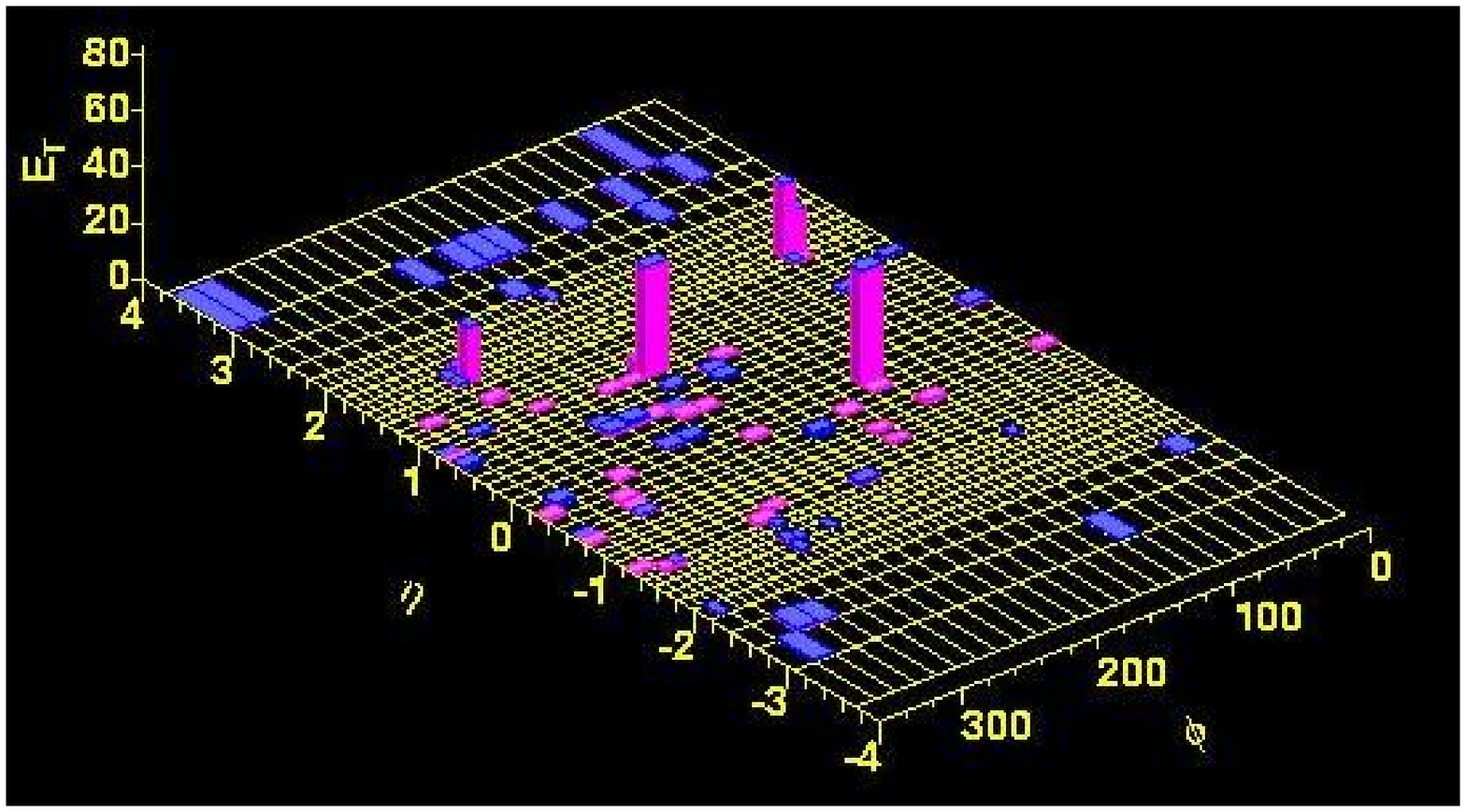}

\vspace{0.1in}
\epsfxsize=2.6in \epsfysize=1in \epsffile{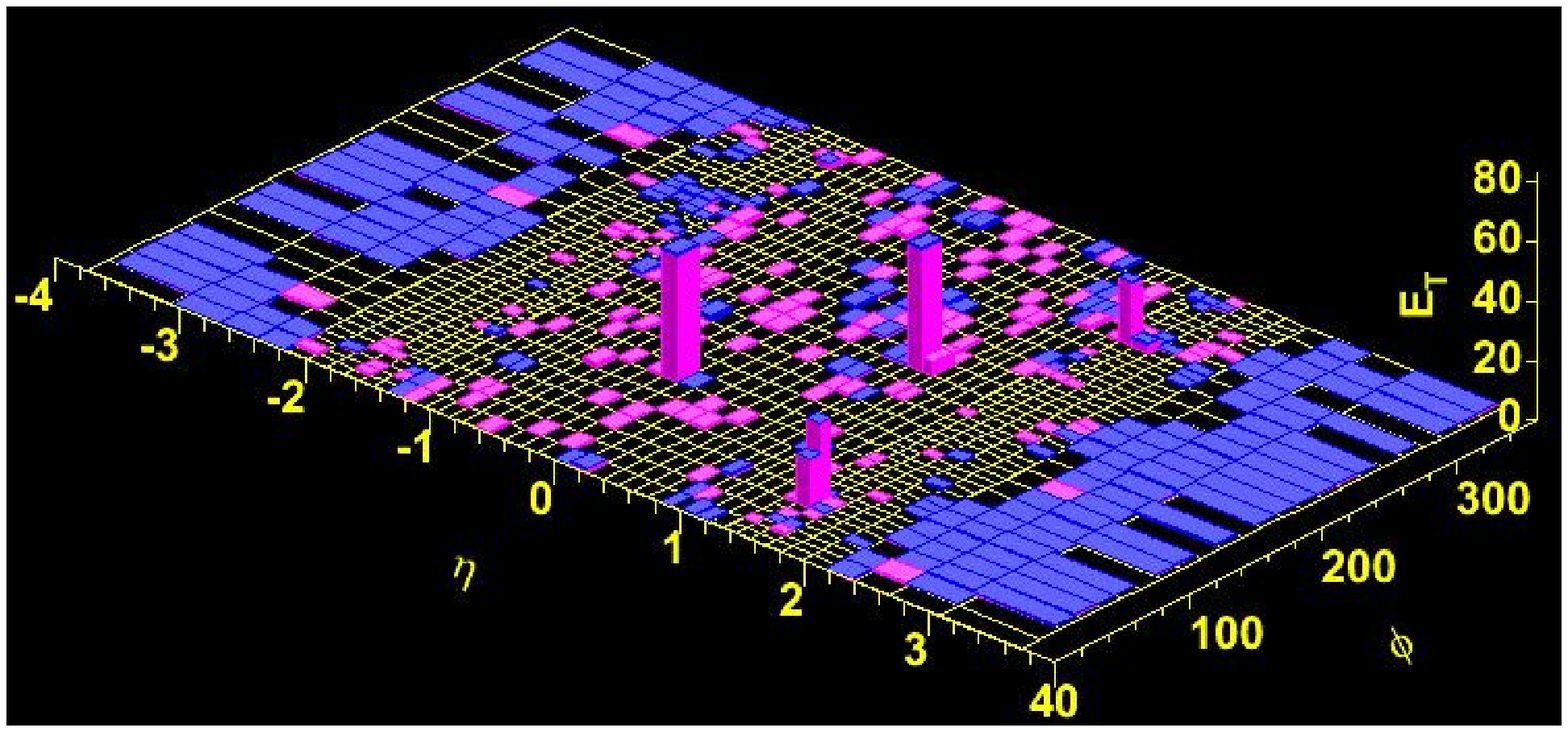}

\vspace{0.1in}
\epsfxsize=2.6in \epsfysize=1in \epsffile{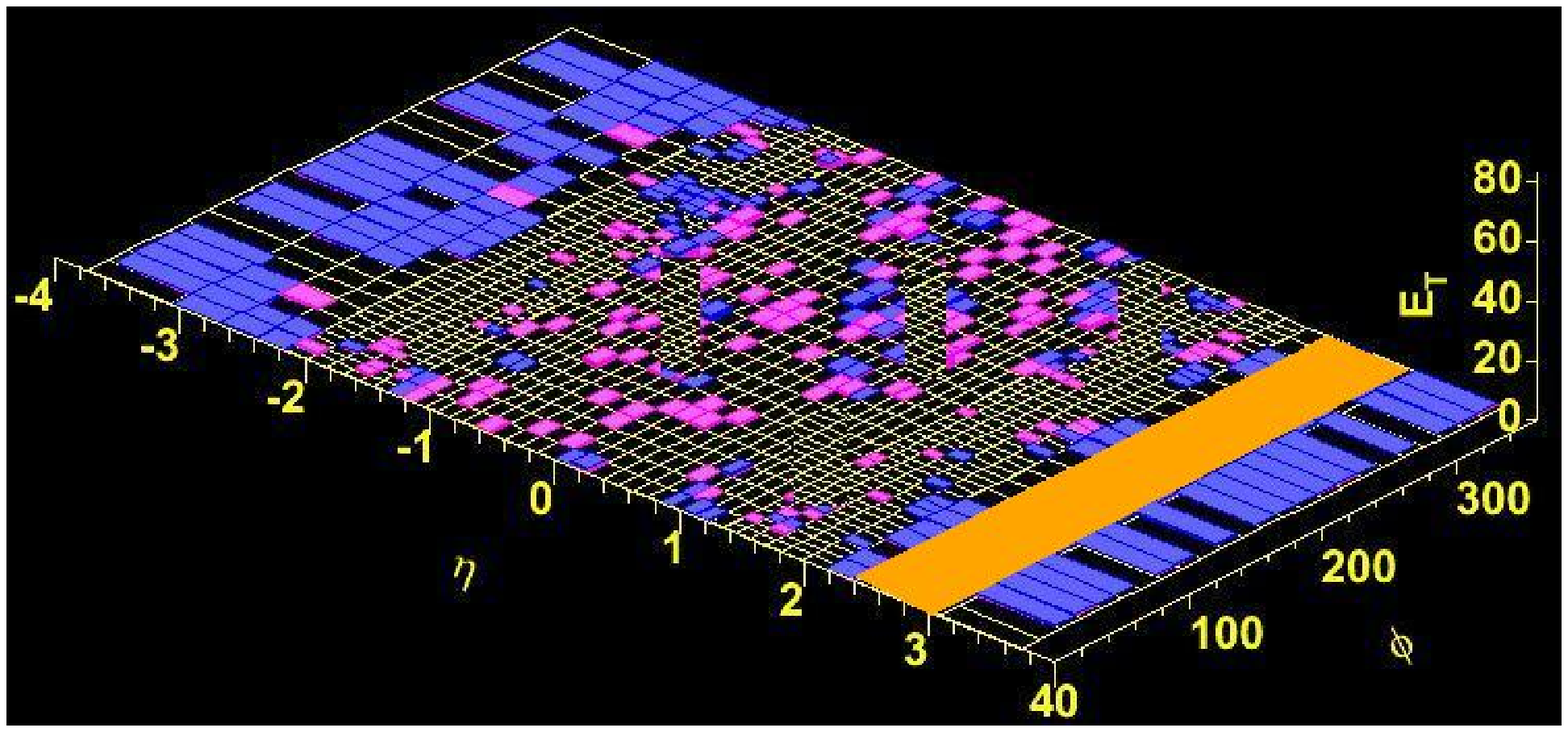}
}
\parbox{0.5in}{$~$

\vspace{0.2in}
(c)

\vspace{0.9in}
(d)

\vspace{0.9in}
(e)}


Figure 16. A High Multiplicity $Z^0Z^0$ Event at the Tevatron 
(a) event profile with $p_T > $ 200 MeV (b) event details given in \cite{hkg} (c) (mirror reflected) lego plot with $E_T > $ 500 MeV (d) $E_T >$ 100 MeV (e) $Z^0Z^0$ production region.
\end{center}

\subhead{7.2 Similar Events Could Have Been Missed}

First we note that pile-up has eliminated the possibility to use event displays to discover events of the kind of Fig.~16 during the 
subsequent part of Run 2.
Very regrettably, from our perspective, the luminosity increase has resulted in multiple vertices appearing in every event display.
An example is provided by the high mass CDF event shown in Fig.~17 in which it is clear, from Fig.~17(a), that there are at least two additional vertices. In this case, it is surely impossible to distinguish calorimeter energy deposits from central region soft particles associated with the vertex of interest and non-central
soft particles associated with the other vertices.

The event shown in Fig.~17 is one of the two most spectacular high mass CDF events discussed below. It has appeared in all the published analyses\cite{CDF4l1,CDF4l2,CDF4l7}
and is listed\footnote{The kinematic details of this event have evolved with each appearance. Increases in the smaller Z boson mass and the largest lepton
$p_{\perp}$ have increased $M_{ZZ}$ from 311.9 GeV to 329 GeV.} with $M_{ZZ}$ = 329 GeV in Fig.~19. The earliest published 
lego plot\cite{CDF4l1} of associated particles is shown in Fig.~17(b).
Accepting that this does not simply reflect the multiplicity of small $p_{\perp}$ particles associated with the vector boson production vertex it is, nevertheless,
worrisome that there is no sign of the high $E_T$ jets that the more sophisticated tracking of \cite{CDF4l7} has detected as associated with this vertex.
\begin{center}
\epsfxsize=2.3in\epsffile{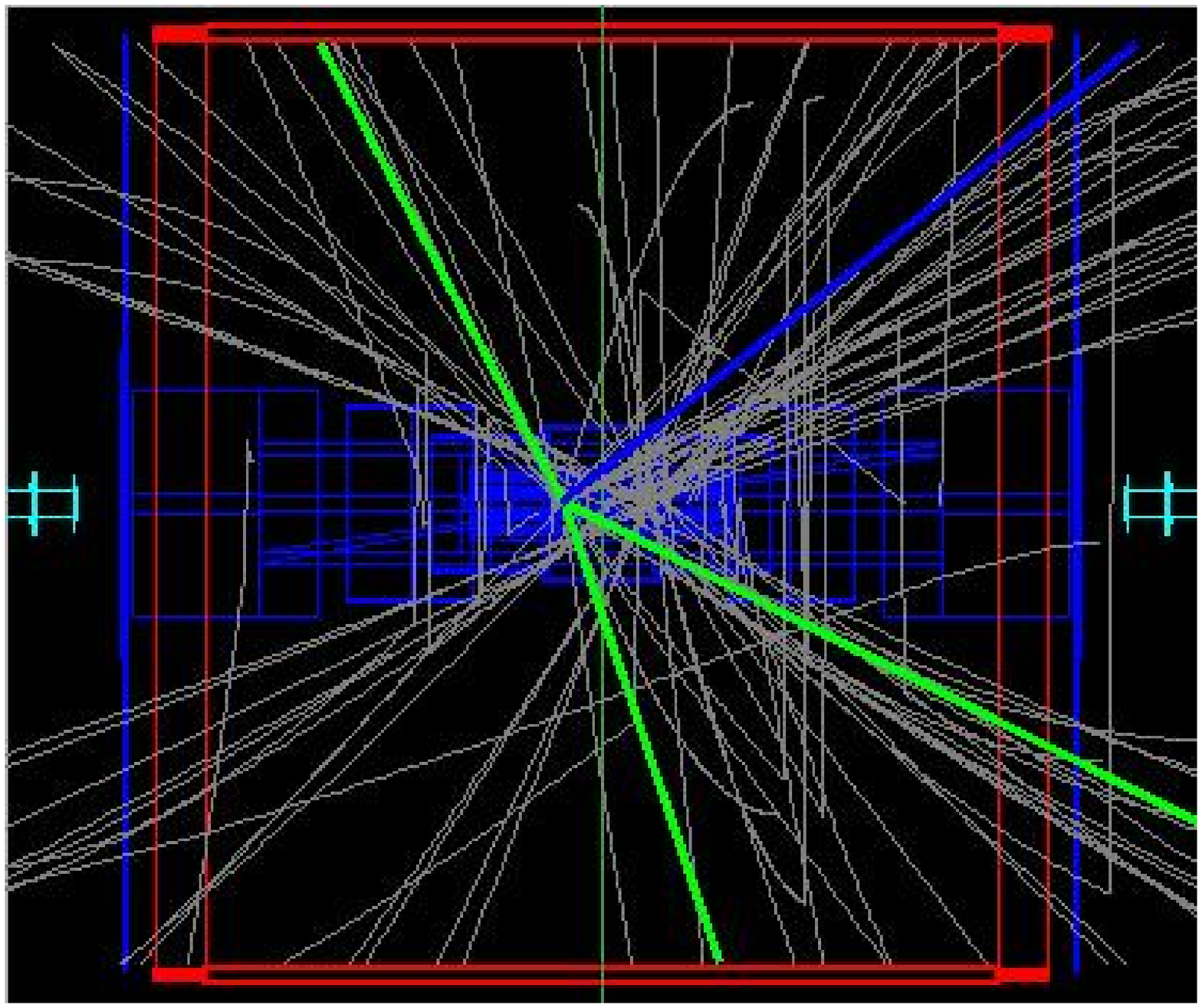}
\hspace{0.3in}
\epsfxsize=2.8in\epsffile{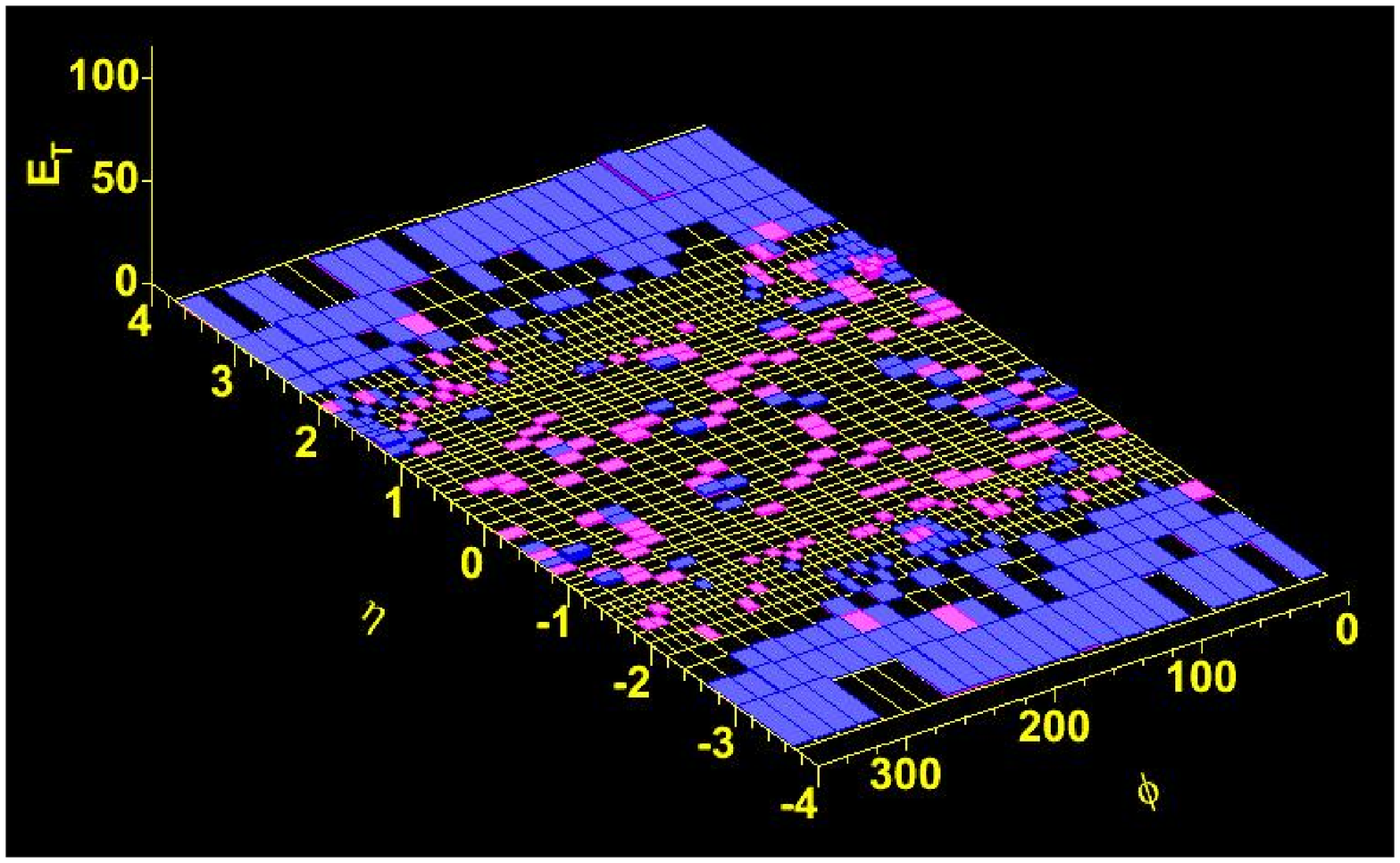}

(a) \hspace{2.8in} (b) 

Figure 17. A High Mass Event Seen by CDF
(a) the r-z display \hspace{2.8in} (b) the earliest lego plot

\end{center}

It is instructive, therefore, to consider just what can be learnt from the properties of the ``unpolluted'' event displayed in Fig.~16.
The event was discovered serendipitously and, unfortunately, at the time it was considered impossible to search for other similar events. The associated (small $E_T$) multiplicity was simply not recorded as an interesting quantity in vector boson events (although it could have been) and so it could not be searched on. A major re-recording of 
events would have been needed. Also, it might be expected that, in general, similar events with a large associated multiplicity would have even worse containment problems for the leptons than those that led to the rejection of the Fig.~16 event. In addition, this event has other properties that imply 
very similar events could easily not be recognized.

From Fig.~16(a), and the event details given in Fig.~16(b), it is clear that the production angles for the original Z pair are further in the plug than either of the electrons and so were close to the edge of the detector. The two electrons were only just detectable inside the 
plug and so (hypothetically) if the Z's had not decayed, producing wide angle electron pairs, they would have gone ``forward''  as produced
``sextet pions''. Without calculating the complete details, it is obvious that
the production directions for the Z's must have been, roughly, in the yellow area shown in Fig.~16(e) (with $\eta$ close to 3). There is a high multiplicity of additional small $p_{\perp}$
particles covering the entire central rapidity region, apart from
from an interval close to the production process. (This could well be an effect of
gluon exchange linking the sextet and triplet states mentioned in Section 3.)  
Given the large energy deposit 
in the forward region that includes the Z pair production, it is possible that there was also 
another small $p_{\perp}$ state on the very forward side of the produced pair, as anticipated in Section 5. 

While Standard Model production is a parton model process that is, necessarily, predominantly
in the central rapidity region, we do not expect this to be the case for the enhanced (sextet pion) production of longitudinal pairs. There need only be a sufficient rapidity interval on either side of the production. 
For Z pairs (or W pairs) produced even further forward than in Fig.~16, at least some of the produced leptons would not be detected and so, obviously, such events would not have been seen by CDF. Note also that the limitations on muon detection are such that
if muons replaced the plug electrons in the current event, they would not have been detected and they certainly would not have been detected for further forward produced pairs.

Hadronic decays that produce quark jets will not be identified for either Z or W pairs and the neutrinos, that are a major presence in W pair decays and also can be produced by Z pair decays, will not be detected at all
when the boson pair is produced towards a forward region. 
Clearly, there could be a large cross-section for (towards) the forward 
direction production of vector boson pairs that has been missed entirely at the Tevatron. Such events could, however, be a significant component
of the high multiplicity and small $p_{\perp}$ cross-sections seen by the
LHC detectors and discussed in Section 4. The question, which we discuss further in the next Section, is how well the vector boson component can be detected
at the LHC.

\subhead{7.3 Monte Carlo Models For Vector Boson Pair Production}

There were lepton plus dijet events seen\cite{UA1W} at UA1 that suggested electroweak vector boson pairs were being produced with a cross-section 
almost\cite{ksj} two orders of magnitude 
above the Standard Model value. It was natural to expect that the Tevatron 
would amplify this discrepancy. Indeed, in the very first CDF run (Run 0)
there was a very clear ZZ $\to e^+e^-$ + 2 jets event that also seemed to forecast a large cross-section. 

Subsequently, it was realized that the central region W + 2 jets
cross-section produced by QCD is so large at the Tevatron that separation of the vector boson pair cross-section in which one boson decays to dijets is very difficult. Only recently, has it been claimed\cite{wwjj} that 
this cross-section can be extracted and that it is consistent with the Standard Model.
Unfortunately, this extraction involves the heavy use of the Monte Carlo models which have now been shown to dramatically underestimate the high
multiplicity production of small transverse momentum particles. If this particle production is sometimes associated with large rapidity vector boson pair production,
as we have just discussed, then these models will not be reliable for extracting the signal being searched for from the background. (Indeed, the vector bosons will, most likely, be missed much of the time.) 
It seems likely that this problem is sufficiently severe that it interferes with the task of separating new physics 
corresponding to an excess cross-section for W + 2 jets. 
It is possible that the new physics that has been 
seen in a recently published analysis\cite{W2j} is simply (part of) an
excess vector boson pair cross-section of the kind 
that I am arguing for.

\subhead{7.4 The $\eta_6$ in the Z Pair Cross-Section}

Very recently, CDF has published\cite{CDF4l7,CDFEPS} the results of a search for high mass resonances decaying into a Z pair. As we noted above, some remarkable events have been accumulated. As shown in Fig.~18(a), there is a cluster of four events that are
\begin{center}
\epsfxsize=2.3in\epsffile{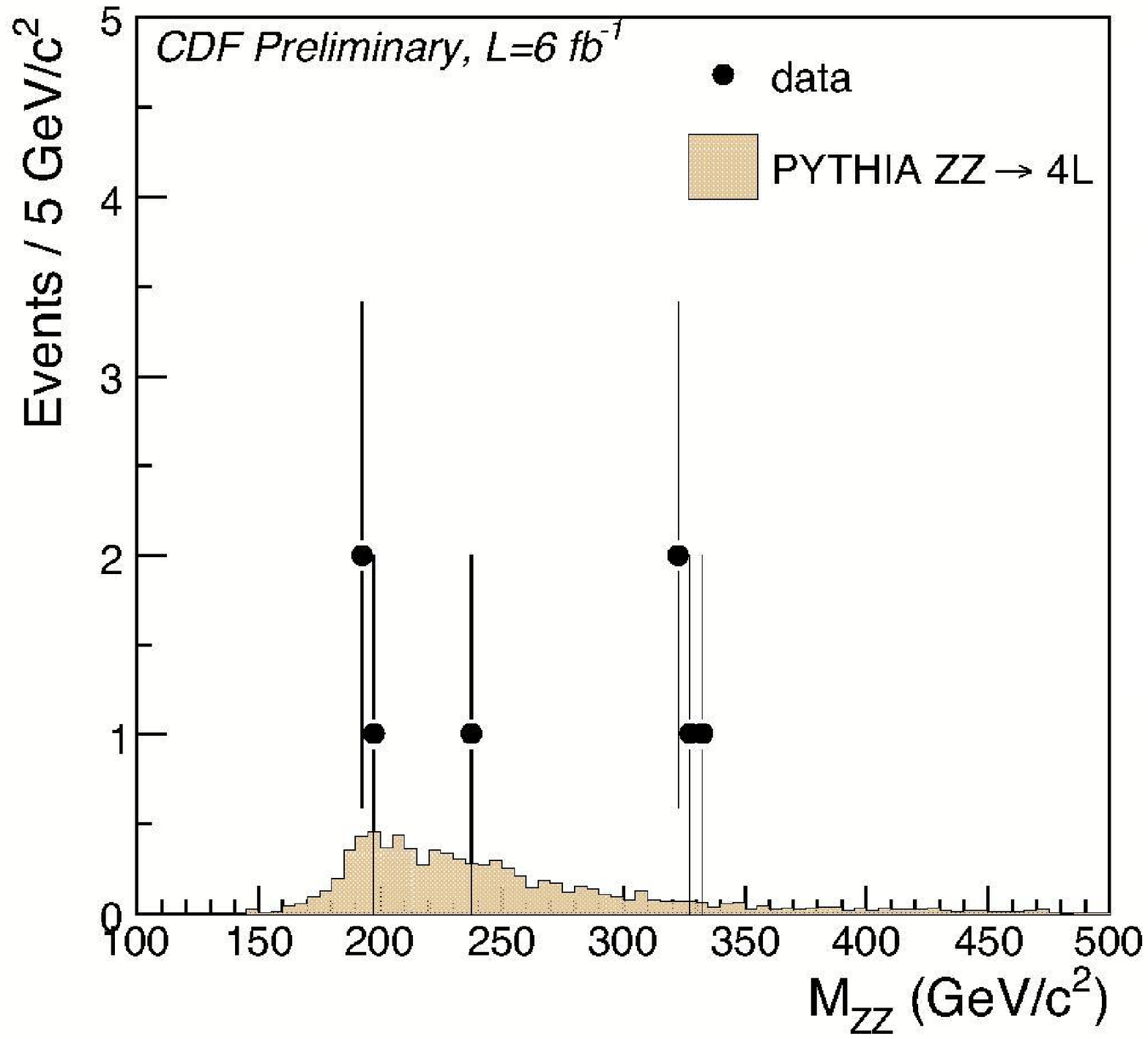}
\hspace{0.5in}
\epsfxsize=1.9in\epsffile{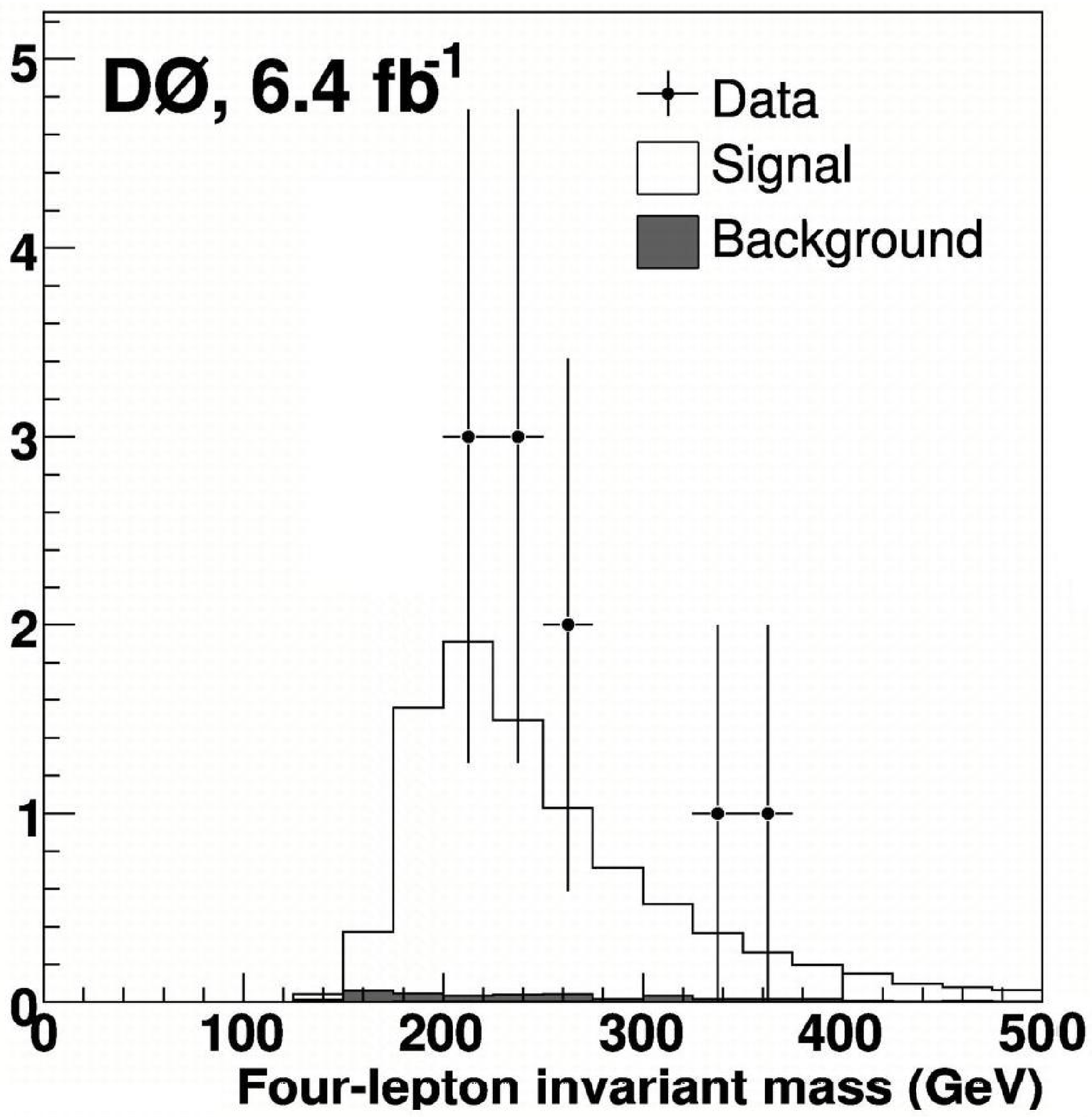}

(a) \hspace{2.8in} (b)

\epsfxsize=2.2in\epsffile{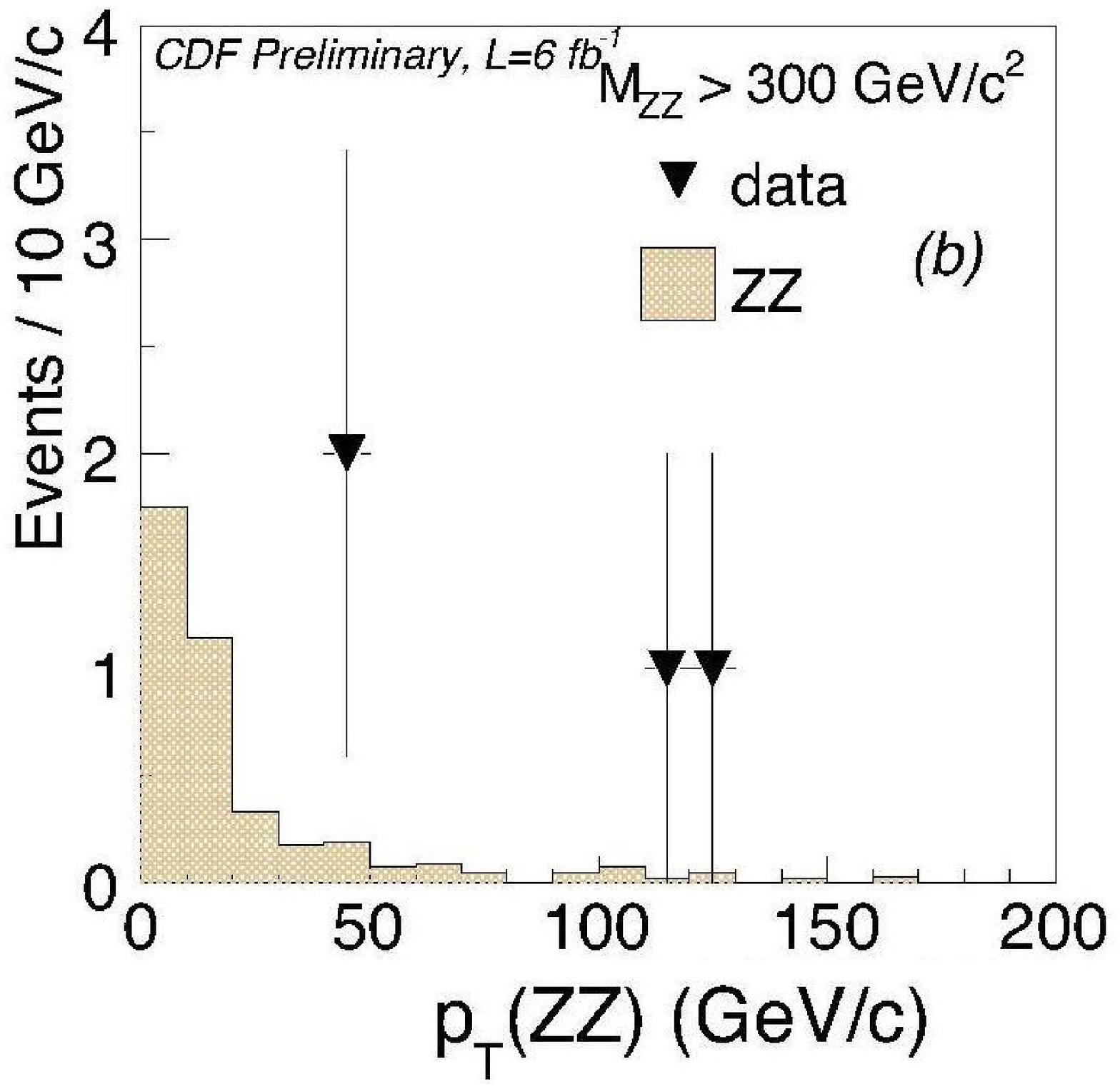}
\hspace{0.5in}
\epsfxsize=2.4in\epsffile{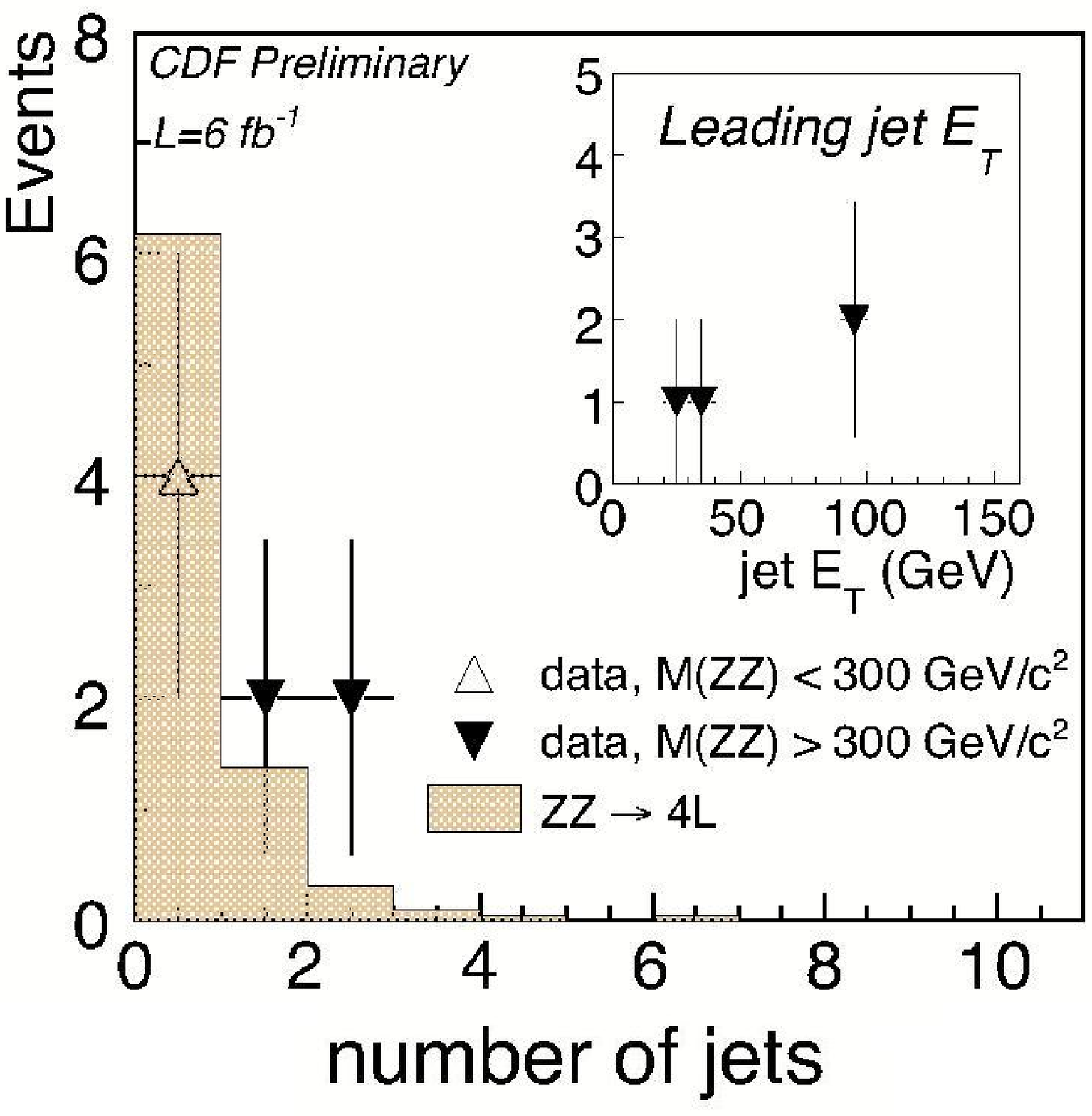}

(c) \hspace{2.8in} (d)

Figure 18. ZZ $\to$ Four Charged Leptons
~(a) CDF mass spectrum ~(b) D0 mass spectrum$ (c) p_{\perp}$(ZZ) for the high mass CDF events (d) Jets in CDF events (d)  
\end{center}
very close in mass, with $m_{ZZ} \sim 328 \pm 7$~GeV.
Moreover, this mass is at, or very close to, the top-antitop quark threshold.
The threshold mass is, of course, not unambiguously defined but the latest ``pole mass'', determined from the cross-section, gives $m_{t\bar{t}} \sim 332 \pm 14$ Gev. In Fig.~18(b), a comparable D0 plot\cite{CDFEPS} shows two similar high mass events, one close to the top threshold and one just above this threshold. Taken together, these events suggest that, if there is a (broad) resonance, then it could indeed be at the
top threshold. 

In the next Section, we will discuss high mass events seen by ATLAS and CMS which encourage, even if they do not yet confirm, the same conclusion. 

As can be seen from Fig.~18(c), the high mass CDF events are also characterized by an ``electroweak scale'' $p_{\perp}$(ZZ) which is balanced by either one or two high $E_T$ jets, as shown in Fig.~18(d). The table in Fig.~19 gives details of all the eight events found.
The two most spectacular events, with ZZ masses of 329 and 325 GeV, have $p_{\perp}$(ZZ) $>$  100 GeV, have one or both Z's with $p_{\perp}$(Z) $>>$ 100 GeV, and have two
large $E_T$ jets. The four muon event, with a mass of 329 GeV, corresponds to the event display in Fig.~17. The accompanying jets have, presumably, been located via
the superior tracking utilised in \cite{CDF4l7}. The more elaborate event selection procedure has also led to the inclusion of the Fig.~16 event which, as we noted earlier, is the first event in the Fig.~19 table. 
\begin{center}
\epsfxsize=5.7in\epsffile{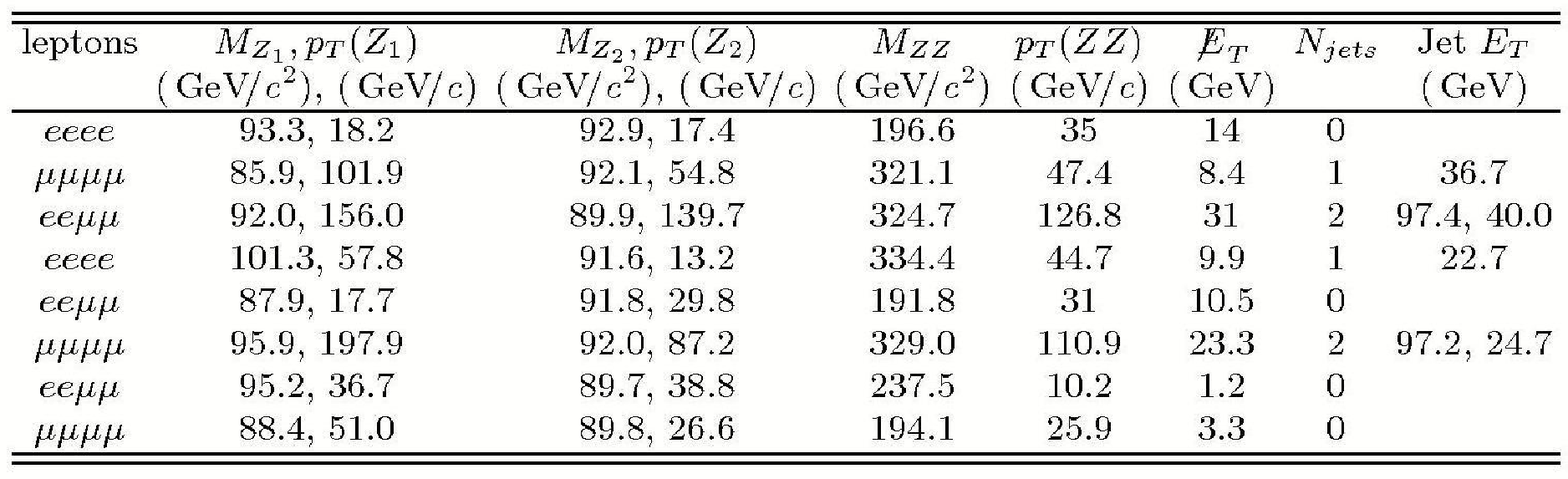}

Figure 19. Kinematic Details for the Eight Events Found in \cite{CDF4l7}.
\end{center}

It is clearly plausible, if not likely, that the large mass events are produced by new electroweak scale dynamics. In Section 11 we will propose that top-antitop production is via the $\eta_6$ sextet quark resonance. The appearance of this resonance, at the right mass, in the ZZ cross-section would be a major confirmation of this proposal. That the $\eta_6$ is a pseudoscalar could, perhaps, be related to  why CDF do not see corresponding jet and neutrino events and could even lead to 
a smaller $pp$ cross-section, relative to the $p\bar{p}$ cross-section. 

\mainhead{8. Sextet Quark States at the LHC}

Given that the LHC energy is above the knee and that, moreover, the discussion in Section 6 suggests that, most likely, this energy is also above the threshold for the production of sextet states in proton-proton interactions, we might expect to see more ZZ events like that seen by CDF. Even though, the cross-section may still be relatively small it would seem, 
at first sight, that increased rapidity coverage and improved detection of small $p_{\perp}$ particles should lead to the recording of a significant number of events with properties approaching
those desired.  Unfortunately, the rapidity range available 
for the production of forward vector boson pairs has significantly increased relative to 
the Tevatron. As a result, there could easily be a large rapidity cross-section that would be almost entirely missed just because of current detector capabilities. 

Nevertheless, even with the limited rapidity coverage, we might still expect to see an excess cross-section composed of ``non Standard Model'' ZZ events, were it not for the insidious problem of pile-up. Before discussing this problem we consider the ``first ZZ event'' seen by CMS.

\subhead{8.1 The First CMS $Z^0Z^0$ Event}

The first $Z^0Z^0$ event recorded by CMS created a big stir when it was published. Fortunately, it was recorded in the very early days before the big luminosity build-up began. Consequently, we have beautiful event displays without the contamination of additional vertices and we can see that  
the four muon event\cite{cmszz}, shown in Fig.~20,
is remarkably clean compared to what would be expected for a Standard Model event. 
\begin{center}
\parbox{3.2in}{\epsfxsize=2.9in\epsffile{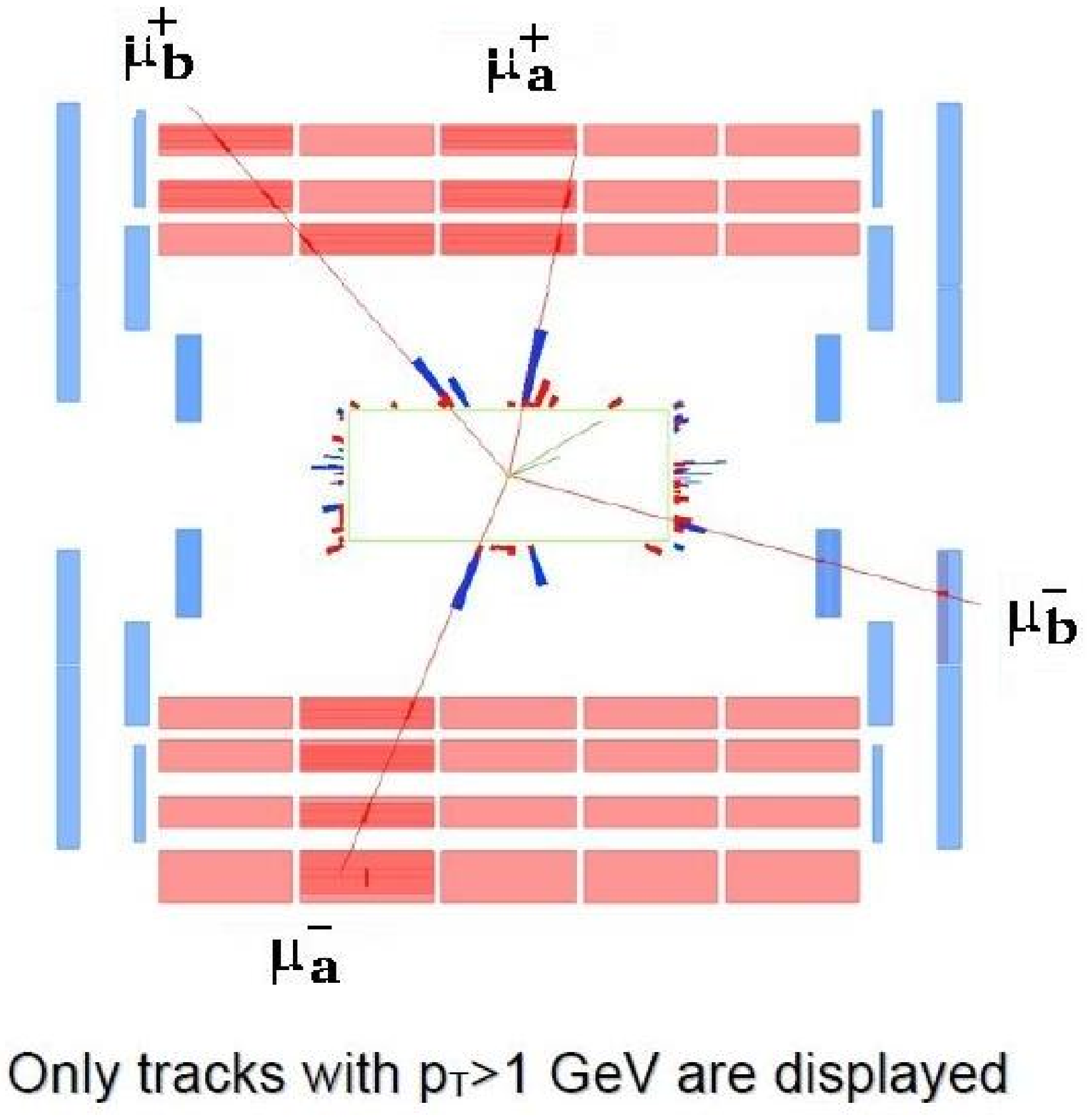}}
\parbox{2.6in}{\epsfxsize=2.5in\epsffile{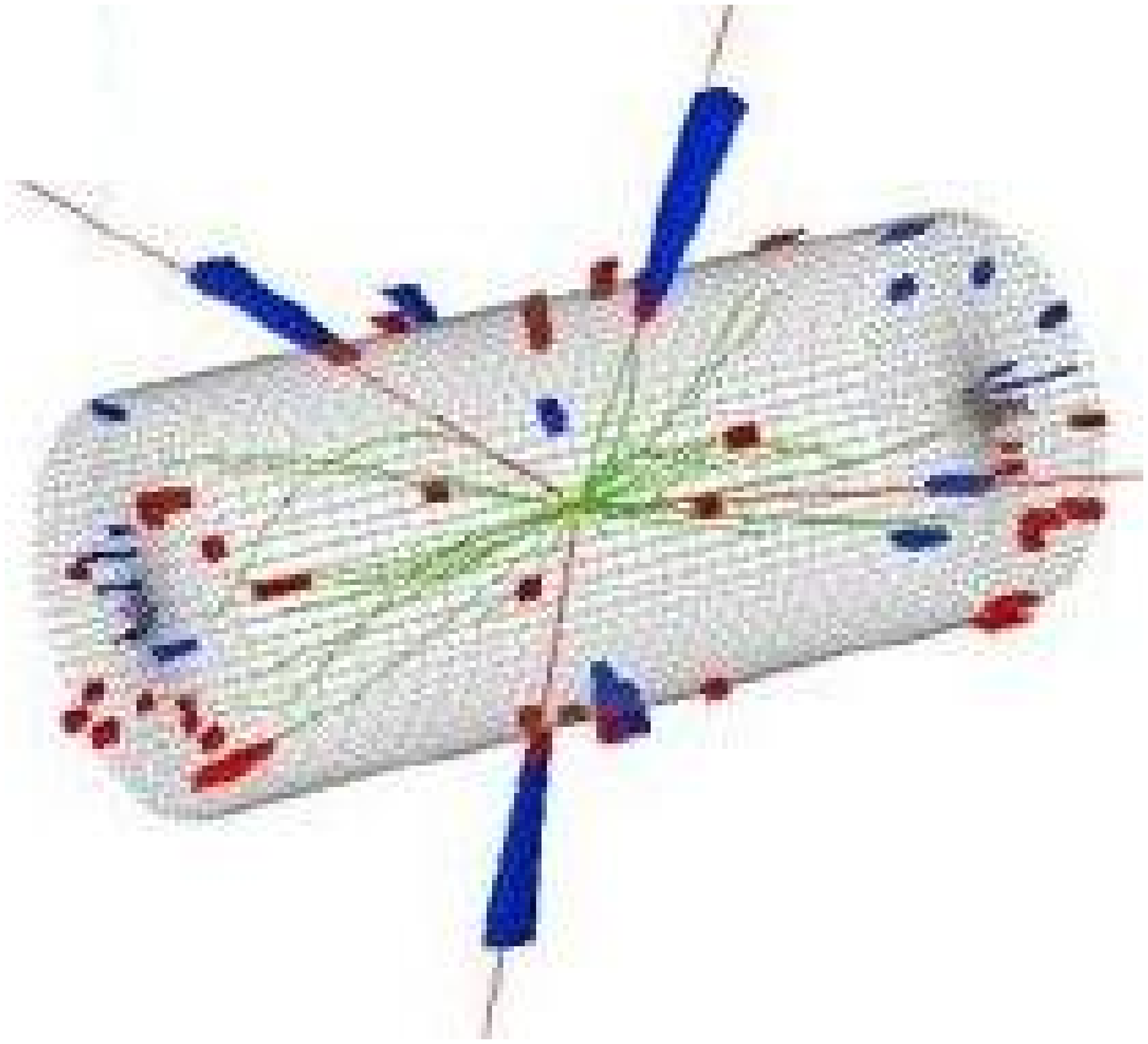}
}

(a) $p_{\perp} > 1$ Gev \hspace{2.3in} (b) No $p_{\perp}$ cut-off

Figure 20. The First CMS $Z^0Z^0$ Event - a Four Muon Event
\end{center}

Imposing a 1 GeV $p_{\perp}$ cut-off, as in Fig.~20(a), only 
two additional (relatively small $p_{\perp}$) particles remain.  
After removal of the
cut-off there are, as can be seen in Fig.~20(b), twenty additional
low $p_{\perp}$ particles, with an average $p_{\perp}$ of 0.6 GeV, that split into two components with momenta in the two forward directions. The multiplicity
and $p_{\perp}$ of the associated event are, therefore, close to the average
for the minimum bias events discussed in Section 4. 

From Fig.~20(a), it is clear that both Z bosons were produced with very central rapidities. Consequently,
the multiplicity of the two low $p_{\perp}$ components is comparable with the multiplicity of particles in the immediate rapidity region surrounding the 
ZZ pair production in the Tevatron event of Fig.~16, discussed in the previous Section. 
Presumably, there could be more low $p_{\perp}$ particles in the adjacent rapidity
intervals that would make the two observed components part
of higher multiplicity configurations similar to that seen in the Tevatron event. Since the $p_{\perp}$(ZZ) is also very low, $\sim$ 3 GeV, there is no evidence that this was a hard scattering event. Clearly, we would like to conclude that the event is indeed a sextet pion production event of the kind that we are looking for. 

Even though we don't know, from the published information, how big the adjacent multiplicities were for the Fig.~20 event, it is clear that this event and the 
Tevatron event share many characteristics. If 
electroweak symmetry breaking via the sextet quark sector is indeed at work, there should be very many more events of this kind at the LHC and, moreover
this number should increase rapidly with energy. 

\subhead{8.2 Many More Events?}

Recent ZZ event plots from CMS and ATLAS, published as part of the Higgs search, are shown in Fig.~21. (I have removed events below the ZZ threshold.)
\begin{center}
\epsfxsize=2.75in\epsffile{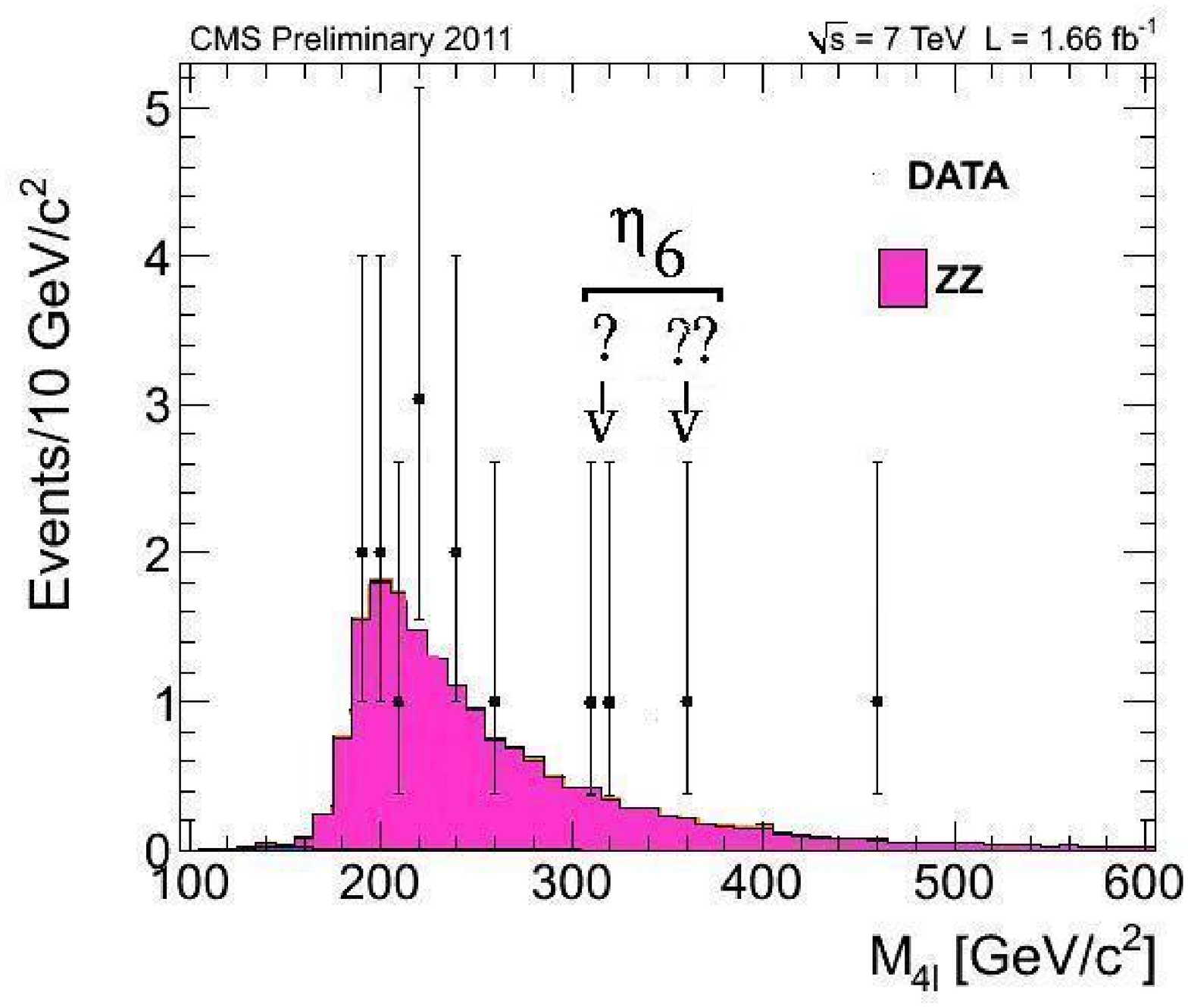}
\epsfxsize=3.15in\epsffile{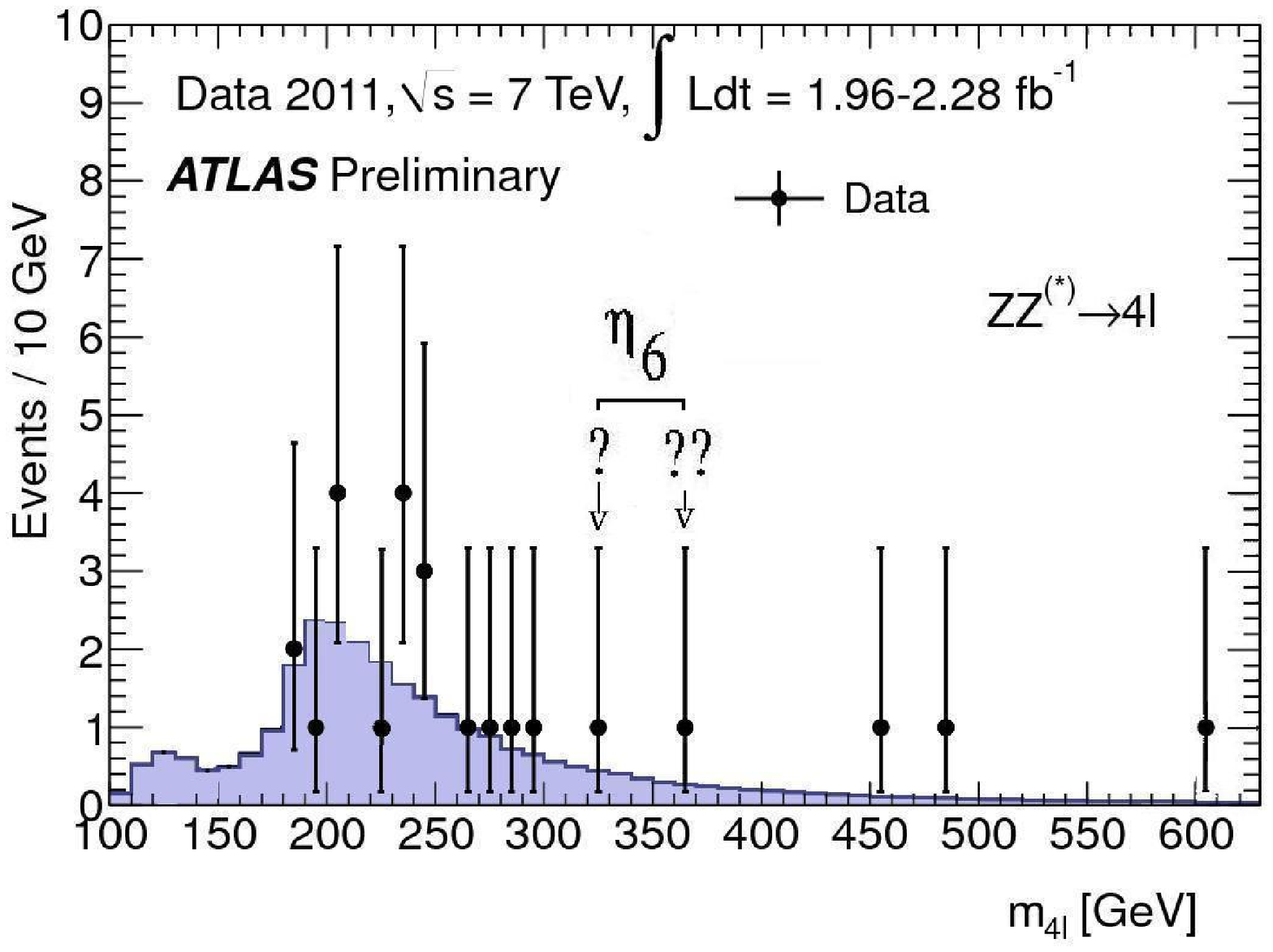}

(a) \hspace{3in} (b)

Figure 21. ZZ Events Recorded by (a) CMS and (b) ATLAS  
\end{center} 
The accumulated luminosity for ATLAS is close to 2 fb$^{-1}$
while that of CMS is not much lower.. 
The accumulated CMS luminosity when the Fig.~20 event was recorded was only $\sim$2-3 pb$^{-1}$ and so, naively, we would expect that this event should be part of a large cross-section that would have accumulated $\sim$ 500-1000 additional events at this point. In Fig.~21, we see only 14 additional events. Consequently, as with the CDF event of Fig.~16, the event of Fig.~20 is either a spectacularly rare event 
that, a priori, should not have been seen, or it is part of a cross-section that is being missed altogether.

Events that cleanly produce two central Z bosons, as in Fig.~20, will
certainly be relatively rare. There will, presumably, be events in which all the charged leptons are contained within the central detectors but, one or more of them suffer from isolation problems. In general, if the vector bosons are produced across a
wide part of the rapidity axis (extending well outside of the central detectors) then, obviously, even when only charged leptons are produced, they will not all be detected and the nature of the event will surely not be recognized, even before we take the pile-up into consideration.

\subhead{8.3 Pile-Up}

Assuming that W pairs producing one or more neutrinos outside the central region have no chance of detection, the crucial question would appear to be whether there is any possibility to detect events which combine the production of large numbers of small $p_{\perp}$ particles with the production of Z boson pairs
across at least some part of the rapidity axis?
Unfortunately, the huge increase in pile-up would seem to imply that the answer must be no. Even when the Z bosons can be detected, it is surely impossible to uncover 
detailed properties of the accompanying soft hadrons. An example of a CMS event with pile-up is shown in Fig.~22(a). The complexity is, as expected, much worse than produced by the Tevatron pile-up in Fig.~17(a). The contrast with the event displays shown in Fig.~20 is striking, not surprisingly.

At first glance, the left side of the event display in Fig.~22(a) shows five vertices in the event. However, a close study of the upper right $\rho$-$\phi$ 
plot shows at least as many as ten vertices. The frequency of the number of vertices appearing in the full data set is shown in Fig.~22(b). The task of associating even the largest 
$p_{\perp}$ particles to the originating vertex is clearly a major tracking plus timing problem and so it is surely impossible to determine any properties of associated high multiplicities of soft (or even semi-soft) hadrons via calorimeter deposits (or tracking).
\begin{center}
\epsfxsize=3.2in\epsffile{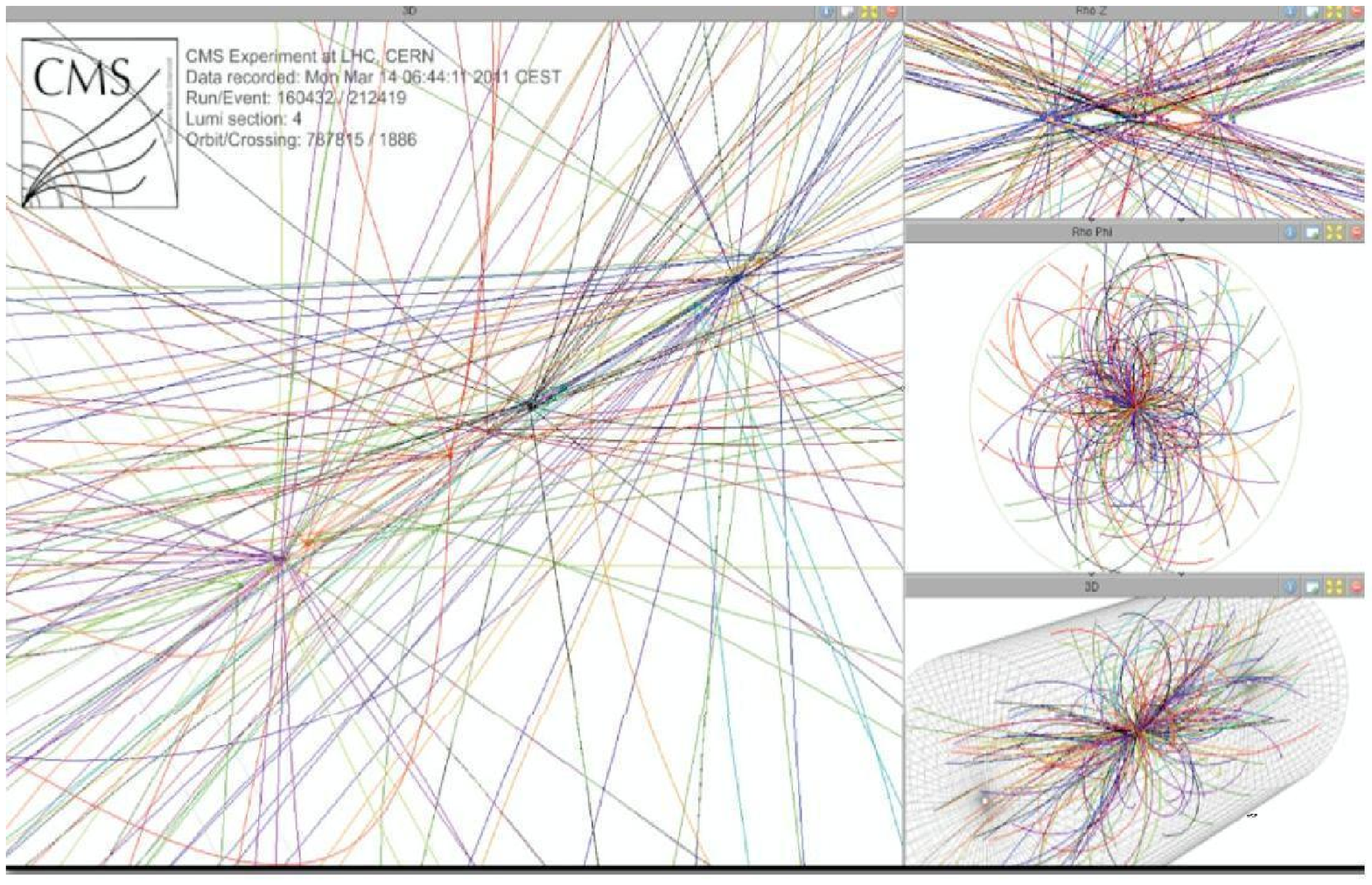}
$~~~~$ \epsfxsize=2.2in\epsffile{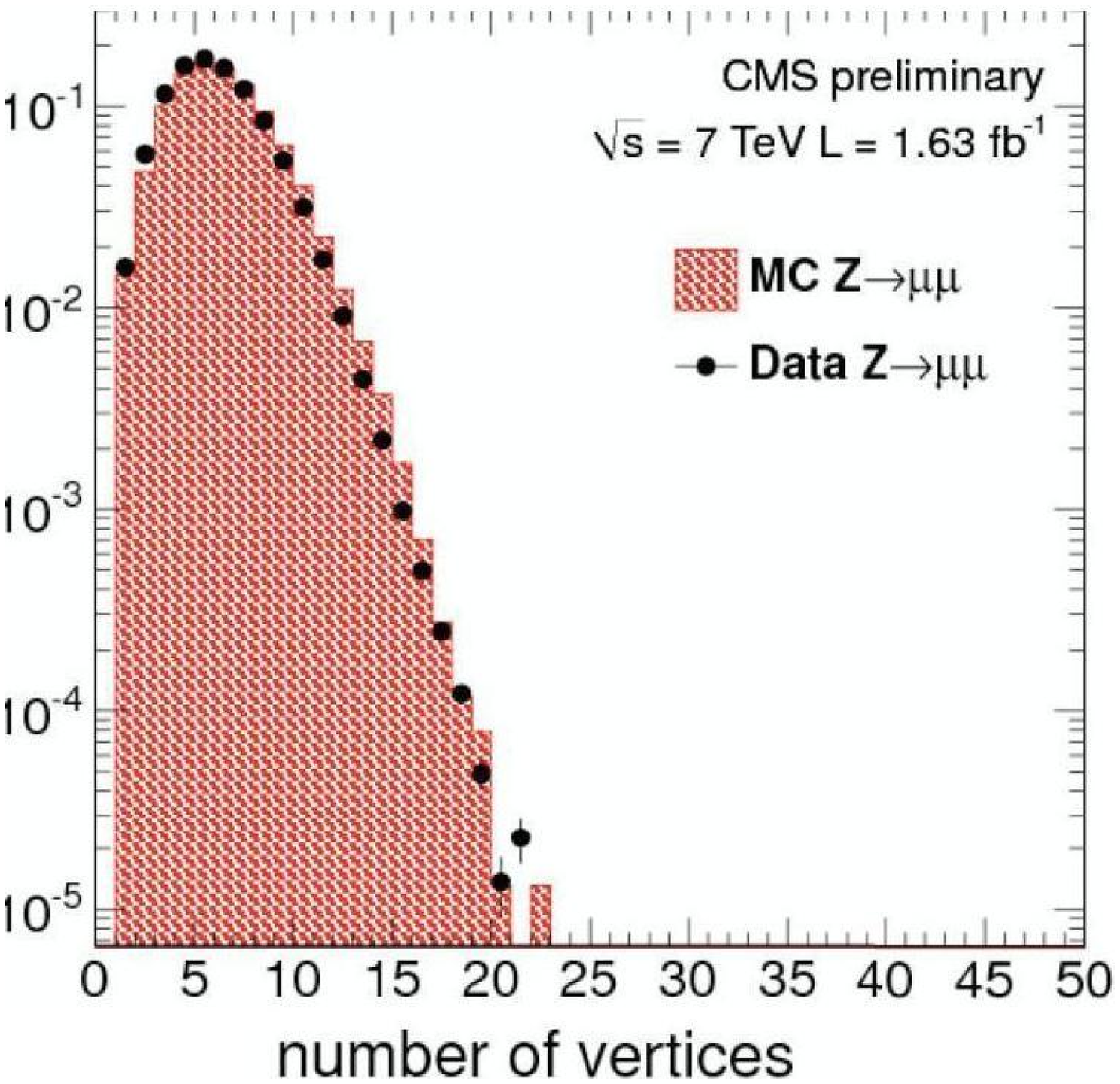}

$~~~~~~~~~~~$ (a) \hspace{2.5in} (b)

Figure 22. CMS Pile-Up (a) an Event Display (b) Number of Vertices
\end{center}

\subhead{8.4 A Discovery That Might Have Been? }

Ironically, if the LHC had been run at the luminosity which produced the event of  
Fig.~20 and if, hypothetically, additional events occurred at the frequency implied by this event's 
occurence, CMS would have accumulated almost as many events. The events would clearly be ``Beyond the Standard Model'' and we would have beautiful event displays within which associated soft hadron multiplicities could be studied. 

Of course, the non-existence of the Higgs would not have been ``discovered''. 
Moreover, if the current theory paradigm is to be overturned, as I have already emphasized (and will discuss further later) is essential for the acceptance of QUD, the Higgs non-discovery surely has the most immediate significance.

\subhead{8.5 The ATLAS and CMS Events and the $\eta_6$. }
 
The best evidence that we can expect for sextet symmetry breaking, as long as the luminosity is extremely large, is that there is an excess Z pair cross-section that is not entirely buried by the pile-up
and that contains events that are, in some way, beyond the Standard Model. Ideally, this cross-section should be, close to, rapidity independent and also favor lower transverse momentum. Most importantly, of course,
we would very much like to see the $\eta_6$ in this cross-section.

From Fig.~21, we see that both ATLAS and CMS have recorded events that extend over a considerably larger mass range than is expected in the Standard Model.
In particular, ATLAS and (to a lesser extent) CMS see very high mass events, with $M_{ZZ} >$ 450 GeV,  that should be very rare in the Standard Model. Clearly, this could signal the existence of a high mass cross-section that
is indeed ``Beyond the Standard Model''. Both experiments also have events that are above the Standard Model mass range and are in the neighborhood of the $t\bar{t}$
threshold that, as we have indicated in Fig.~21, might be early
indication of an $\eta_6$ signal.

Kinematic details for almost all of the CMS events are given in Fig.~23. The first listed event is, as we noted earlier, the ``first ZZ event'' that is displayed in Fig.~20. That it has the smallest $p_{\perp}$(ZZ) of all the events seems to be in accord with our interpretation of it's significance. The two largest rapidity events also have small $p_{\perp}$(ZZ). It is particularly interesting that the two events with 
masses closest to the $t\bar{t}$ threshold have very large $p_{\perp}$(ZZ), comparable with the large mass CDF events. As we remarked, in discussing these events, it is surely plausible, if not likely, that all of these events are produced by new electroweak scale dynamics responsible for the $\eta_6$
resonance.

In the large majority of events that contain vector boson decays 
accompanied by a high multiplicity
of additional soft particles, the decays will
involve quarks or (much less often) neutrinos, and so will have very little chance of being recognized. Events of this kind could, however, be 
contributing significantly to the unexpected excess of events with large momentum leading particles recorded in Fig.~12(a).
\begin{center}

\epsfxsize=4.9in\epsffile{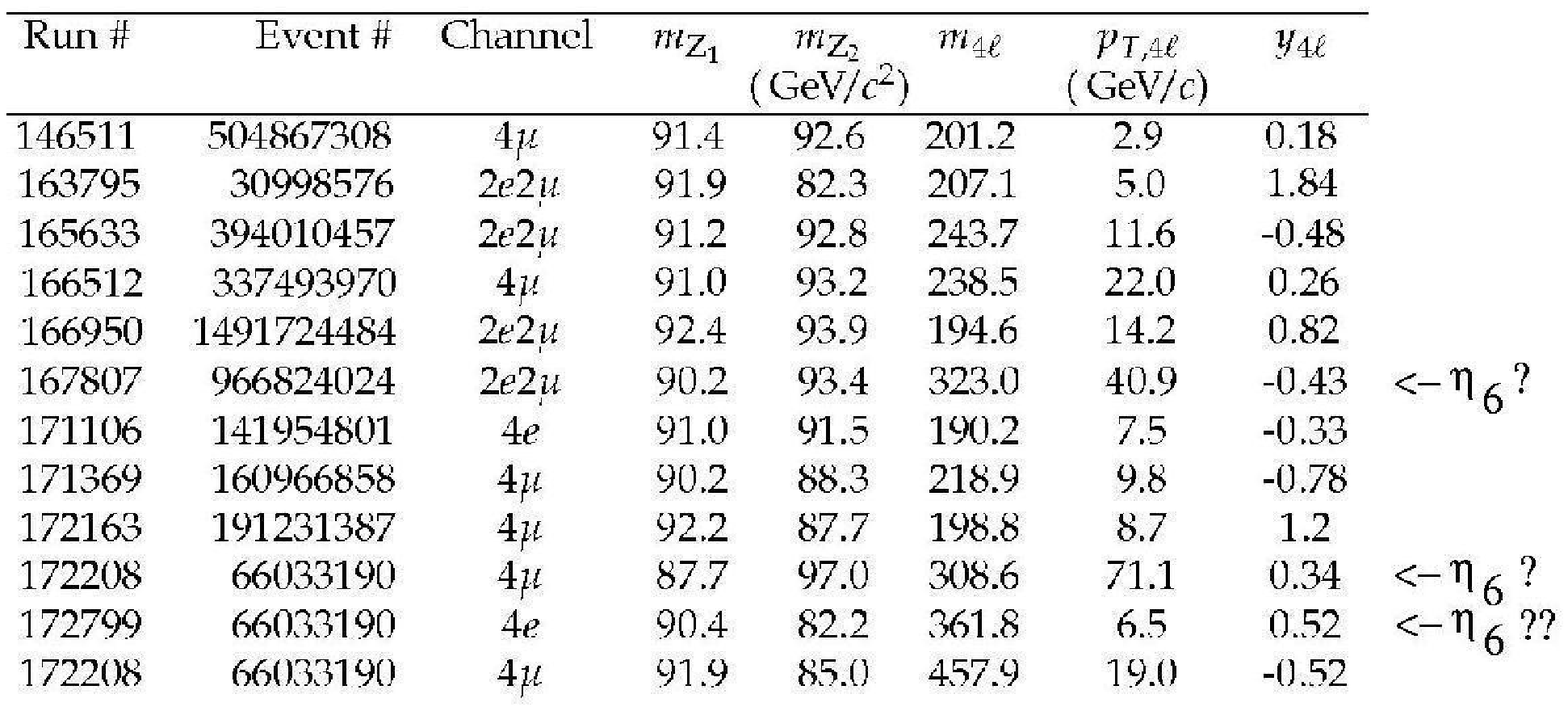}

Figure 23. CMS Event Details
\end{center}

\subhead{8.6 Additional Evidence}

Additional evidence for sextet electroweak
symmetry breaking, besides multiple $Z^0$'s with high associated multiplicities,
may be harder to obtain. In principle, 
$N_6\bar{N}_6$ pair production, accompanied by 
large rapidity hadron states could be seen. However, this is clearly 
impossible if the low energy $N_6$ hadronic 
cross-section, for collisions in a calorimeter, is zero - as we have argued is an essential part of the identification of neusons as dark matter. Even if it were not,
the confusion created by the accompanying particles, together with 
the fact that, in the pile-up, missing energies of several hundred GeV (carried by neutrinos) will be relatively common, would make the detection impossible. 
A priori, proson-antiproson ($P_6\bar{P}_6$) pairs should be identifiable via their
electromagnetic interaction 
- assuming the $P_6$ is not too unstable, although
a massive charged particle with a large production cross-section
will not be immediately identified with the sextet sector. 

\subhead{8.7 The Double Pomeron Cross-Section} 

I have argued, for several years now, that the double pomeron cross-section could actually provide the most definitive evidence for the existence of the sextet sector. This could still be the case, if the collaboration between
TOTEM and CMS, that has continued to be mentioned as a long term goal, could be realized. The CMS detector has to be active during a special purpose, low luminosity, TOTEM run.

With the pomerons detected via Roman pots, 
the environment is clean and 
well controlled. 
As (triplet quark) pion pairs dominate the double pomeron
cross-section at low mass, $W$ and $Z$ pair production should dominate the cross-section at the electroweak mass scale (in general, with some number of associated small $p_{\perp}$ hadrons). Consequently, 
some spectacular events should be expected,  
in which protons are tagged and (essentially) only 
large $E_T$ charged leptons are seen in the central detector.

The observation of a significant double-pomeron cross-section for $W$ and $Z$ pairs
would unambiguously imply that the longitudinal components of the $W$ and the $Z$  
have direct strong interactions. The only known possibility for this
is the existence of the sextet sector. 
There could also be events that appear to be 
double pomeron production of a single $Z^0$, with one of the pomerons actually
being a photon. The cross-section would be much smaller, but this would 
clearly be direct evidence for the sextet sector that would be difficult to
provide an alternative explanation for.

A direct search for ``dark matter'' would obviously be highly desirable..
The cross-section for double-pomeron production
of stable $N_6\bar{N}_6$ pairs (with a pair mass
$~ \centerunder{\raisebox{0.5mm}{${\scriptstyle >}$}}{${\scriptstyle \sim}$}
~1~TeV$) might, just possibly,
be large enough that it will be definitively seen by the forward pot
experiments when the LHC energy is maximized. It will be a spectacular process to look for. Tagged protons would determine
that a very massive state is produced, while no charged particles are seen
and there is also (almost) no hadronic activity in the central calorimeter.
Of course, if the $P_6$ is relatively stable, 
and not too different in mass from the $N_6$, it would be much simpler
to first detect $P_6\bar{P}_6$ pairs.
  
\mainhead{9. Massless QCD$_S$}

I refer to the the special version of QCD obtained by adding two sextet flavors to six triplet flavors as\footnote{The $S$ denotes either ``sextet'', or ``saturated'' (asymptotic freedom), or simply ``special''.} QCD$_S$.

\subhead{9.1 The Critical Pomeron}

Initially ($\sim$ 30 years ago), I believed\cite{arw81} that QCD$_S$ had all the features needed to produce Critical Pomeron high-energy behavior,
independently of the quark masses. 
Asymptotic freedom saturation implies, firstly, that no further quarks can be added to move the theory closer to the critical surface. It also allows    
an asymptotically-free scalar field to be added, and smoothly decoupled, that produces a color superconducting phase that contains the first dynamical element of the Supercritical Pomeron, i.e. a single reggeized massive vector particle. I argued that,
after the summation of reggeon infra-red divergences in the superconducting phase, a pomeron regge pole that is exchange degenerate with the vector reggeon (as a supercritical pomeron condensate requires)
would be produced by 
\begin{center}
 {\it ``an SU(2) color zero cloud of uniformly soft gluons with $\tau \neq C$ that 
accompanies the vector reggeon without screening it's own infra-red singularity''}
\end{center}
\noindent I also argued that the cloud/condensate should be present in all bound-states. If the supercritical theory could be constructed, the decoupling of the vector reggeon, as the full SU(3) color symmetry is restored, would give the critical theory.

After many years of trying various formulations and different approaches
aimed at correctly reproducing the supercritical theory, it eventually became 
clear\cite{arw05} that, ``an SU(2) color zero cloud of uniformly soft gluons'' must indeed appear in both the pomeron and all bound-states, but it
has to result from the color symmetry breaking of a wee gluon condensate that is already present in unbroken QCD$_S$. Unavoidably, this requires a massless quark infra-red fixed point, together with the
anomaly dynamics involving reggeon interaction anomaly poles that I describe next. As a corollary, it  appeared that the Critical Pomeron could occur only in massless QCD$_S$, making it's appearance in a massive hadron theory seem very unlikely, if not impossible!

\subhead{9.2 The QCD$_S$ S-Matrix}

At first sight, any physical relevance for massless QCD$_S$ seems problematic. This theory is maximally inside the ``conformal window'' where it is commonly anticipated that there are no particle states. 
In fact, this anticipation is based on the existence of off-shell scattering amplitudes to which a renormalization group analysis can be applied and which,
a priori, should contain the physical spectrum. In principle, 
a QCD$_S$ bound-state S-Matrix produced by anomaly dynamics, for which there are no off-shell amplitudes, could exist without contradiction. 

An additional problem for massless QCD$_S$ is, however, that it has both a very large triplet quark chiral symmetry and a sextet quark chiral symmetry. Consequently, a possible S-Matrix would have to contain many massless Goldstone bosons, with whatever infra-red problems this would lead to. Fortunately, a resolution of this problem
is provided by the embedding of QCD$_S$
in QUD. As we discuss in the next Section, QUD is both a massless field theory with all the infra-red fixed-point properties needed to obtain the Critical Pomeron and an S-Matrix theory which 
has no exact chiral symmetries and so contains only masssive particles. 

A priori, the construction of a ``non-perturbative'' S-Matrix that has no starting off-shell approximation is a forbiddingly difficult challenge.
Very fortunately, QCD$_S$ and QUD share some special properties that allow multi-regge theory to be used (only in outline, so far) to construct high-energy states and amplitudes.
I first applied the following procedure to QCD$_S$, before realizing that the existence of a QUD S-Matrix is, as a matter of principle, straightforward compared to the
issues involved if we needed to actually construct an S-Matrix for QCD$_S$.
Nevertheless, because of the much simpler algebraic structure and because the results can be carried over directly to QUD, we begin our description of how infra-red anomalies produce S-Matrix amplitudes by discussing QCD$_S$.

\subhead{9.3  Di-Triple-Regge Amplitudes} 

I construct high-energy bound-state amplitudes from perturbative reggeon diagrams containing reggeized quarks and gluons. Reggeon diagrams are transverse
momentum diagrams that also contain angular momentum propagators for the reggeons involved. Each diagram sums the  high-energy behavior of an infinite sum of feynman diagrams. 

My construction is carried out in the di-triple regge kinematic region. In this multi-regge region there are sufficient large light-cone momenta that ``universal wee partons'', produced by $k_{\perp}$ infra-red divergences of the reggeon diagrams, can simultaneously play a vacuum-like role in all the reggeon channels producing bound-states and also in channels producing an interaction exchange. Initial reggeon masses and a $k_{\perp}$ cut-off are essential. Their removal creates the divergences that produce the wee partons of the massless theory. Moreover, it is the manner of this removal, {\it most importantly the color symmetry is first restored to SU(2),} that crucially resolves the (light-cone) Gribov ambiguity\cite{gr}. 

Because many internal feynman diagram lines are placed on-shell by the multi-regge limit, effective reggeon interactions are generated that contain triangle diagrams. 
It is crucial that the infra-red triangle anomaly occurs only in reggeon interactions that connect reggeons in different (rapidity and transverse momentum)
reggeon channels. As a consequence of the
removal of fermion masses, triangle anomaly diagrams generate anomaly poles\cite{arw10} produced by chirality 
transitions corresponding to zero momentum Dirac sea shifts involving
positive to negative (or vice versa) zero energies. 

At first sight, chirality is conserved in zero mass triangle diagrams
- producing a well-known conflict between the axial-vector anomaly and vector
current conservation. As a result, it appears that the infamous problem of the regularization 
of $\gamma_5$ amplitudes enters the reggeon diagram amplitude construction process. Fortunately, and sufficiently for our purposes, it can be shown\cite{arw10} that vector current conservation plus the axial anomaly implies unique massless infra-red anomaly pole chiral amplitudes containing chirailty transitions. The pattern of the fermion mass removal determines the reggeon interactions in which the anomaly poles occur. 

In color zero amplitudes there is a cancelation of all infra-red divergences produced by the removal of gluon massses, apart from gluon divergences that couple via anomaly pole chirality transitions. 
The divergent wee gluon reggeons  carry zero transverse momentum and, because thay couple via anomalies, necessarily have opposite color and space parities. 
Therefore, we refer to them as ``anomalous wee gluons''. 
The infra-red fixed-point safeguards the divergence in large classes
of diagrams of the form shown in Fig.~24, that (after the divergence is subtracted) provide the physical amplitudes of the theory. 
\begin{center}
\epsfxsize=4.8in\epsfbox{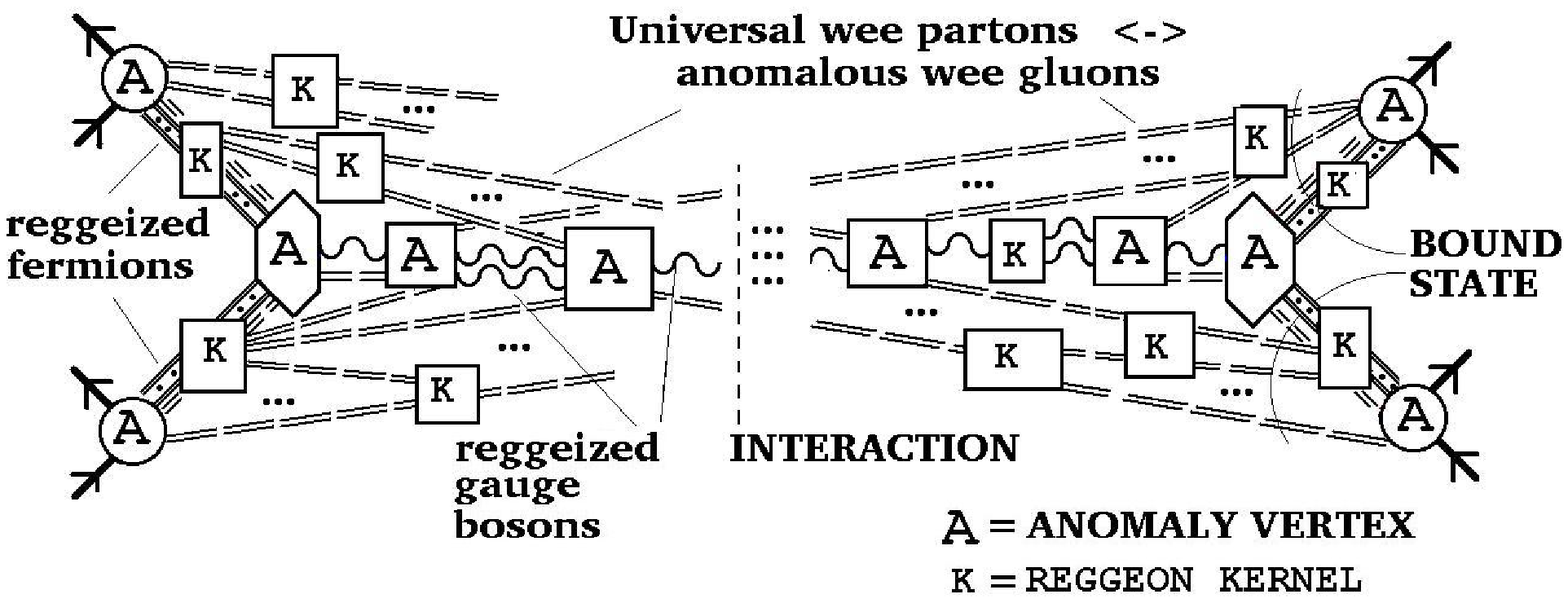}

Figure 24 A Di-Triple Regge Amplitude in Both QCD$_S$ and QUD
\end{center}
Note that, in Fig.~24, the anomaly vertices occur only in external vertices, vertices connecting the bound-state and interaction channels, and vertices
connecting the wee partons in the bound-state channels to interctions in the interaction channel.

All bound-states contain chiral Goldstone boson anomaly poles from the external vertices. Using the intermediate state of the triangle diagram, an anomaly pole can be described as a quark pair state with one of the pair in a zero momentum, negative energy, state. 
Alternatively, using the final state produced, it can be described
as a physical reggeon state containing physical quark reggeons plus anomalous wee gluons. In the first case, bound-states can be described as having only a quark content, while in the second case 
the anomalous wee gluons appear universally in both bound-states and interactions
and obviously play a vacuum-like role. Mesons contain only anomaly poles,
while baryons contain an additional quark reggeon. 

\subhead{9.4 SU(3) Color Restoration}

Only SU(3) color zero states survive when the full color symmetry is restored
and the SU(2) subgroup
is effectively randomized (averaged over) within SU(3).
Anomalous wee gluons survive in SU(2) color subgroups 
and combine with dynamical reggeons to 
produce a color zero projection in each channel. That the Pomeron becomes Critical
can be established by a direct 
construction of the Supercritical Pomeron before the SU(3) symmetry is restored.

The simplest contribution to a ``pion'' scattering amplitude involving
``pomeron'' exchange is shown in Figure 25. As illustrated, the wee gluons appear in both the
pions and the pomeron via 
anomaly vertices involving zero-momentum quark chirality transitions.
In addition, for the pion, there is a longitudinal gluon exchange allowed by the Gribov
ambiguity. The full pion coupling to the pomeron also contains a perturbative
coupling of the dynamical quark and gluon reggeons that are involved.

Clearly, the construction can not be extended down to the finite energy region and so the high-energy 
amplitudes have to be regarded as boundary conditions that, in effect, determine
the spectrum and interactions of the full finite momentum theory.
\begin{center}
\epsfxsize=5.7in\epsfbox{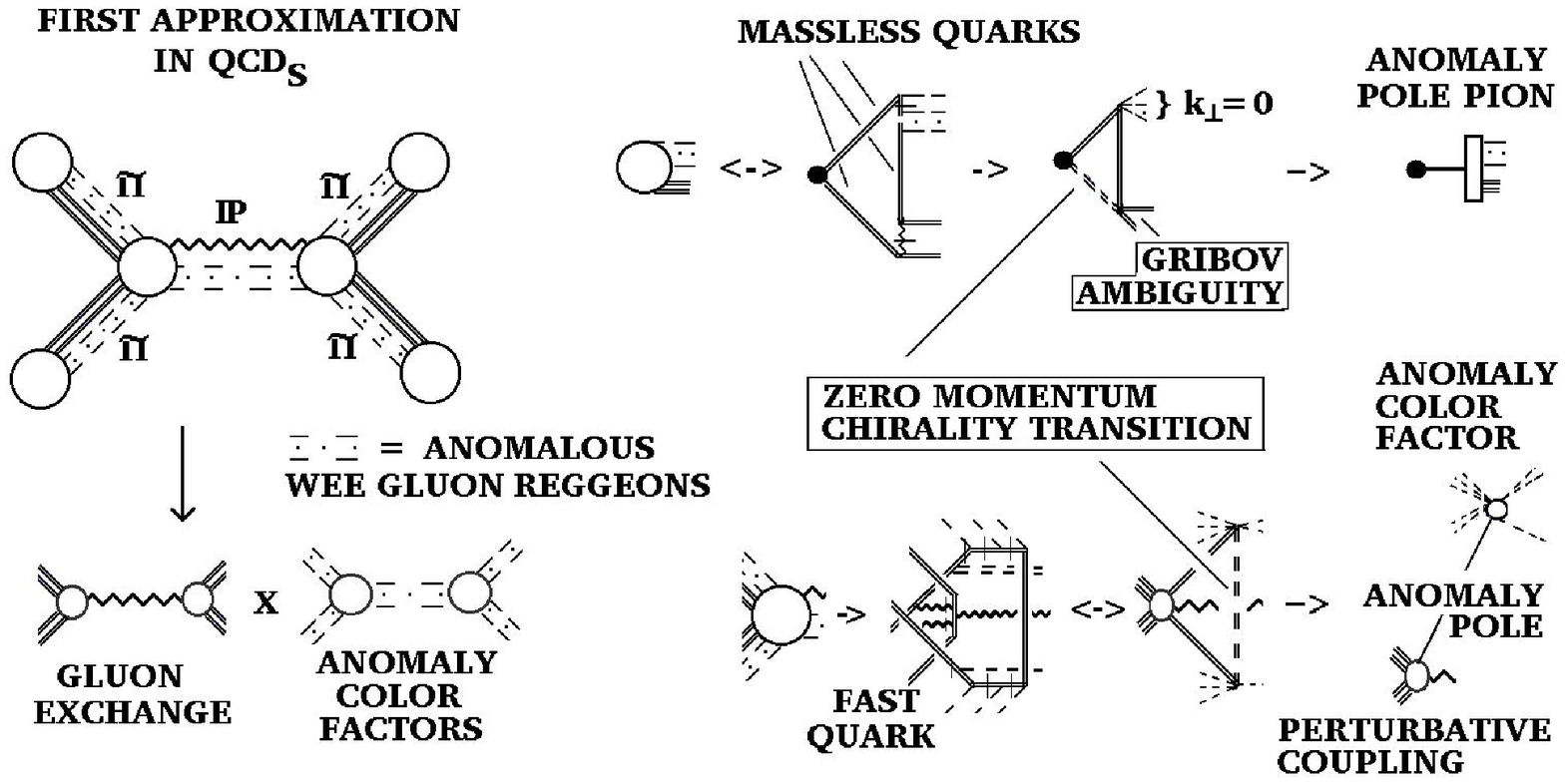}  

Figure 25. Pion Scattering in Massless QCD$_S$
\end{center}
The final outcome can be summarized as follows.
{\it \begin{itemize} 
\item{The only bound-states are pseudoscalar mesons 
and baryons, formed 
\newline (separately) from
 triplet and sextet quarks.}
\item{There are {\bf NO} hybrid sextet/triplet states,} 
\item{There are {\bf NO} glueballs, {\bf NO} BFKL pomeron, and {\bf NO} odderon.}
\item{\it All reggeon states and interactions contain 
infinite sums of anomalous wee gluons coupled via {\bf anomaly color factors}.}
\item{The pomeron is a factorized (isolated) regge pole - a gluon reggeon plus anomalous wee gluons - that becomes Critical via interactions.}
\item{Anomaly color factors
imply {\bf sextet} high-energy cross-sections are much larger
than {\bf triplet} cross-sections}.
\end{itemize}}

The above results are at variance with conventional expectations for high-energy QCD.
There is a dramatic selection of just a minimal part of the degrees of freedom of the underlying field theory and so there are many fewer states (than requiring just confinement and chiral symmetry breaking) and the interaction is much simpler. 
It has to be emphasized, however, that both features are 
{\it strongly suggested\cite{cm,ce} by experiment!}

If sextet pions become the longitudinal components of massive electroweak bosons, the sextet baryons (prosons, antiprosons, neusons and antineusons) 
are the only new states. As we will return to in the last Section, the $\eta_6$ aquires an electroweak scale mass by mixing with the (daughter of the) pomeron.  The $N_6$ neuson is stable because electric charge makes the $P_6$ proson heavier. The very strong, very short range, QCD self-interaction implies the $N_6$'s could form ``dark matter clumps''. 

\subhead{9.5 Large Rapidity Multiparticle States}
 
Fig.~25 illustrates how regge region bound states and interactions are generated in full amplitudes. For the purposes of this paper, it is necessary 
to extend the arguments to the generation of the imaginary parts of amplitudes via large rapidity intermediate states, as illustrated in Fig.~26.
\begin{center}
\epsfxsize=5.5in\epsfbox{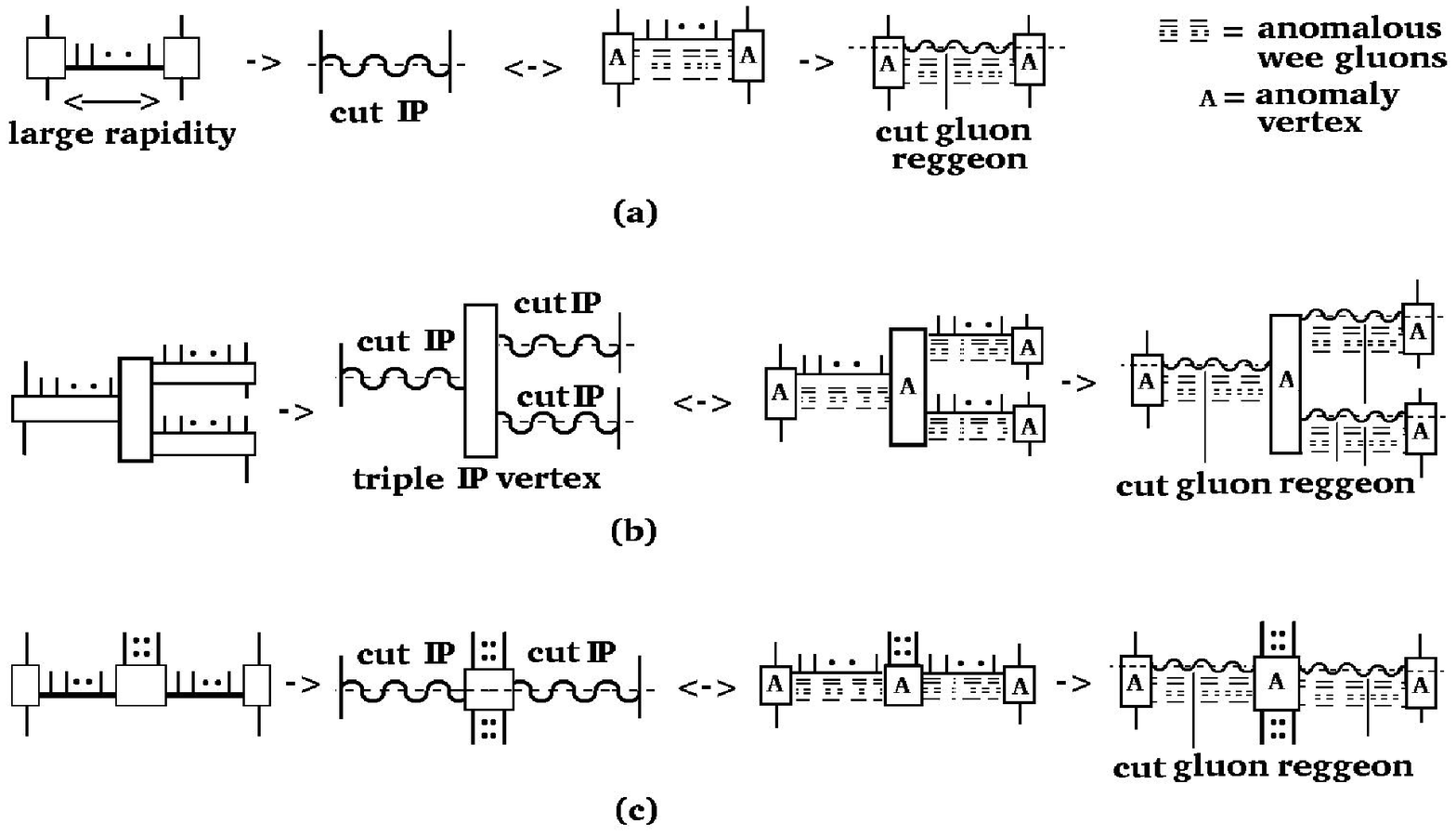}  

Figure 26. Anomalous Wee Gluons in Large Rapidity Soft Hadron States 
\newline (a) the Total Cross-Section (b) Triple Pomeron Processes (c) the
Central Plateau 
\end{center}
The arguments in 
the early Sections of this paper rely on the build-up of the cut pomeron as illustrated in Fig.~26(a). The corresponding formation of triple pomeron
amplitudes is illustrated in Fig.~26(b) and the corresponding central region production is illustrated in Fig.~26(c). In leading logs
the cut gluon reggeon is simply reproduced via the well-known
(and extensively established\cite{vs}) reggeon bootstrap. It is this bootstrap that determines that the gluon reggeon is an isolated regge pole. The anomalous wee gluons have
the highly desirable effect of transforming the reggeon bootstrap into a multiperipheral bootstrap for the isolated regge pole pomeron. 

The main point illustrated by Fig.~26 is that the anomalous
wee gluon divergence also occurs in high multiplicity large rapidity production amplitudes. Even though we can not (straightforwardly) use multi-regge theory to obtain details of individual hadrons in the produced state,
large anomaly couplings are possible if gluon exchanges provide the final coupling
to the fermion loop containing the anomaly. There is, however, even more that is not understood about anomaly couplings of this kind than the couplings that appear in amplitudes of the form of Fig.~24.
For the present, we can say only that consistency demands that 
processes take place as illustrated in Fig.~26. This is an  essential component of the argument that sextet states, most particularly the sextet pions that are the longitudinal components of the electroweak vector bosons,
should couple strongly to the large rapidity, small transverse momenta, states that are produced so prolifically at the LHC. 

\mainhead{10. The QUD Bound-State S-Matrix.}  

Initially, Kyungsik Kang and I discovered\cite{kw} QUD simply by looking for a unified theory that contained the sextet quarks needed for electroweak symmetry breaking. At first we assumed that we had discovered a conventional ``Grand Unified Theory'' and tried for some time to find a dynamical framework that would relate QUD to the Standard Model in a conventional manner. Only after I finally understood the anomaly dynamics of QCD, did I realize that the same dynamics could provide a physical
solution for QUD that I was astonished to find might actually be the Standard Model. I also realized, with great satisfaction, that, although Critical Pomeron production via anomaly dynamics appears to require massless quarks, effective quark masses within bound-states - that do not disturb the dynamics - would be produced by the embedding of QCD$_S$ in QUD. Within QUD, we could have not only massless quarks and leptons that give rise to infra-red anomaly couplings, but also massive constituent 
quarks and massive physical leptons.

\subhead{10.1 QUD}

There is just one unitary gauge theory, that is anomaly free and asymptotically free, that contains both the electroweak and QCD interactions, and has a 
fermion representation that contains
the sextet sector required for electroweak symmetry breaking. It is\cite{kw}
\begin{center} {\bf QUD  $\equiv ~~$ 
SU(5) gauge theory with left-handed   
massless $~~~~~~~~~~~~~~~~$fermions in the representation}
$~~{\bf R \equiv 5 \oplus 15 \oplus 40 \oplus 45^*}$.
\end{center}
Under $SU(3)_C\otimes SU(2)_L\otimes U(1)$
\begin{center}
\openup0.75\jot
{\footnotesize ~~~${\bf 5=[3,1,-\frac{1}{3}]^{3}
+[1,2,\frac{1}{2}]^{ 2}~,~~~~~ 15=[1,3,1]+
[3,2,\frac{1}{6}]^{1}+ \{6,1,-\frac{2}{3}\}^{\#}~,}$}
{\footnotesize $ {\bf 40=[1,2,-\frac{3}{2}]^{3}
+[3,2,\frac{1}{6}]^{2}+
[3^*,1,-\frac{2}{3}]~+~[3^*,3,-\frac{2}{3}]+
\{6^*,2,\frac{1}{6}\}^{\#}+[8,1,1]~,}$}\\
{\footnotesize ${\bf
 45^*=[1,2,-\frac{1}{2}]^{1}+[3^*,1,\frac{1}{3}]
+[3^*,3,\frac{1}{3}]+[3,1,-\frac{4}{3}]
+[3,2,\frac{7}{6}]^{3}+
\{6,1,\frac{1}{3}\}^{\#} +[8,2,-\frac{1}{2}]}$}
\end{center}
As requested, the sextet quarks \{...\}$^{\#}$ have just the right quantum numbers for sextet ``pions'' to provide the longitudinal components of the electroweak vector bosons. Remarkably, both
the triplet quark and lepton sectors, for which no request was made, are amazingly close to the Standard Model. 

Under the same decomposition, the gauge bosons give 

\vspace{0.1in}
\centerline{\footnotesize $ {\bf 24=[1,1,0] ~+ ~[1,3,0]
~+~[3^*,2,\frac{5}{6}] ~+~
[3,2,-\frac{5}{6}]~+~[8,1,0]}$}

\noindent and so they can directly transform triplet quarks, but not sextet quarks, to leptons. 
This is why massive quarks would be generated (by self-energy corrections) 
as a component of a (Standard Model?) effective lagrangian if 
the bound-state massive physical leptons that we obtain below could be introduced by ``integrating out'' the elementary massless leptons.

QUD is real (vector-like) with respect to $SU(3) \otimes U(1)_{em}$ and actually contains QCD$_S$ in it's entirety. There are three generations 
of quarks with charges (2/3,~-1/3) - denoted by superscripts {\bf 1,2,3} - and  corresponding antiquarks with charges (-2/3,~1/3). 
The $SU(2)_L \otimes U(1)$ quantum numbers do not quite match those of the Standard Model quarks and so it is crucial that, as we will come to, the physical $SU(2)_L$ symmetry is distinct from (although related to) the QUD subgroup. The presence of the higher charge quarks and antiquarks will be important for the top physics discussed in the next Section.  
For the three generations of leptons, similarly labeled by superscripts,
the $SU(2)_L \otimes U(1)$ quantum numbers are also not quite right. However, the lepton anomaly is correct - allowing the physical lepton spectrum to emerge within an S-Matrix in which {\it all the elementary leptons and quarks
are confined and massless}. The octet quarks play a fundamental role in the emergence of bound-state lepton and hadron generations with Standard Model quantum numbers. 

\subhead{10.2 Di-Triple-Regge Amplitudes in QUD}

To provide a minimal background for the discussion of the QUD hadronic spectrum in the next Section, we provide a brief recap of what is already only an outline multi-regge construction of QUD states and amplitudes in \cite{arw10}. Again the di-triple regge region is utilised and, as for QCD$_S$, we start with masses for all reggeons 
and a cut-off $\lambda_{\perp}$, the manner of the removal of which is crucial for the production of anomaly vertices.

A combination of {\bf 5$\oplus$5$^*$} 
and {\bf 24} scalar VeV's has to be used to give masses to all the fermions. (The {\bf 5} couples the 15 to the 40$^*$, the {\bf $5^*$} couples the 40 to the 45, and the {\bf 24} couples the 5 to the 45.)
This identifies particle/antiparticle pairs and so determines the chirality transitions that produce triangle diagram anomaly poles in the massless theory. 
For the gauge bosons we use
only {\bf 5$\oplus$5$^*$} 
VeVs, so that a smooth massless limit can be anticipated via complementarity.
The fermion mass scalars are decoupled first, leaving 
chirality 
transitions that break SU(5) to SU(3)$_C\otimes$U(1)$_{em}$, {\it in anomaly vertices only.}
The subsequent successive decoupling of gauge boson scalars gives
global reggeon symmetries $SU(2)_C \rightarrow  SU(4) \rightarrow  SU(5)$.
The last scalar to be removed is asymptotically free, allowing $\lambda_{\perp} \to \infty$ before the SU(5) limit.

The SU(2)$_C$ limit gives amplitudes of the form already shown in Fig.~24. 
The anomaly vertices {\bf A} break the SU(5) gauge symmetry to SU(3)$_C\otimes$U(1)$_{em}$ and contain anomaly poles. The normal reggeon kernels 
{\bf K}  survive when the gauge boson SU(5) symmetry is restored only if they couple reggeon states that have an SU(5) singlet projection that allows them to contribute to infra-red finite amplitudes. At this stage, the anomaly pole bound-state hadrons are massless Goldstone bosons
associated with the separate chiral symmetries of the quark sectors. Via the $5\oplus5^*$ chirality transitions, 
reggeon states containing SU(2)$_C$ anomalous wee gluons produce chiral Goldstones ($\pi_C$'s),that are ``$q\bar{q}~$ mesons", or ``$qq$ nucleons'', or ``$\bar{q}\bar{q}~$ nucleons", with the quarks $q$ being {\bf 3's, 6's,} or {\bf 8's} under SU(3)$_C$. The {\bf 8's} have no SU(3)$_C$ anomaly, but they contain complex SU(2)$_C$ chiral doublets that produce anomaly poles when only SU(2)$_C$ is restored.

Very importantly, the higher charged triplet quarks and antiquarks do not have any chiral symmetry and so they do not form massless chiral Goldstone bosons. Presumably, they either form very massive resonances or do not contribute at all to the physical spectrum.

\subhead{10.3 Massive Electroweak Bosons}

The leading interaction exchanges contain an SU(2)$_C$ singlet massive vector boson (a gluon or a photon) accompanied by anomalous wee gluons. As for QCD$_S$,
the couplings to bound-states contain both an anomaly vertex involving wee gluons 
and a perturbative coupling of dynamical
fermions to the exchanged boson, just as in Fig.~25. Elementary left-handed $W^{\pm}$ and $Z^0$ exchanges, accompanied by wee gluons, are exponentiated to zero via fermion
loop interactions, but the
5$\oplus$5$^*$ chirality transitions provide crucial wee-gluon 
vertex couplings to the ${\pi_C}'s$ ($\sim {\pi_{6}}'s$)
and provide a mass 
that survives the SU(5)
symmetry restoration, while also providing a 
sextet flavor quantum number that prevents the exponentiation. The perturbative coupling of the $W^{\pm}$ and $Z^0$ is retained, but {\it only for doublets} that 
provide the necessary anomaly coupling to the $\pi_{6}$. In the process, the SU(2)$_L$ symmetry becomes, effectively, the SU(2) sextet flavor symmetry.
The {\bf 24} chirality
transitions provide very important wee gluon triple pomeron vertices
with the symmetry properties needed to produce the Critical Pomeron. 

\subhead{10.4 SU(5) Color Restoration}

As first SU(4), and then SU(5), color is restored, the reggeon amplitudes 
for all left-handed gauge bosons that become massless are 
exponentiated to zero. The left-handed bosons survive, however, in the reggeon kernels. As a result, finite amplitudes are produced when the direction of the symmetry breaking due to the initial mass-producing scalars is randomized within SU(5).

The SU(4) singlet fermion reggeon states all have 
{\it octet quark} components ($\pi_8$ or $\eta_8$). They are

\vspace{0.1in} 
\centerline
{{\it Leptons ~-~} $\pi_L~ + ~\pi_8$
~+~ elementary lepton $\to$ 3 generations.}

\vspace{0.1in}
\centerline{{\it ~Mesons ~-~} $\pi_{3,6}~+~ \eta_8~$, ~~ {\it Baryons} $~\leftrightarrow~$ additional quark.$~~~~~$}
\noindent After the SU(5) symmetry is restored, 
the octet quarks form a real SU(3) representation and so the corresponding
infra-red anomaly pole residues vanish in all amplitudes. However, the $\lambda_{\perp} \to \infty$ limit (taken before the 
SU(5) limit) introduces a $k_{\perp} = \infty$ octet anomaly contribution 
in all vertices. In bound-states, dynamical fermion reggeons combine with the 
$k_{\perp} = \infty$ anomaly contribution, and adjoint representation anomalous wee bosons, to give an SU(5) singlet projection that leads to infra-red finite amplitudes. 
Combinations of three elementary fermions, two of which produce an anomaly pole, provide the $SU(2)\otimes U(1)$ representations

\vspace{0.1in}
\centerline{${\bf (2,-\frac{1}{2})_L ~},~~ or ~{\bf (2,\frac{1}{2})_R ~}, ~~ or
 ~{\bf ~~(1,1)_L ~}, ~~or ~{\bf (1,-1)_R} $}

\noindent and so, as a result, there are both leptons and hadrons that form Standard Model generations. However, as we will describe in Section 11, there are only two 
Standard Model hadron generations. The third generation is more complicated. 

In first approximation, the pomeron is a gauge boson reggeon accompanied by odd signature anomalous wee gauge bosons and the photon is a gauge boson reggeon accompanied by even signature anomalous wee gauge bosons. Consequently,
the massless photon is the odd-signature partner of the even signature Critical Pomeron. However,
there is no ``triple-photon'' vertex and the photon does not have the anomaly couplings to hadrons that make the pomeron interaction so much stronger.

\subhead{10.5 The QUD S-Matrix and the Standard Model}

Details of the triplet hadronic sector will be discussed in the next Section, as
part of the discussion of top quark physics. The interpretation of top quark physics is significantly different to the Standard Model. Nevertheless,
I will argue that it may indeed be consistent with the observed 
experimental results while also being considerably more attractive philosophically. 
Consequently, the possibility is very real that the S-Matrix of QUD could provide an amazingly economic underlying unification for the Standard Model in which
\begin{enumerate} 
\item{{\it All elementary fermions are confined.} The zero-momentum chirality 
transitions of the Dirac sea produce {\bf SU(5) $\to$ SU(3)$\otimes$U(1)$_{em}$}
{\it symmetry breaking}.} 
\item{{\it There is no Higgs~!!} All particles, including (Majorana) neutrinos, are bound-states with dynamical masses.}
\item{There are only {\it Standard Model interactions}.}
\item{{\it Physical lepton and hadron states} are equivalent to those of the {\it Standard Model}.}
\end{enumerate}

Beyond the known generations
and the sextet quark sector that, potentially, solves the 
other outstanding mysteries of dark matter and electroweak symmetry breaking, there is only the lepton-like octet quark sector, that is buried in all states in an 
infinite-momentum (light-cone) subtraction role that produces 
leptons and hadrons in Standard Model form, and the exotically charged quarks
that play an important role in top physics. 

Clearly, there is much to be understood and an enormous amount of detail that
has to be given, before we can unambiguously determine how the full anomaly-based QUD multi-regge S-Matrix is built up via the complete set of reggeon diagrams that are a generalization of Fig.~24. In particular, a much more explicit description needs to be given of the symmetry restoring limits. However, it should be clear  
that, in principle at least, all the mixings and mass generation can 
be studied diagramatically, once the anomaly interactions are properly 
categorized.

Although the physics of the QUD S-Matrix is both novel and radical, 
it is consistent with all established Standard Model physics 
and explains many puzzles. Unfortunately, the
multi-regge theory that I use to uncover it is so
erudite that general interest may well require that the physical phenomena discussed in this paper are seen to require, of necessity, the explanations I provide. Moreover, as I have already emphasized, because of it's uniqueness, if it fails to reproduce any established element of the Standard Model S-Matrix,
it is necessarily wrong. 

\subhead{10.6 The Theory Paradigm}

There is clearly a major rewrite of the current theory paradigm that is involved in my advocacy of QUD. I have not directly addressed how it could be that a very weak coupling massless field theory can produce Standard Model coupling scales.
The infra-red fixed-point implies that $\alpha_{\scriptscriptstyle QUD}$ is very small ($\sim 1/120$). While this surely implies small neutrino masses, it also implies that  
Standard Model couplings can not possibly be obtained via QUD evolution (although
the underlying symmetry could produce a high-energy tendency towards unification of the Standard Model couplings in the QUD S-Matrix).
Instead, because multi-regge S-Matrix amplitudes 
are selected by an infra-red divergence, all physical states and amplitudes 
contain infinite sums of wee gauge bosons involving     
anomaly color factors (that can, presumably, be expressed as integral formulae) 
that enhance all interaction
strengths. 
This is how it is possible for elementary leptons to be confined while only small masses are generated by the strength of the underlying interaction.
(The SU(3) interaction is strongly amplified by both color anomaly factors and the triple-pomeron interaction.) 
It follows, however, that QUD can only be physically applicable as an S-Matrix theory without off-shell amplitudes. Again, as for
QCD$_S$, this is possible because of the role played by infra-red anomalies that
appear in the process of generating on-shell physical amplitudes. 

The current theory paradigm anticipates that all physical theories are full quantum field theories with off-shell amplitudes in which the unification of couplings at short distances, via evolution, is a major ingredient. This paradigm continues to dominate the field
even though no four-dimensional field theory has yet been shown to exist outside of perturbation theory (despite a millenium prize being offered for a proof). The existence of only an S-Matrix which is, moreover, accessible semi-perturbatively at infinite momentum, is a much lesser demand. Historically it has been demonstrated that the alternative paradigm, that particle physics is an S-Matrix theory\cite{arw00}, is both self-consistent and physically viable. It is hard, however, to imagine a unification with any form of quantum gravity, even though Einstein gravity could be induced\cite{bh}.

\mainhead{11. Top Quark and Jet Physics}

In the initial stage of the construction of reggeon diagram bound states, as described in the last Section, the chiral symmetry of the six conventionally charged triplet quarks will, in lowest order, produce a large degenerate multiplicity of 
Goldstone bosons. As the complexity of the
amplitudes is built up and masses are aquired via the interactions, the 
differing quantum numbers will separate the states involving 
the various quarks that will, therefore, acquire a wide range of effective constituent masses.
That the massless triplet quarks that appear in $R$ are not aligned
in generations is very good for the ultimate emergence of a physically realistic particle mass spectrum, but it means that at first sight they appear to be very different to the commonly identified massive quarks of the Standard Model. Fortunately, as I will now argue, when the full implications of the bound-state construction are taken into account this is not the case. 

\subhead{11.1 The Triplet Quarks and Antiquarks}

The conventionally charged quarks can be listed as follows. There are two ``Standard Model'' generations
\begin{center}
\parbox{5in}{~SU(2) doublet $~~\longleftrightarrow ~~
(3,\frac{2}{3}),~ (3,-\frac{1}{3}) ~\equiv~ [3,2,\frac{1}{6}]
{\large \bf ~\in  15}$ ~~~~~~ \{G1\}
\newline $~$
\newline $~$ SU(2) doublet $~~\longleftrightarrow ~~(3,\frac{2}{3}),~ (3,-\frac{1}{3}) ~\equiv~ [3,2,\frac{1}{6}] {\large \bf ~\in  40}$
~~~~~~ \{G2\}}
\end{center}
\noindent and also an unconventional ``third generation''
\begin{center}
\parbox{4in}{~
\newline $~$ SU(2) singlet $~~~\longleftrightarrow ~~
~~~(3,-\frac{1}{3}) ~\equiv ~[3,1,-\frac{1}{3}] {\large \bf ~\in 5}$
\newline $~$
\newline $\in~$ SU(2) doublet $~~\longleftrightarrow ~~(3,\frac{2}{3}) ~\in [3,2,\frac{7}{6}] {\large \bf ~\in 45^*},$
} \parbox{0.5in}{\huge \} } \parbox{0.3in}{ $~~$
\newline \{G3\} }
\end{center}

Similarly, the conventionally charged antiquarks can be listed as follows. There are two, ``almost identical generations'' which, at first sight, do not look like Standard Model antiquarks 
\begin{center}
\parbox{5in}{$\in~$ SU(2) triplet $~\longleftrightarrow ~(3^*,-\frac{2}{3}),~(3^*,\frac{1}{3})\in [3^*,3,-\frac{2}{3}] {\large \bf ~\in   40}$, ~~~~~~ \{AG1\}
\newline $~$ 
\newline $\in~$ SU(2) triplet $~\longleftrightarrow 
~(3^*,-\frac{2}{3}),~ (3^*,\frac{1}{3}) \in [3^*,3,\frac{1}{3}] {\large \bf ~\in  45^*}$, ~~~~~~ \{AG2\}} 
\end{center}
There is also a ``conventional third generation''
\begin{center}
\parbox{5.9in}{
$~$ SU(2) singlets $\leftrightarrow (3^*,-\frac{2}{3}) \equiv [3^*,1,-\frac{2}{3}] {\large \bf ~\in  40}$, ~$(3^*,\frac{1}{3})\equiv [3^*,1,\frac{1}{3}] {\large \bf ~\in  45^*}$ ~ \{AG3\}}
\end{center}

\subhead{11.2 The Top Quark}

The most remarkable feature of the quark listing is the presence 
of a ``top quark'' in the ``unconventional third generation'' \{G3\}. The charge 2/3 quark forms an SU(2)$_L$ doublet with an exotic quark belonging to a set of four exotics. 
\begin{center}
$~$ \parbox{0.8in}{\bf exotics \huge{\{ }} \parbox{4.9in}{quarks $~\leftrightarrow ~
(3,\frac{5}{3}) ~\in [3,2,\frac{7}{6}] {\large \bf ~\in 45^*}, ~~~
(3,-\frac{4}{3}) ~\equiv ~[3,1,-\frac{4}{3}] {\large \bf ~\in 45^*}$

\vspace{0.1in}
antiquarks $\leftrightarrow ~(3^*,-\frac{5}{3}) \in [3^*,3,-\frac{2}{3}] {\large \bf ~\in  40}, ~~(3^*,\frac{4}{3}) \in [3^*,3,\frac{1}{3}] {\large \bf ~\in  45^* }$
}
\end{center} 
As a result, the top quark will have a physical electroweak coupling to the exotic quark sector 
which, as we noted earlier, has no chiral symmetry and so will not form, initially massless, bound states. Consequently, the mixing with this sector will, presumably, destabilize any low mass bound states involving the top 
quark. Before discussing the fate of the ``bottom'' quark that is also 
present in the unconventional third generation \{G3\}, we first discuss crucial properties of the antiquarks.

\subhead{11.3 Mixing of the Physical Antiquarks}

At first sight, the two ``almost identical generations'' of antiquarks,
\{AG1\} and \{AG2\}, appear to have  wrong weak interaction quantum  numbers. Each pair is part of an 
SU(2) triplet containing an exotic quark that does not contribute to bound states.
However, as elaborated in the previous Section, the SU(2)$_L$ carried by 
the physical electroweak bosons is such that they
couple only to doublet SU(2) fermions and so the antiquark
pairs actually do have the right (singlet) quantum number. Moreover, although the SU(2) subgroup of the gauge symmetry is not the weak interaction symmetry group, it is still actively part of the full SU(5) reggeon interaction kernels. Consequently, the triplet quantum numbers of both antiquark pairs will actually result in a mixing of the two generations, \{AG1\} and \{AG2\}, in a manner that could produce the desired mixing of the two lowest mass generations of physical antiquarks. 

The quarks corresponding to the two mixed generations of antiquarks are determined by the initial particle/antiparticle matchings and {\it so will not simply be} the two ``Standard Model generations'', \{G1\} and \{G2\}, listed above. 
Although the ``bottom'' quark which appears in the ``unconventional third generation'' \{G3\} has no SU(2) quantum numbers, because of the 
initial {\bf 24} VeV, it's {\it antiparticle is a combination of the antiquark} in the ``conventional third generation'' \{AG3\} and the second of the ``almost identical generations'', i.e. \{AG2\}. Consequently, the physical charge -1/3 quark states corresponding to the physical antiquarks that we have identified 
as forming the two lightest generations, are linear combinations
of those in the ``Standard Model generations'', \{G1\} and \{G2\}, and the ``bottom'' quark from \{G3\}.

\subhead{11.4 The Physical Bottom Quark}

Assuming that the mixed antiquark generations \{AG1\} and \{AG2\} correspond to the physical antiquarks of the lightest mass  generations, the ``bottom'' antiquark in the conventional third generation \{AG3\} will be the physical bottom antiquark. The physical bottom quark will be the corresponding antiparticle and so will also
be a linear combination of quarks in the ``Standard Model generations'' 
\{G1\} and \{G2\} and the \{G3\} quark initially identified as ``bottom'' above.

Consequently, even though it has no direct isodoublet partner, the physical 
bottom quark will have an isodoublet coupling to the weak interaction.
An implication of the associated generation structure of the physical quarks and antiquarks is that a linear combination of the 
physical charge -1/3 quarks has no coupling to the weak interaction.
Obviously, it is important to determine what the experimental
consequences of this would be. 

That the bottom quark is not part of a simple third generation forming a Standard Model doublet, as in the Standard Model, is possible only because of the different anomaly cancelation that is involved. In the underlying SU(5)
theory the anomaly cancelation involves both sextet and octet quarks in addition to
the triplet quarks. If there is an anomaly cancelation for the physical SU(2)$_L$
symmetry, in the QUD S-Matrix, then it must necessarily  involve both triplet and sextet quarks. Therefore, it may not be surprising if, as we discuss next, a sextet resonance is substituted for 
the direct top quark production that is required for anomaly cancelation in the Standard Model.

\subhead{11.5 Top Production Via the $\eta_6$}

If we turn to the sextet chiral symmetry then, in addition to the ``sextet pions''
that become the longitudinal component of the electroweak vector bosons, we might
also expect to find a sextet flavor singlet pseudoscalar, the $\eta_6$, that is a Goldstone boson resulting from the axial U(1) symmetry. If the sextet pions are compared with the Standard Model Higgs scalars that give the vector bosons their masses, then the $\eta_6$ compares directly with the left-over scalar, the ``Higgs'', that is searched for as the missing element of the Standard Model. It can, therefore, be regarded as the answer, in the QUD S-Matrix, to the 
frequently repeated argument that if there is no Standard Model Higgs boson, then there must be something else that plays a similar role. The $\eta_6$ plays a simiilar role to the Higgs in terms of it's relationship to the longitudinal components of the electroweak vector bosons, It does not, however, duplicate the role of the Higgs in providing general masses.

In fact, the chirality
transistions break the U(1) symmetry and, as a manifestation of this breaking,
the anomalous color parity of the pomeron allows the $\eta_6$ to mix with the first pomeron daughter - associated with the non-leading exchange\cite{ks} of a reggeized gluon. As a result, the $\eta_6$ appears only as an electroweak scale resonance and the
daughter of the pomeron does not produce a large low energy interaction
(that would contradict the arguments of Section 2 based on the absence of  
such an interaction between the triplet and sextet sectors).

The mixing with the pomeron (daughter) implies that the quark component of the $\eta_6$ will effectively
be produced primarily via gluon production, just as is conventionally 
assumed to be the case for the Standard Model top quark. Moreover, the $\eta_6$ will also decay in part via pair production of the non-resonance forming top quark discussed above (always with anomalous wee gluons in attendance). It is natural, therefore, to propose that the observation
of a $t\bar{t}$ ``threshold'' at the Tevatron is
actually the observation of the $\eta_6$ (i.e. the ``sextet sector
Higgs''). 

It is commonly argued that the (Standard Model) top quark, once produced, has no time to interact and form a hadronic resonance before it decays. This implies that, necessarily, it is directly produced. However, since it is only detected by isolating potential final state events and eliminating backgrounds, there is no evidence for or against the involvement of a resonance. The appearance of a resonance at the $t\bar{t}$ threshold mass in the Z pair cross-section would be the first direct evidence.
A key consequence of the involvement of the $\eta_6$ could be that, since it is a pseudoscalar
carrying negative parity, interference with the positive parity background
would produce an asymmetry of the kind observed\cite{CDFt} by CDF.

Both theoretically and philosophically, it would surely be attractive if
an electroweak scale mass, i.e. 330 GeV, is explained
as the (dynamical) mass of a sextet
quark/antiquark bound state (implying that a sextet baryon should have a mass of $\sim $ 500 GeV, as we have assumed), rather than as twice the value of a 
lagrangian parameter of the triplet quark sector. 
The logical paradox that the mass of a 
colored, confined, state is a well-defined physical observable, 
would also be avoided altogether. 

\subhead{11.6 Non-Perturbative $\eta_6$ Decay Modes}

Since many experimental features would
be similar to the conventional picture, another key signal 
could be the observation of, one or more, non-perturbative QCD decay modes
for the $\eta_6$.
To discuss these decay modes, we start by exploiting the parallel between the \{$\pi^{\pm}_6,\pi^0_6,\eta_6$\} sextet
states, corresponding to \{$W^{\pm},Z^0,\eta_6$\},
and the familiar \{$\pi^{\pm},\pi^0,\eta$\} triplet quark states.
Although the width is most likely large, we obviously take 
$m_{\eta_6} \sim 2 m_{top} \sim$ 330 GeV.
In this case, the relative couplings and masses of
the vector mesons, and the photon, imply that the 
primary non-perturbative decay mode should be (in parallel with 
$\eta~\to~ \pi^+~\pi^-~\pi^0$) 
$$
\eta_6~~\to~~ W^+~W^-~Z^0 
\auto\label{dk1}
$$
which, when $Z^0 \to b\bar{b}$, would give the same final state as $t\bar{t}$. 
The next most significant mode 
$$
\eta_6~~\to~~ Z^0~Z^0~Z^0 
\auto\label{dk2}
$$
(in parallel with $\eta~\to~ \pi^0~\pi^0~\pi^0$) 
should have a smaller branching ratio, because of the larger $Z^0$ mass. 
In fact, (\ref{dk2}) would  
be indistinguishable from (\ref{dk1}) when the $Z^0$'s decay hadronically, as they
do most of the time. 

If the parallel between the \{$\pi^{\pm}_6,\pi^0_6,\eta_6$\} sextet
states components of \{$W^{\pm},Z^0,\eta_6$\} and the
 \{$\pi^{\pm},\pi^0,\eta$\} triplet quark states were complete then 
the decay mode 
$$
\eta_6~~\to~~ Z^0~Z^0
\auto\label{dk21}
$$
on which we focus in Sections 7 and 8 
would parallel $ \eta \to \pi^0 \pi^0$ and so would 
be forbidden, as a non-perturbative QCD decay, by parity. However,
because the final states are vectors and not pseudoscalars this argument does not hold. The decay can also proceed as an electromagnetic decay via an anomaly, in parallel with  $\eta \to \gamma \gamma$. However, for the decay to appear as a significant component of the ZZ cross-section, the non-perturbative process should, presumably, dominate.

Because the $\eta_6$ mass is so large, 
decay modes that necessarily require an electromagnetic coupling, such as
$$
\eta_6~~ \to~~ W^+~W^-~\gamma~, ~~~Z^0 ~Z^0~\gamma ~, ~~~ Z^0 ~\gamma~\gamma~, 
~~~ \gamma~\gamma
\auto\label{dk3}
$$
would be expected to have smaller branching ratios but should 
be present at some level. 

\subhead{11.7  Jet Physics at the Top Mass Scale}

If the top quark events are produced by the $\eta_6$ then ``$~m_{top}~$''
would be the sextet dynamical mass scale above which $\alpha_s$ should not
evolve. In this case there should surely have been a jet excess at 
the Tevatron which, at least in part, can be interpreted\cite{CDF} as non-evolution 
of $\alpha_s$ beyond $E_T \sim$``$~m_{top}~$''. 
The increasing entry of sextet sector states into the dynamics
implies that the ``excess'' should continue to grow as $E_T$ increases and that
there should be an 
enrichment of jets with $M_{jet} \approx M_{W/Z}$.

As illustrated in Fig.~27, early presentations of CDF data did, indeed, have the 
\vspace{0.01in}
\parbox{2.8in}{
\epsfxsize=2.6in
\epsffile{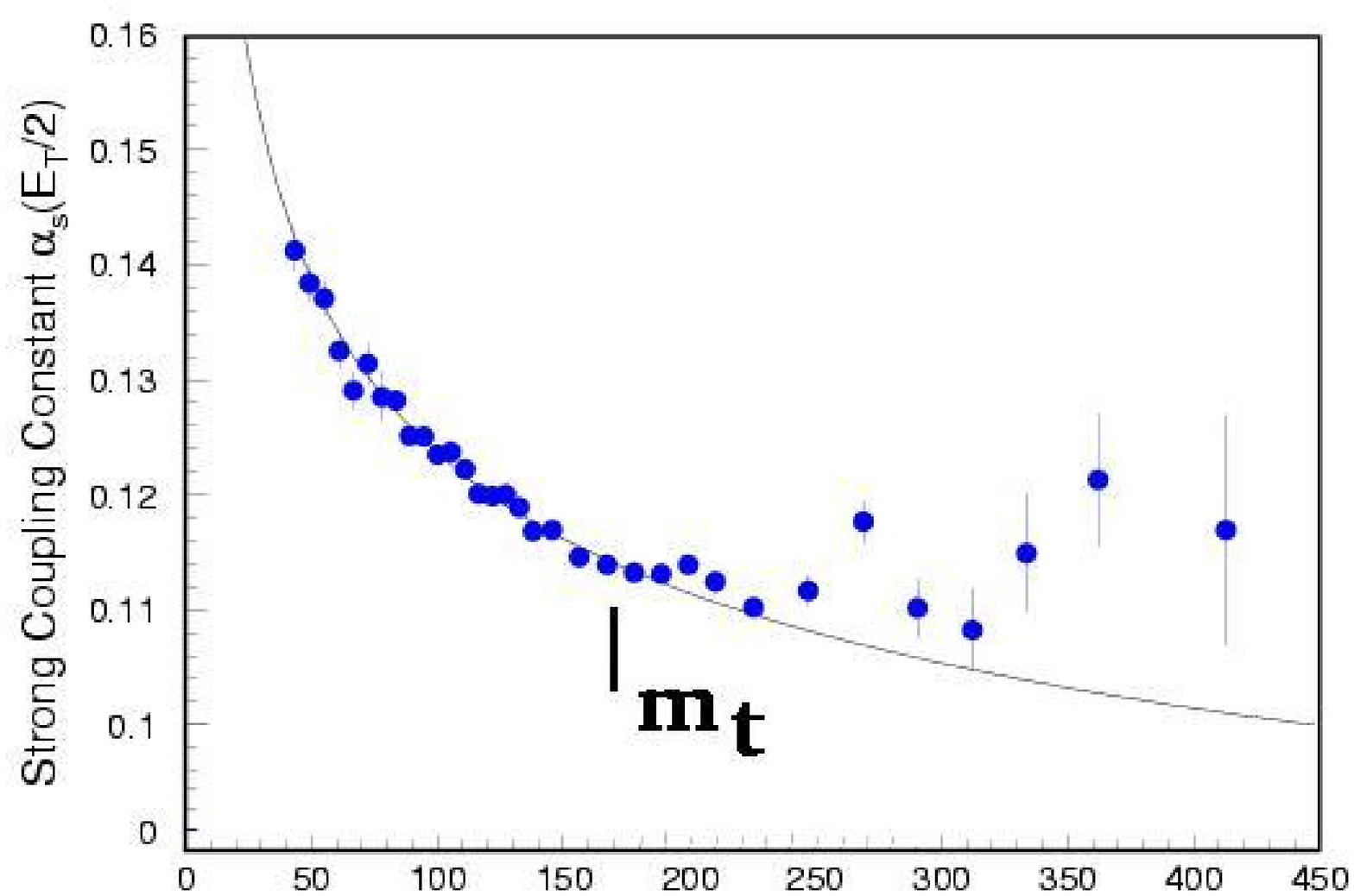}}
\parbox{3in}{\epsfxsize=2.8in
\epsffile{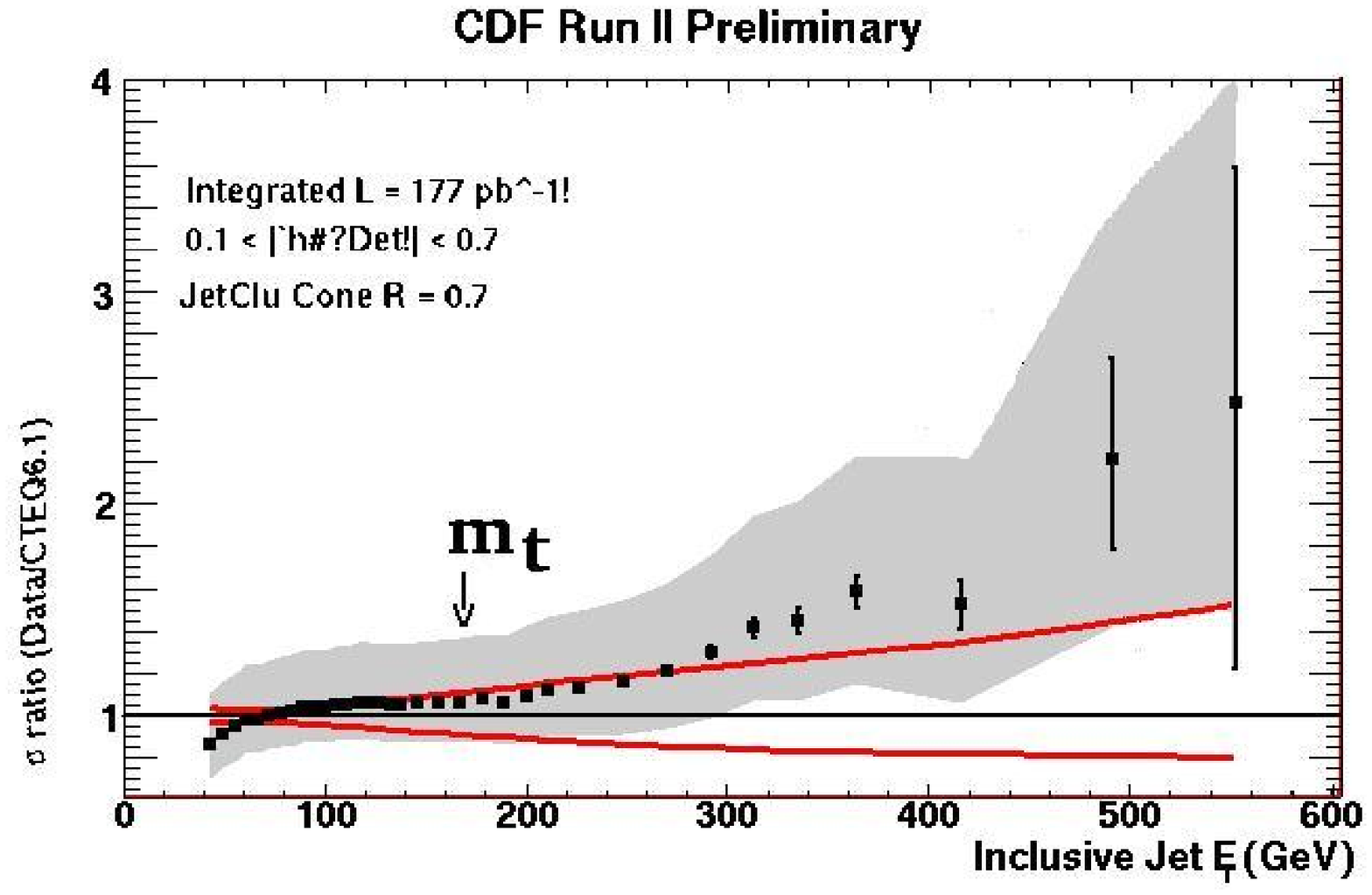}
\newline $~$}

\begin{center}
(a) \hspace{2.7in} (b)

Figure 27. Early CDF data suggesting new physics appears above $E_T \sim m_t$ 
\newline (a) Run 1 measurement\cite{CDF} of $\alpha_S$ (b) Run 2 preliminary jet cross-section\cite{CDFe}.
\end{center}
interpretaton that new QCD physics is entering at $E_T \sim m_{top}$, just as I 
anticipate should be the case. Unfortunately, as is well known, more sophisticated jet algorithms were searched for and combined with appropriately modified gluon distributions until the conclusion was reached that 
there is no significant jet excess. 

It is interesting that Fig.~27(b) appeared in all early talks on the Run 2 jet cross-section, but it was excised from almost all published versions of the talks
and also from all later talks. It was also declared that since there is no independent measurement of the gluon distribution above $E_T \sim $ 100 GeV,
extraction of $\alpha_S$ at higher $E_T$ is not possible. Indeed, if such an analysis has been carried out, it has not been published. Of course, it can 
be argued that if there really is new physics in the QCD jet cross-section, it will eventually emerge at high-enough energy, no matter what jet formalism is used. So, should it show up in the LHC cross-section? This brings us back to the ``QCD phenomenology'' issue raised at the beginning of the paper.

It has, apparently, become accepted that it is necessary\cite{dmg}
for non-perturbative elements to be included in all applications of perturbative QCD
at the LHC, As a result, ``non-perturbative'' hadronization corrections that 
might be obscuring large $E_T$ behavior due to new physics are routinely included in jet cross-section analyses. Fig.~28 shows an early study\cite{soy} of the LHC ``QCD jet cross-section'',  
in which an appropriate hadronization factor is included.
It is apparent, from Fig.~28(a), that the hadronization effect does not disappear until very large $p_{\perp}$.
Correspondingly, if the pure NLO results are uniformly moved down to coincide with the data at lower $p_{\perp}$, as in Fig.~28(c), it appears that, in fact, an excess above $p_{\perp} \sim m_{top}$ does emerge just as in Fig.~28(b), even though the data that is used is preliminary and has very large errors.

\begin{figure}
\begin{center}
\epsfig{file=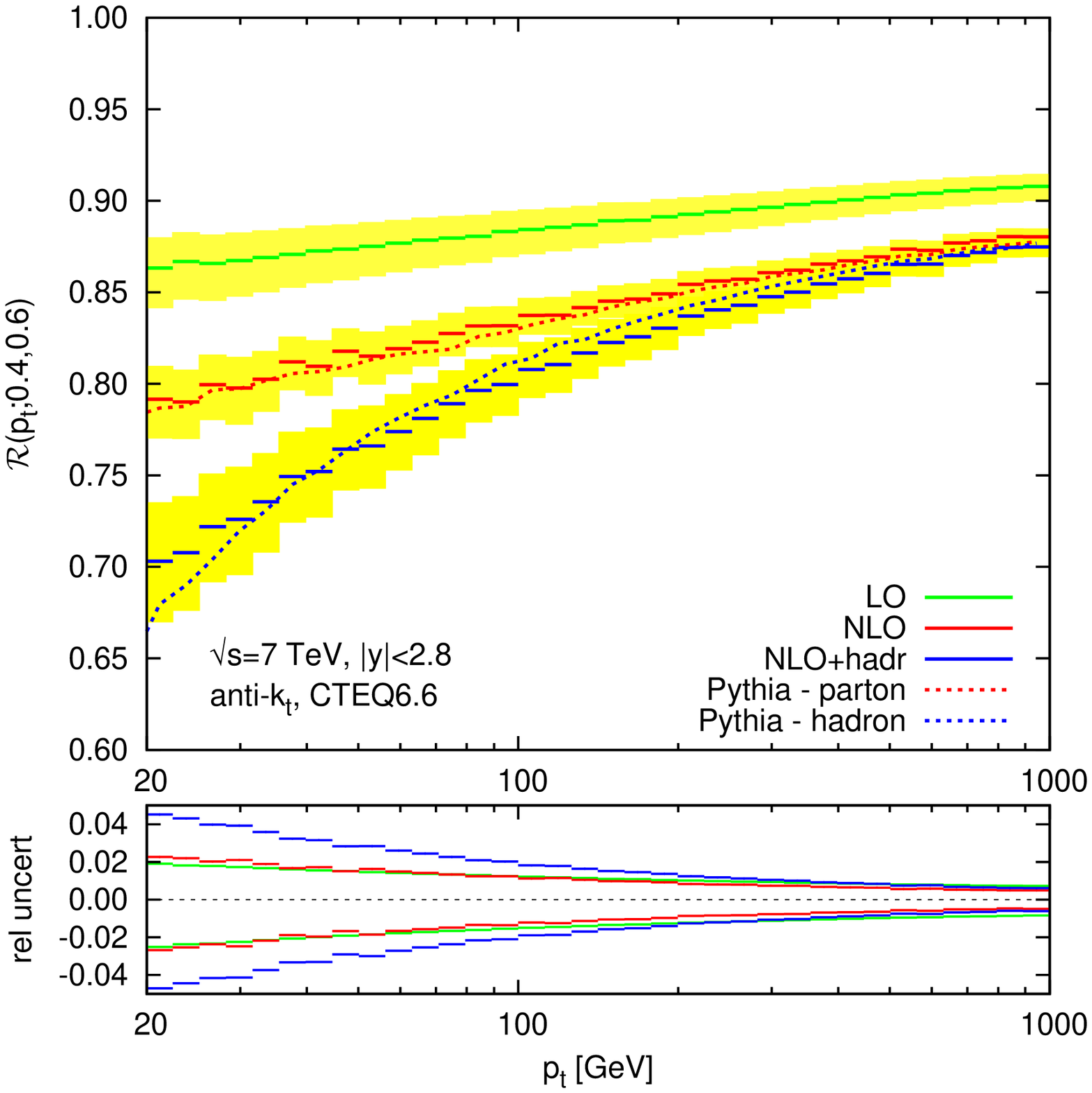, width=2.3in, angle=270}
\epsfig{file=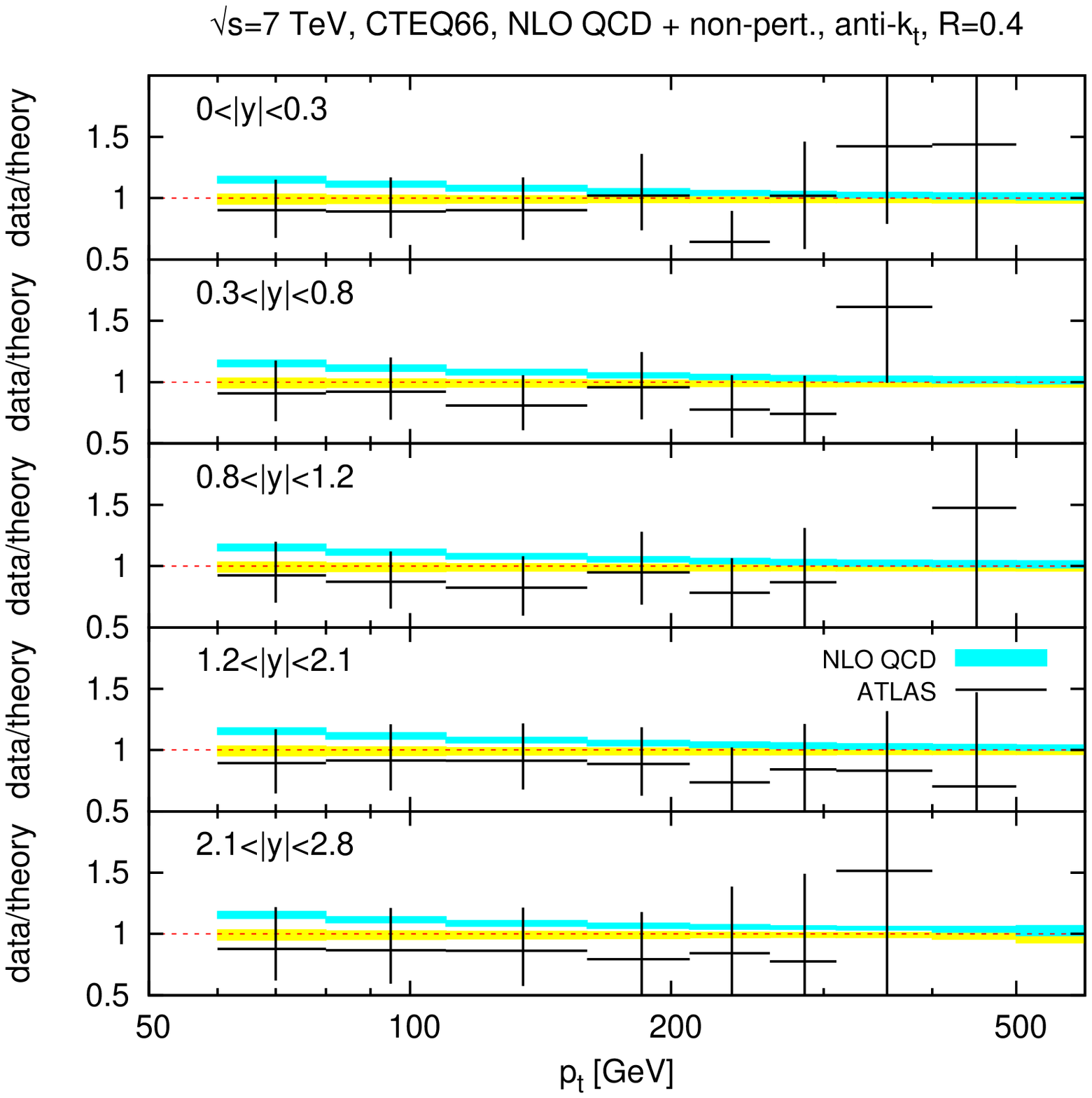, height=2.3in, angle=270}

$~$ 

(a) \hspace{2.4in} (b)

\epsfig{file=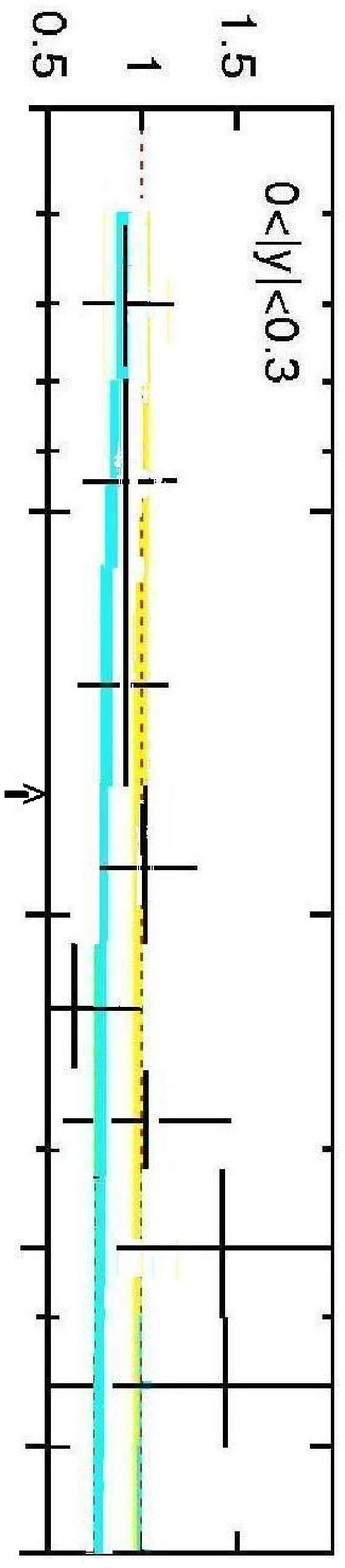, width=0.8in, angle=90}

\vspace{-0.1in}
\centerline{$m_{top}$}

$~$

(c)

$~$

Figure 28.  Preliminary ATLAS Jet Cross-Sections Compared With QCD + Non-Perturbative Hadronization (a) Comparison of LO, NLO and NLO + Had computations
(b) Full Comparison with Data (c) Comparison With Pure QCD  
\end{center}
\end{figure}

It is essential to emphasize that, without an understanding of the interplay 
between confinement and perturbation theory, there can not be a fundamental basis
for computing non-perturbative corrections to jet cross-sections. The ``parton model'' assumptions used in such an endeavor go way beyond the realm of proven factorization theorems. Consequently, to study jet physics in a 
manner that could, potentially, discover new QCD physics, it is essential to have a formalism in which a 
generalized (Feynman) parton model is derived simultaneously with confinement.
This is what is promised by QUD. 

That the BFKL pomeron is absent and that, instead, the pomeron is dominated by small $p_{\perp}$ physics, implies that the corrections to leading-order perturbative results will be correspondingly different. In particular, small $p_{\perp}$ high multiplicity states will be much more important. As a result, it may very well be that new QCD physics will show up very obviously 
when $E_T > m_{top}$. This could be the case, not only in all jet cross-sections, but also in all cross-sections involving both jets and electroweak vector bosons. 


\end{document}